\documentclass[fleqn]{emulateapj}

\usepackage{graphicx}	
\usepackage{amsmath}	
\usepackage{amssymb}	
\usepackage{bm}		
\usepackage[hidelinks=true]{hyperref}
\hypersetup{
  colorlinks   = true, 
  urlcolor     = blue, 
  linkcolor    = blue, 
  citecolor   = blue 
}

\usepackage[latin1]{inputenc}
\usepackage[T1]{fontenc}

\usepackage{newtxtext,newtxmath}

\shorttitle{GOGREEN: Evidence for quiescent disk excess in clusters at $1.0<z<1.4$}
\shorttitle{Chan et al.}

\begin{document}

\title{The GOGREEN Survey: Evidence of an excess of quiescent disks in clusters at $1.0<z<1.4$}
\author{Jeffrey C.C. Chan$^{1}$, Gillian Wilson$^{1}$,  Michael Balogh$^{2,3}$,  Gregory Rudnick$^{4}$, Remco F. J. van der Burg$^{5}$,  Adam Muzzin$^{6}$, \\
Kristi A. Webb$^{2,3}$, Andrea Biviano$^{7,8}$, Pierluigi Cerulo$^{9}$, M. C. Cooper$^{10}$, Gabriella De Lucia$^{7}$, Ricardo Demarco$^{9}$, \\
Ben Forrest$^{1}$, Pascale Jablonka$^{11,12}$, Chris Lidman$^{13}$, Sean L. McGee$^{14}$, Julie Nantais$^{15}$, Lyndsay Old$^{16}$, Irene Pintos-Castro$^{17}$, \\ 
Bianca Poggianti$^{18}$, Andrew M. M. Reeves$^{2,3}$, Benedetta Vulcani$^{18}$, Howard K.C. Yee$^{17}$,  and Dennis Zaritsky$^{18}$}

\email{E-mail: jeffreyrdcs@gmail.com}

%
\affil{$^{1}$Department of Physics \& Astronomy, University of California, Riverside, 900 University Avenue, Riverside, CA 92521, USA \\
$^{2}$Department of Physics and Astronomy, University of Waterloo, Waterloo, Ontario N2L 3G1, Canada \\
$^{3}$Waterloo Centre for Astrophysics, University of Waterloo, Waterloo, Ontario, N2L3G1, Canada \\
$^{4}$Department of Physics and Astronomy, The University of Kansas, Malott room 1082, 1251 Wescoe Hall Drive, Lawrence, KS 66045, USA\\
$^{5}$European Southern Observatory, Karl-Schwarzschild-Str. 2, 85748, Garching, Germany\\
$^{6}$Department of Physics and Astronomy, York University, 4700 Keele Street, Toronto, Ontario, ON MJ3 1P3, Canada\\
$^{7}$INAF-Osservatorio Astronomico di Trieste, via G. B. Tiepolo 11, 34143, Trieste, Italy\\
$^{8}$IFPU - Institute for Fundamental Physics of the Universe, via Beirut 2, 34014 Trieste, Italy \\
$^{9}$Departamento de Astronom\'ia, Facultad de Ciencias F\'isicas y Matem\'aticas, Universidad de Concepci\'on, Concepci\'on, Chile \\
$^{10}$Department of Physics and Astronomy, University of California, Irvine, 4129 Frederick Reines Hall, Irvine, CA 92697, USA\\
$^{11}$Laboratoire d'astrophysique, \'Ecole Polytechnique F\'ed\'erale de Lausanne (EPFL), 1290 Sauverny, Switzerland  \\
$^{12}$GEPI, Observatoire de Paris, Universit\'e PSL, CNRS, Place Jules Janssen, F-92190 Meudon, France \\
$^{13}$The Research School of Astronomy and Astrophysics, Australian National University, ACT 2601, Australia \\
$^{14}$School of Physics and Astronomy, University of Birmingham, Edgbaston, Birmingham B15 2TT, England \\
$^{15}$Departamento de Ciencias F\'isicas, Universidad Andres Bello, Fernandez Concha 700, Las Condes 7591538, Santiago, Regi\'on Metropolitana, Chile\\
$^{16}$European Space Agency (ESA), European Space Astronomy Centre (ESAC), E-28691 Villanueva de la Ca\~nada, Madrid, Spain\\
$^{17}$Department of Astronomy and Astrophysics, University of Toronto 50 St. George Street, Toronto, Ontario, M5S 3H4, Canada\\
$^{18}$INAF-Padova Astronomical Observatory, Vicolo dellÕOsservatorio 5, I-35122 Padova, Italy\\
$^{19}$Steward Observatory and Department of Astronomy, University of Arizona, Tucson, AZ, 85719}



\begin{abstract}
We present results on the measured shapes of $832$ galaxies in 11 galaxy clusters at $1.0<z<1.4$ from the GOGREEN survey. We measure the axis ratio ($q$), the ratio of the minor to the major axis, of the cluster galaxies from near-infrared \textit{Hubble Space Telescope} imaging using S\'ersic profile fitting and compare them with a field sample. We find that the median $q$ of both star-forming and quiescent galaxies in clusters increases with stellar mass, similar to the field.  Comparing the axis ratio distributions between clusters and the field in four mass bins, the distributions for star-forming galaxies in clusters are consistent with those in the field.  Conversely, the distributions for quiescent galaxies in the two environments are distinct, most remarkably in $10.1\leq\log(M/{\rm M}_{\odot})<10.5$ where clusters show a flatter distribution, with an excess at low $q$. Modelling the distribution with oblate and triaxial components, we find that the cluster and field sample difference is consistent with an excess of flattened oblate quiescent galaxies in clusters. The oblate population contribution drops at high masses, resulting in a narrower $q$ distribution in the massive population than at lower masses. 
Using a simple accretion model, we show that the observed $q$ distributions and quenched fractions are consistent with a scenario where no morphological transformation occurs for the environmentally quenched population in the two intermediate mass bins. Our results suggest that environmental quenching mechanism(s) likely produce a population that has a different morphological mix than those resulting from the dominant quenching mechanism in the field.

\end{abstract}




\section{Introduction}
\label{sec:Introduction}
It is well-established that the environment of a galaxy plays a crucial role in its evolution. In the local Universe, the galaxy population in high-density environments comprises mainly galaxies that have ceased forming stars. The dominance of quiescent galaxies in groups and clusters, as reflected by the higher quiescent fraction at fixed stellar mass compared to the field \citep[e.g.,][]{Baloghetal2004, Baldryetal2006, Wetzeletal2012}, suggests that there are physical processes that correlate with the environment to suppress star formation. These quiescent galaxies are composed of mostly early-type objects, as opposed to the late-type morphologies seen in star-forming galaxies \citep[e.g.,][]{Dressler1980, Postmanetal2005, Holdenetal2007, Bassettetal2013}. This implies a morphological change must have taken place at a certain evolutionary stage. Despite the focused effort in recent decades, the physical processes that drive the quenching of star formation and the morphological transformation of the galaxies in dense environments are not yet fully understood.

Detailed studies of the properties of the galaxy population in clusters and groups in the local Universe and low redshifts have revealed various mechanisms that can contribute to environmental quenching \citep[see, e.g.,][for reviews]{BoselliGavazzi2006, BoselliGavazzi2014}. For example, the cut-off of the cold gas accretion from the cosmic web during infall into a massive halo can gradually quench the star formation of a galaxy, as the fuel slowly runs out \citep[``strangulation'' or ``starvation'',][]{Larson1980, Baloghetal1997, Baloghetal2000}.  Quenching can also occur due to rapid removal of the cold gas in the galaxies when it passes through the intracluster medium (ICM) \citep[``ram pressure stripping'',][]{GunnGott1972} or due to interactions between galaxies with other group or cluster members \citep[``galaxy harassment'', e.g.,][]{Mooreetal1998}. Nevertheless, the relative importance of each of these mechanisms are still not well understood, in part because the efficiency of these mechanisms depends on both the properties of the galaxies (e.g., gas content, star-formation rate) and the environment they are in (e.g., halo mass, ICM density). Environmental quenching at low redshift is shown to be largely separable from quenching driven by mechanisms that act internally in the galaxy \citep[i.e., mass-quenching, e.g.,][]{Pengetal2010}.  One interpretation is that environmental quenching operates independently and does not depend strongly on stellar mass \citep[but see also][]{DeLuciaetal2012, Wetzeletal2013, Fillinghametal2015}.  Nevertheless, there is growing evidence that the situation is very different at $z \gtrsim 1$. Recent works reported a mass dependence in the environmental quenching efficiencies at redshift $z \gtrsim 1$ \citep{Cooperetal2010b, Baloghetal2016, Kawinwanichakijetal2017, Fossatietal2017, Papovichetal2018, PintosCastroetal2019, vanderBurgetal2020}, which suggests that the effects from both classes are no longer separable. This points to a possible change in the dominant environmental quenching mechanism at high redshift \citep{Baloghetal2016}.

The studies of environmental quenching efficiency that were mentioned above mostly rely on measuring the stellar mass function of the star-forming and quiescent galaxy population as a function of redshift and environment. The relative fraction of the two populations provides important constraints on galaxy evolution in different environments.  Additional, complementary, information about the morphologies or structural properties of the galaxy population is also often considered. Many of the proposed environmental mechanisms have unique implications or predictions on the morphology of the galaxies. The most striking example of them all is the stripped gas tails produced by gas removal processes such as ram-pressure stripping, which are easily recognizable by their peculiar morphologies in $H\alpha$ imaging and spectroscopy \citep[e.g.,][]{Gavazzietal2001, Fumagallietal2014b, Yagietal2015, Sheenetal2017}.  Significant efforts have been put in searching for galaxies that exhibit tails reminiscent of a debris trail, known as ``jellyfish'' galaxies, in groups and clusters at low redshifts \citep[e.g.,][]{Poggiantietal2016, McPartlandetal2016, RobertsParker2020}.  Similarly, galaxies with peculiar morphologies, such as merging pairs, tidal features, and truncated or warped disks, are often treated as the proof of the existence of the corresponding mechanisms (i.e., mergers, harassment, stripping).  The structural properties of galaxies have also provided crucial insights into the evolutionary path of the galaxy population. For example, studies of the quiescent galaxy population in clusters and the field at high redshift have shown that they are on average more compact than their local counterparts of the same mass \citep[e.g.,][]{Trujilloetal2006b, Newmanetal2012, vanderWeletal2014, Chanetal2018, Matharuetal2019}, suggesting that they must have undergone significant evolution in size but only mild growth in mass \citep[but also see][]{Valentinuzzietal2010b, Cooper2012, Lani2013, Poggiantietal2013, Delaye2014}. Repeated minor mergers have been shown to be the primary mechanism that gives rise to the observed size evolution and the inside-out growth of the galaxies \citep[e.g.,][]{Naabetal2009, vanDokkumetal2010, Shankaretal2013, Suessetal2019a}, although the effect of continual arrival of larger quenched galaxies may also play a role \citep[i.e., progenitor bias, e.g.,][]{vanDokkumetal2001, Sagliaetal2010, Carolloetal2013, Poggiantietal2013, Bellietal2015, Matharuetal2020}.

The projected axis ratio (ellipticity) distribution of the galaxy population has also long been used to study their intrinsic structural properties and shapes \citep[e.g.,][]{Sandageetal1970, Franxetal1991, TremblayMerritt1996}. Although individual axis ratios do not carry much information as they are degenerate with the inclination angle, their distribution can be used to infer the intrinsic shape distribution under the assumption of random viewing angles. From studies of the last few decades, it is established that the majority of star-forming galaxies in the local Universe are flattened, oblate systems \citep[e.g.,][]{Ryden2004, PadillaStrauss2008}. \citet{vanderWeletal2014b} showed that this is also true for the more massive $\log(M / {\rm M}_{\odot}) > 10.0$ star-forming galaxies at high redshift, up to $z\sim2.5$.  On the other hand, the projected axis ratio distribution of the early-type population in the local Universe requires a two-component model, which comprises a triaxial set and an oblate set of objects, to well describe its properties \citep[e.g.,][]{TremblayMerritt1996, Holdenetal2012}.  The exceptions are the massive quiescent galaxies with $\log(M / {\rm M}_{\odot}) > 11.0$, where they are preferentially round and can be described by a single triaxial population \citep[e.g.,][]{vanderWeletal2009}.  \citet{Changetal2013} extended such axis ratio analysis to quiescent galaxies at $1<z<2.5$ in the field and found that the fraction of oblate galaxies relative to the total population evolves over redshift.  For massive quiescent galaxies with $\log(M / {\rm M}_{\odot}) > 11.0$, the oblate fraction is almost three times higher at $z>1$.

This two-component picture is also supported by the observed stellar kinematics of the low-$z$ quiescent galaxies \citep[i.e., the slow rotators and fast rotators, e.g.][]{Emsellemetal2011}. Recent integral field spectroscopy studies have shown that the intrinsic shape of a galaxy is correlated to the degree of rotational support. For example, \citet{Weijmansetal2014} found that fast rotators have flattened intrinsic shape distributions similar to spiral galaxies, while slow rotators are likely to be mildly triaxial  \citep[see also][]{Corteseetal2016, Pulsonietal2018}. With a larger sample, \citet{Fosteretal2017} showed that galaxies with higher ``spin'' parameter \citep{Emsellemetal2011} have more flattened intrinsic axis ratios and more likely to be axisymmetric systems.

The projected axis ratio distribution has also been used to study the formation of lenticular galaxies (S0s), which are abundant in local galaxy clusters. For example, \citet{Vulcanietal2011} studied the axis ratio distributions of a sample of early-type galaxies in intermediate redshift clusters ($z\sim0.6$) and compared them to those in local clusters; they found that there are fewer flattened objects in the intermediate redshift sample due to a lower fraction of S0 galaxies. Similar studies also found that the S0 fraction drops rapidly with increasing $z$. By $z\sim0.5$ the fraction of S0 galaxies is found to be $<10\%$ \citep[e.g.,][]{Fasanoetal2000, Postmanetal2005}.  The high occurrence of S0 in low-$z$ clusters, although not fully understood, is generally believed to be due to environmental effects \citep[e.g.,][]{Justetal2010, Johnstonetal2014, Kelkaretal2017}. Since the mass dependence of environmental quenching efficiencies emerges at $z \gtrsim 1$, it is therefore interesting to extend the axis ratio distribution studies to even higher redshift.

In this paper, we investigate the axis ratio distributions of the galaxies in 11 clusters of the Gemini Observations of Galaxies in Rich Early ENvironments survey \citep[GOGREEN;][]{Baloghetal2017,Baloghetal2021} at $1.0<z<1.4$.  This recently completed survey is an imaging and spectroscopic survey targeting 21 known high-redshift overdensities that are representative of the progenitors of the clusters we see today.  The deep spectroscopy and imaging of GOGREEN allows us to study the axis ratio distributions of an unprecedentedly large sample of cluster galaxies in this redshift range. The goal of this work is to study the effect of environment on galaxy structures by comparing the axis ratio distributions of cluster galaxies to those in the general field.  The field comparison sample is taken from the CANDELS \citep{Groginetal2011, Koekemoeretal2011}, and 3D-\textit{HST} Treasury programs \citep{Brammeretal2012, Skeltonetal2014}.

This paper is organized as follows. In Section~\ref{sec:Data} we describe the data set used in this work and present the derivation of structural parameters and other quantities.  We present the results of the axis ratio distributions and describe the procedure and results of the axis ratio modeling in Section~\ref{sec:Results}.  We then explore the relationship between environmental quenching and morphological transformation and discuss the results in Section~\ref{sec:Discussion}. In Section~\ref{sec:Conclusion}, we draw our conclusions.

Throughout the paper, we assume the standard flat cosmology with $H_{0} = 70$~km~s$^{-1}$~Mpc$^{-1}$, $\Omega_{\Lambda} = 0.7 $ and $\Omega_{m} = 0.3$.  Magnitudes quoted are in the AB system \citep{OkeGunn1983}. The stellar masses in this paper are computed with a \citet{Chabrier2003} initial mass function (IMF).

\section{Sample and Data}      
\label{sec:Data}

The cluster sample used in this work is from the GOGREEN survey \citep{Baloghetal2017, Baloghetal2021}. The GOGREEN sample consists of 21 overdensities at $1.0<z<1.5$ spanning a wide range of halo masses, including three clusters from the South Pole Telescope (SPT) survey \citep{Brodwinetal2010, Foleyetal2011, Stalderetal2013}, nine clusters from the Spitzer Adaptation of the Red-sequence Cluster Survey \citep[SpARCS,][]{Wilsonetal2009, Muzzinetal2009a}, of which five were followed up extensively by the Gemini Cluster Astrophysics Spectroscopic Survey \citep[GCLASS,][]{Muzzinetal2012}, and nine group candidates selected in the COSMOS and Subaru-XMM Deep Survey (SXDS) fields.

In this study we focus on 11 GOGREEN clusters at $1.0 < z < 1.4$ that have complete spectroscopic and photometric catalogues at the time of this work\footnote{The one cluster in the twelve GOGREEN clusters that was not included is SpARCS 1033, as deep $K$-band imaging has not yet been obtained at the time of this work.}.  Eight of the clusters were discovered using the red-sequence or the stellar-bump technique \citep{Wilsonetal2009, Muzzinetal2009a, Demarcoetal2010a}. The remaining three clusters were discovered via the Sunyaev-Zeldovich effect signature \citep{Bleemetal2015}. The properties of the clusters are summarised in Table~\ref{tab_data_summary}. 

The main spectroscopic dataset of GOGREEN was obtained from a Gemini Large and Long Program (GS LP-1 and GN LP-4; PI Balogh) using the Gemini Multi-Object Spectrographs (GMOS) on Gemini-North and South. The  large program allows us to obtain unbiased spectroscopy of galaxies of all types down to stellar masses of $M_{*} \gtrsim 10^{10.3}~{\rm M}_{\odot}$, with the faintest targets having exposure time up to 15 hours. Five of the GOGREEN clusters that are part of GCLASS also have similar GMOS spectroscopy data for the bright galaxies in the clusters. These data have been incorporated into the GOGREEN spectroscopy sample. For the specifics of the targeting selection and data reduction, we refer the reader to the survey and data release papers \citep{Baloghetal2017, Baloghetal2021}.

In addition to deep spectroscopy, GOGREEN has obtained deep multi-band imaging ($UBVRIzYJK$ and IRAC $3.6~\mu$m) for the sample. We begin by describing in detail the derivation of structural properties of galaxies, the key focus of this paper, from \textit{HST} imaging in Section~\ref{subsec:HST observations and data reduction} and ~\ref{subsec:Structural parameters}.  The deep multi-wavelength photometric data also allows us to characterize galaxy SEDs and derive photometric redshifts, stellar population parameters, and rest-frame colors. A brief summary of the derivation of these properties is given in Section \ref{subsec:Photometric catalogue}, \ref{subsec:Spectroscopic and photometric redshifts} and \ref{subsec:Stellar mass estimates and rest-frame colors}. We refer the reader to \citet{vanderBurgetal2020} for more details.

\begin{table*}                               
  \caption{Summary of the gogreen cluster sample used in this study}   
  \centering
  \label{tab_data_summary}
  \begin{tabular}{lcccccccc}
  \hline
  \hline  
  Name &  RA$^{\rm{BCG}}_{\rm{J2000}}$ &  Dec$^{\rm{BCG}}_{\rm{J2000}}$  &   Redshift & $\sigma_v$\textsuperscript{a} & $M_{200}$\textsuperscript{a} &  $R_{200}$\textsuperscript{a} &  $N_{\rm{mem,HST}}$\textsuperscript{b}           \\
           &                 &     &           & (kms$^{-1}$)      &   ($10^{14} M_{\sun}$)   &  (Mpc) &      &                              \\
  \hline
  SpARCS J1051+5818  & 10:51:11.23 & $+$58:18:02.7    & $1.035$ &  ${689 \pm 36}$   &   ${2.51^{+0.65}_{-0.55}}$  &  ${0.88 \pm 0.07}$    &51  &                \\
  SPT-CL J0546--5345    & 05:46:33.67 & $-$53:45:40.6    & $1.067$ &  ${977 \pm 68}$  &   ${6.11^{+1.52}_{-1.30}}$  &  ${1.17 \pm 0.09}$     &151 &              \\ 
  SPT-CL J2106--5844    & 21:06:04.59 & $-$58:44:27.9   & $1.132$ &  ${1055 \pm 83}$  &   ${7.65^{+2.02}_{-1.72}}$   &  ${1.23 \pm 0.10}$  & 145 &              \\       
  SpARCS J1616+5545   & 16:16:41.32 & $+$55:45:12.4    & $1.156$ &  ${782 \pm 39}$   &   ${3.29^{+0.69}_{-0.60}}$  &  ${0.92 \pm 0.06}$   & 85  &                \\ 
  SpARCS J1634+4021   & 16:34:37.00 & $+$40:21:49.3    & $1.177$ &  ${715 \pm 37}$   &   ${2.66^{+0.60}_{-0.52}}$  &  ${0.85 \pm 0.06}$   & 57 &        \\ 
  SpARCS J1638+4038   & 16:38:51.64 & $+$40:38:42.9    & $1.196$ &  ${564 \pm 30}$   &   ${1.52^{+0.42}_{-0.36}}$  &  ${0.71 \pm 0.06}$   & 49 &            \\  
  SPT-CL J0205--5829     & 02:05:48.19 & $-$58:28:49.0    & $1.320$ &  ${678 \pm 57}$  &   ${2.22^{+0.89}_{-0.70}}$  &  ${0.76 \pm 0.09}$   & 61 &         \\ 
  SpARCS J0219-0531   & 02:19:43.56 & $-$05:31:29.6    &  $1.325$ &  ${810 \pm 77}$   &   ${2.51^{+1.33}_{-0.98}}$  &  ${0.79 \pm 0.12}$   & 47 &            \\
  SpARCS J0035-4312   & 00:35:49.68 & $-$43:12:23.8    &  $1.335$ &  ${840 \pm 52}$   &   ${4.14^{+1.00}_{-0.87}}$  &  ${0.93 \pm 0.07}$   & 90 &            \\ 
  SpARCS J0335-2929   & 03:35:03.56 & $-$29:28:55.8    &  $1.368$ &  ${542 \pm 33}$   &   ${1.60^{+0.65}_{-0.51}}$  &  ${0.67 \pm 0.08}$   & 47 &            \\     
  SpARCS J1034+5818  & 10:34:49.47 & $+$58:18:33.1    &  $1.385$ &  ${250 \pm 28}$   &   ${0.08^{+0.03}_{-0.03}}$  &  ${0.24 \pm 0.03}$   & 49 &            \\ 
    \hline
  \multicolumn{9}{p{.775\textwidth}}{\textsuperscript{a} The cluster mass $M_{200}$ and $R_{200}$ (radius where the mass overdensity is 200 times the critical density at the cluster redshift) are derived from a scaling relation with the velocity dispersions $\sigma_v$. See \citet{Bivianoetal2021} and \citet{Oldetal2020} for details.} \\
  \multicolumn{9}{p{.775\textwidth}}{\textsuperscript{b} $N_{\rm{mem,HST}}$ is the number of galaxies that are spectroscopically and photometrically selected as cluster members, are within the \textit{HST} image FOV, and have a good structural fit.} \\
\end{tabular}
\end{table*}

\subsection{\textit{HST} observations and data reduction}          
\label{subsec:HST observations and data reduction}
We make use of the near-infrared \textit{HST}/WFC3 F160W imaging of the GOGREEN clusters to quantify the structural properties of the galaxies. The \textit{HST}/WFC3 F160W images were obtained in a Cycle 25 program (GO-15294; PI: Wilson) dedicated to studying galaxy morphologies. Each cluster was targeted with a $1 \times 2$ mosaic of WFC3 pointings centered on the cluster, covering a region of $136'' \times  233''$. At the redshift of the GOGREEN clusters, this corresponds to a $\sim 1.1 \times 1.9$ Mpc rectangular region on the sky. Each pointing has 1-orbit depth. We constrained the \texttt{ORIENT} to within $20^{\circ}$ of the GMOS mask orientation to maximize the overlap between the imaging and the GMOS spectroscopy.

The data are reduced and combined using \textsc{Astrodrizzle} (version 2.1.22) \citep{Gonzagaetal2012}. All the calibrated frames (\texttt{\_flt.fits}) downloaded from the Mikulski Archive for Space Telescopes (MAST) archive are first examined to check the quality of the cosmic rays and bad pixels identification by the \texttt{calwf3} pipeline. We find hot stripes that span across the field of view (FOV) in two of the frames (e.g., due to satellite trails), which are not fully flagged by the pipeline. We mask these regions generously in the data quality array of the flt files. In addition, a total of seven frames in five clusters show a smooth background gradient, presumably due to earthshine. To remove the gradient, we follow a similar approach described in \citet{Windhorstetal2011}. Sources on the flt image are first masked, then the gradient is fitted with a fifth-order bivariate spline function and is subtracted from the frame before drizzling.

For the final drizzling, we adopt a pixel scale of $0.06''$ pixel$^{-1}$, a square kernel, and a~\texttt{pixfrac} of 0.8. We produce weight maps using both inverse variance map (\texttt{IVM}) and error map (\texttt{ERR}) weighting for different purposes. The \texttt{IVM} weight maps, which contain all background noise sources except Poisson noise of the objects, are used for object detection, while the \texttt{ERR} weight maps are used for structural analysis as the Poisson noise of the objects is included.  The final images and the weight maps are included in the first GOGREEN public release \citep{Baloghetal2021}.  The characteristic point-spread function (PSF) of each cluster is constructed by median-stacking isolated bright unsaturated stars. Depending on the cluster, 5-22 stars are used in the stack. The full-width-half-maximum (FWHM) of the PSFs are $\sim0.17 - 0.18''$.

\subsection{Structural parameters}                   
\label{subsec:Structural parameters}
We derive structural parameters for all sources in the F160W image of each cluster by fitting them with two-dimensional single S\'ersic profiles \citep{Sersic1968}.  The parameters are derived using a modified version of \textsc{GALAPAGOS} (based on v.2.3.1) \citep{Bardenetal2012, Haussleretal2013} with \textsc{GALFITM} (v.1.2.1).  The five independent parameters of the S\'ersic profile, namely the total luminosity ($L_{\rm{tot}}$), the S\'ersic index ($n$), the half-light radius / effective semi-major axis ($R_e$), the axis ratio ($q =b/a$, where $a$ and $b$ are the major and minor axis respectively) and the position angle ($P.A.$), as well as the centroid ($x,y$) of the source are left as free parameters. Source detection from the \textit{HST} image and initial guesses for fitting these parameters were derived by running \textsc{SExtractor}, which is incorporated in the \textsc{GALAPAGOS} run.

We apply fitting constraints of $0.2 < n < 12$, $0.3 < R_{e} < 400$ (pix), $0 < \rm{mag} < 40$, $0.0001 < q < 1$, and $-180^{\circ} < \rm{P.A.} < 180^{\circ}$. The local sky level for each source is fixed to the value determined by GALAPAGOS, which is derived using an elliptical annulus flux growth method. We derive noise maps (RMS noise) from the \texttt{ERR} weight maps output by \textsc{Astrodrizzle} and use them as sigma map input for GALFIT. The noise maps that we generate from \texttt{ERR} weight maps are a more realistic representation of the noise than the internal error estimation in GALFIT, as they include pixel-to-pixel exposure time differences originating from image drizzling and dithering patterns in observations, as well as a more accurate estimation of shot noise. The S\'ersic model is convolved with the characteristic PSF of each cluster.  Nearby objects that are close to the primary source of interest are fitted simultaneously.

We refine the \textsc{GALAPAGOS} configuration parameters, including those that control local sky level estimation and close neighbor treatment, using extensive tests with simulated galaxies. The details and result of these simulations are provided in Appendix~\ref{app:Details of the simulation and the biases of the axis ratio measurements}. In brief, we inject a set of 20000 simulated galaxies (20 at a time), with surface brightness profiles described by a S\'ersic profile, to random locations in the sky region of the F160W image and recover their structural parameters with our science setup. Using this set of simulated galaxies, we then compute the biases of our measurements, modify the \textsc{GALAPAGOS} setup, and re-derive the parameters and biases. This process is iterated a few times to get the best configuration parameters that minimize the biases.

The simulation allows us to characterise the biases in our structural parameter measurements. Among the three S\'ersic structural parameters that describe the shape of a galaxy ($R_e$, $n$, and $q$), the axis ratio $q$ can typically be measured with the highest accuracy. The axis ratio shows an average bias and dispersion of $\sim2\%$ and $\sim12\%$ at F160W $=23$ (AB), which corresponds roughly to $\log(M / {\rm M}_{\odot}) \sim 9.5$, the mass limit we adopted in this work. This is a factor of five (two) better than the average bias (dispersion) we see in S\'ersic index $n$.  We stress that while S\'ersic index is useful for morphological selection, it is not straightforward to compare their distribution between different samples. From our simulations, we find that biases in $n$ depend also heavily on $n$ itself, such that high $n$ values are more uncertain \citep[see also][for a description of systematic uncertainties]{vanderWeletal2012}. These systematic errors can have a detrimental effect on the cluster and field comparison especially for low-mass galaxies, as the expected difference is small at this redshift \citep[see, e.g.,][for the difference in $n$ for a sample of massive galaxies]{Chanetal2018, Matharuetal2019}. Therefore, in this work we focus primarily on the axis ratio distributions.

We also visually inspect outliers that have large sizes for their particular magnitudes or parameters that hit the boundary of the constraints with the procedure similar to \citet{Chanetal2016}.  In cases where sources or nearby objects are not correctly deblended, extra S\'ersic components are added iteratively if necessary to ensure adjacent sources are well-fitted.  Fits that still hit the boundary of our fitting constraints are considered as bad fits and are excluded from subsequent analyses.

\subsection{Photometric catalogue}       
\label{subsec:Photometric catalogue}
We utilize the $K_{s}$-band selected photometric catalogue derived from the multi-band imaging of each cluster. We refer to \citet{vanderBurgetal2020} for details of the  procedure to construct these catalogues. Source detection is performed on the $K_{s}$-band image using \textsc{SExtractor} \citep{BertinArnouts1996}.  Aperture photometry is measured on the PSF-matched images using circular apertures with a diameter of $2''$. To preserve the spatial resolution of the ground-based imaging, aperture photometry of \textit{Spitzer}/IRAC data is measured with a larger aperture of $3''$ and rescaled, following the approach in \citet{vanderBurgetal2013}. The area covered by the catalogues range from $\sim 5' \times 5'$ to $\sim 10' \times 10'$ depending on the cluster. The area considered for this study is, therefore, limited by the \textit{HST} imaging coverage.

\subsection{Spectroscopic and photometric redshifts}           
\label{subsec:Spectroscopic and photometric redshifts}
Spectroscopic redshifts ($z_{\rm{spec}}$) are measured using the Manual and Assisted Redshifting software \citep[\textsc{MARZ},][]{Hintonetal2016}, which utilises a cross-correlation algorithm to match the spectra against a variety of spectral templates. These are supplemented with publicly available $z_{\rm{spec}}$ from various surveys. An exhaustive list of surveys can be found in \citet{vanderBurgetal2020}. Photometric redshifts ($z_{\rm{phot}}$) are derived using \textsc{EAZY} \citep{Brammeretal2008} with standard templates.  In this work we use the peak of the posterior probability distribution of the redshift estimated with \textsc{EAZY} as $z_{\rm{phot}}$. A correction, in the form of a quadratic function, has been applied to the $z_{\rm{phot}}$ to minimise the residual between the measured $z_{\rm{spec}}$ and $z_{\rm{phot}}$ \citep{vanderBurgetal2020}. In this work, we derive an additional correction to the $z_{\rm{phot}}$ of each cluster to better match the $z_{\rm{spec}}$ at the cluster redshift. The correction is taken as the median offset between the $z_{\rm{phot}}$ and the $z_{\rm{spec}}$ of the cluster members (See Section~\ref{subsec:Cluster membership and sample selection} for a description of the cluster membership). The magnitude of this correction is generally small, but in some cases can reach up to $\sim 0.09$.

\subsection{Stellar mass estimates and rest-frame colors}          
\label{subsec:Stellar mass estimates and rest-frame colors}
Stellar masses for all galaxies are inferred from the multi-band photometry using \textsc{FAST} \citep{Krieketal2009}, derived in \citet{vanderBurgetal2020}.  This includes a rescaling factor that is applied to the input aperture fluxes, such that SED fitting gives the total mass of the galaxy. This factor is taken to be the ratio of $K_{s}$-band \texttt{FLUX\_AUTO} measurements to the aperture flux from \textsc{SExtractor} (i.e. $F_{\rm{auto}}$/$F_{\rm{aper}}$).  We use the \citet{BruzualCharlot2003} stellar population synthesis models and assume a \citet{Chabrier2003} IMF, solar metallicity, and the \citet{Calzettietal2000} dust law. The star formation history (SFH) is parameterized as an exponentially declining history $SFR \propto e^{-t / \tau}$, where the timescale $\tau$ ranges between 10 Myr and 10 Gyr. We note that stellar masses derived in this way can typically be $0.2$ dex lower than those derived from non-parametric star formation histories \citep{Lejaetal2019, Webbetal2020}. Nevertheless, as we describe in Section \ref{subsec:Field comparison sample}, the method we used to derive stellar masses in GOGREEN is largely consistent with our chosen field sample and therefore has the advantage of allowing us to compare the stellar masses directly.

In this work, we utilize the rest-frame $UVJ$ color classification to separate the galaxies into star-forming and quiescent.  $UVJ$ classification has become a standard technique in galaxy evolution studies as it can separate ``genuine'' quiescent galaxies from dusty star-forming ones \citep[e.g.,][]{Labbeetal2005, Williamsetal2009, Muzzinetal2013b}.  Rest-frame $U-V$ and $V-J$ colors required for this classification are again derived using \textsc{EAZY}. We adopt the $UVJ$ color classification criteria in \citet{Muzzinetal2013c}.  The best redshift estimate of individual galaxies ($z_{\rm{phot}}$ for those lacking $z_{\rm{spec}}$) is used to measure the rest-frame colors. We also computed the colors by fixing the redshifts of all galaxies in each cluster field to the cluster mean redshift listed in Table~\ref{tab_data_summary}, and confirm this does not change our conclusion.

\begin{figure*}
  \centering
  \includegraphics[scale=0.675]{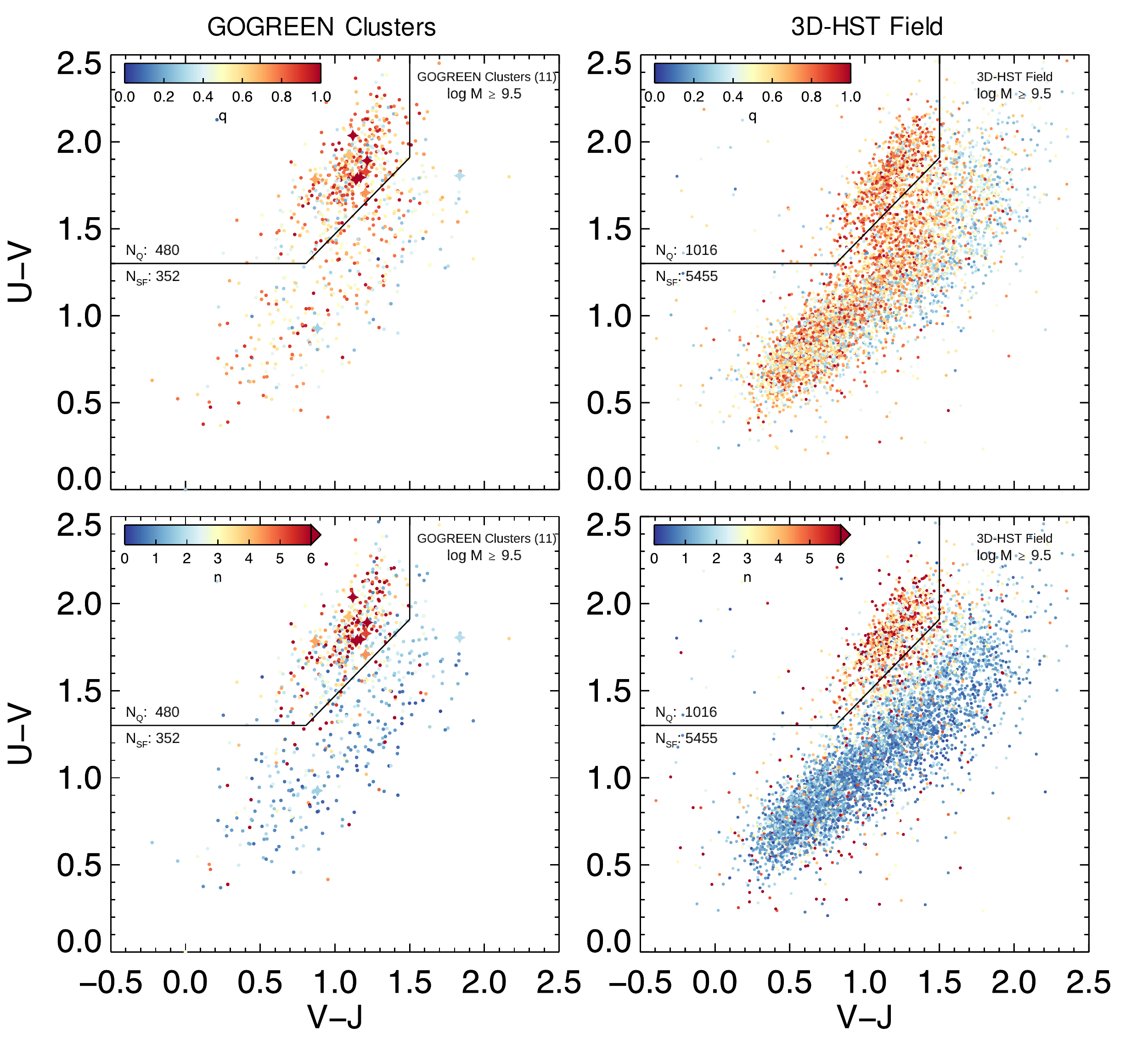}  
  \caption{Rest-frame $UVJ$ diagram of the GOGREEN cluster sample and the field sample used in this study. Top: Data points are color coded by their axis ratio $q$. As expected, both quiescent and star-forming galaxies in clusters and the field exhibit a wide range of axis ratios. Galaxies that have both red $U-V$ and $V-J$ colors can be predominantly seen with small $q$, as the fact that they are viewed edge on result in a higher dust extinction value. Bottom: Data points color coded by their S\'ersic index $n$.  As pointed out in various works, quiescent and star-forming galaxies in both clusters and the field show discernible differences in their $n$ distributions. The number of quiescent ($N_{\rm{Q}}$) and star-forming ($N_{\rm{SF}}$) galaxies in each sample are provided next to the $UVJ$ classification line. The cluster BCGs are marked with a star symbol. The cluster sample shows a higher relative abundance of quiescent galaxies compared to star-forming galaxies than the field.}
  \label{fig_uvj_q_n}
\end{figure*}

\subsection{Cluster membership and sample selection}        
\label{subsec:Cluster membership and sample selection}
We use both the spectroscopic and photometric redshift information of the GOGREEN sample to define cluster membership.  For galaxies with spectroscopic redshifts, we define them as cluster members if they are $\pm 2000$ km~s$^{-1}$ around the mean redshift of the cluster. This corresponds to $\Delta z_{\rm{spec}} \sim 0.015$ at GOGREEN redshifts. Our simple cluster membership selection is slightly different from the ones used in previous GOGREEN papers \citep[e.g.][]{Oldetal2020,Baloghetal2021,Bivianoetal2021},  but fully adequate for the scope of the present analysis, given that we also include members that are selected photometrically.

For galaxies without spectroscopic information, we use their photometric redshifts to determine membership.  Since the main goal of this work is to compare the axis ratio of cluster galaxies with the field, it is important to strike a balance between maximizing the number of cluster members (i.e., sample completeness) and compromising the purity of the sample.  To find the optimal $z_{\rm{phot}}$ selection criteria, we compare the $z_{\rm{spec}}$ of the GOGREEN spectroscopic sample with their $z_{\rm{phot}}$ to estimate the completeness and purity as a function of the $z_{\rm{phot}}$ selection width.  The result of this test is given in Appendix~\ref{app:Completeness and purity of the cluster member selection}. Such a test is possible as the spectroscopic sample is a representative subset of the photometrically selected galaxy population \citep[see Appendix A.2 in][for a discussion]{vanderBurgetal2020}. Photometric cluster members are defined as those with $\Delta z_{\rm{phot}}/(1+z_{\rm{phot}}) \leq 0.06$ around the mean redshift of the cluster. Our test suggests that this selection gives a completeness of $\sim85\%$ and a purity of $\sim80\%$.

We crossmatch the structural parameter catalogue (i.e., detected from the F160W image) with the $K_{s}$-band selected photometric catalogues.  Adopting the abovementioned cluster membership selection and a stellar mass limit of $\log(M / {\rm M}_{\odot}) = 9.5$ results in $860$ cluster members that are within the \textit{HST} image FOV.  This stellar mass limit corresponds to a $\sim80\%$ completeness of the catalogues \citep[see][for details of the completeness characterization]{vanderBurgetal2020}.  After excluding bad structural fits (fits that hit the boundaries of the fitting constraints), our final sample contains $832$ cluster members that have robust structural parameters.  In the left column of Figure~\ref{fig_uvj_q_n}, we show the rest-frame $U-V$ and $V-J$ color distribution of the cluster sample, color-coded by their axis ratios and S\'ersic indices.     

\subsection{Field comparison sample}           
\label{subsec:Field comparison sample}
Although there is a large number of spectroscopically confirmed field galaxies available in GOGREEN, the number of field galaxies within the \textit{HST} image FOV is still too small for a morphology comparison between clusters and the field within GOGREEN.  Expanding this field sample with photometric redshifts is not straightforward, as the number of galaxies declines sharply with the redshift selection. We end up with either a small sample or a sample with low purity (See Appendix~\ref{app:Completeness and purity of the cluster member selection} for a discussion).

The field comparison sample we use in this study is taken from the CANDELS \citep{Groginetal2011, Koekemoeretal2011}, and 3D-\textit{HST} Treasury programs \citep{Brammeretal2012, Skeltonetal2014}.  Among the $\sim99000$ 3D-\textit{HST} grism redshift measurements \citep{Momchevaetal2016}, $\sim5200$ redshifts are within the range of $0.9 < z < 1.5$.  For structural parameters, we use the F160W-band measurements of all five CANDELS/3D-\textit{HST} fields (COSMOS, GOODS-N, GOODS-S, EGS, and UDS) derived by \citet{vanderWeletal2014}.  The CANDELS F160W wide imaging has, on average, one and one-third orbit depth, thus being comparable in depth to the GOGREEN imaging. To ensure our structural parameter measurements are compatible with the \citet{vanderWeletal2014} measurements, we apply our methodology described in Section~\ref{subsec:Structural parameters} to the CANDELS imaging for a sample of galaxies in the redshift range of $0.9 < z < 1.5$. Overall we find that the median ratio and $1\sigma$ uncertainty between our measurements and \citet{vanderWeletal2014} are $1.00 \pm 0.03$ ($0\% \pm 3\%$) down to the mass limit of this work. The result of the comparison is shown in Appendix~\ref{app:Axis ratio comparison between van der Wel et al. 2014 and this work}. We also check that using either set of measurements gives a consistent conclusion. Throughout this work we show the field results using the \citet{vanderWeletal2014} measurements.      

Stellar masses, photometric redshifts, and rest-frame colors are estimated using \textsc{FAST} and \textsc{EAZY} from multi-band photometry \citep{Skeltonetal2014}, in a way that is almost identical to GOGREEN \citep{vanderBurgetal2020}. There are two main differences. Firstly, \citet{Skeltonetal2014} adopted a minimum timescale $\tau$ of $40$ Myr as opposed to $10$ Myr. The second subtle difference is on the definition of the total fluxes, which affects the stellar mass estimates. On top of rescaling the aperture fluxes to \texttt{FLUX\_AUTO} measurements like in GOGREEN (see Section~\ref{subsec:Stellar mass estimates and rest-frame colors}), \citet{Skeltonetal2014} also factored in a correction to account for the missing flux that falls outside the \texttt{AUTO} aperture, determined from measuring the growth curves of the F160W PSFs (the detection band of the 3D-\textit{HST} catalogue). To ensure the stellar masses are comparable, we apply a correction to both the 3D-\textit{HST} and GOGREEN stellar masses, rescaling the stellar masses to the total F160W fluxes of the best-fit S\'ersic profile of the galaxies.  Overall this correction is small; it only increases the stellar mass by $\sim0.02$ (3D-\textit{HST}) and $\sim0.03$ (GOGREEN) dex on average, although in some cases it can exceed 0.1 dex. The fact that GOGREEN galaxies require a slightly larger correction is accordant with the additional missing flux correction that is applied in 3D-\textit{HST}.

We select galaxies that i) are in the redshift range of $0.9 < z < 1.5$, ii) with a stellar mass of $\log(M / {\rm M}_{\odot}) \geq 9.5$, and iii) have robust structural parameters as our field sample.  The selection is done using the $z_{\rm{best}}$ catalogues (v4.1.5), which uses ground-based $z_{\rm{spec}}$ of the galaxies if available, then grism redshift $z_{\rm{grism}}$, and finally $z_{\rm{phot}}$ if the other two are not available. These selection criteria result in a sample of $6471$ galaxies. We verified that applying more sophisticated redshift cuts (e.g., $1.0 < z < 1.4$ for $z_{\rm{spec}}$ and $z_{\rm{grism}}$, $0.9 < z < 1.5$ for $z_{\rm{phot}}$) does not affect our conclusion.     

The right column of Figure~\ref{fig_uvj_q_n} shows the rest-frame $U-V$ and $V-J$ color distribution of the field sample, color-coded by their axis ratios and S\'ersic indices. We have checked for offsets in the rest-frame $U-V$ and $V-J$ colors between the two catalogues by inspecting the color distribution of the 3D-\textit{HST} galaxies and the GOGREEN field galaxies. We find no evidence of any color offset that is larger than 0.05 mag.  To ensure our results are robust, we move the $UVJ$ selection for the clusters in all four directions by 0.05 mag to mimic the effect of potential color offsets and repeat the analyses four times.  All these analyses give consistent results.


\begin{figure*}
  \centering
  \includegraphics[scale=0.620]{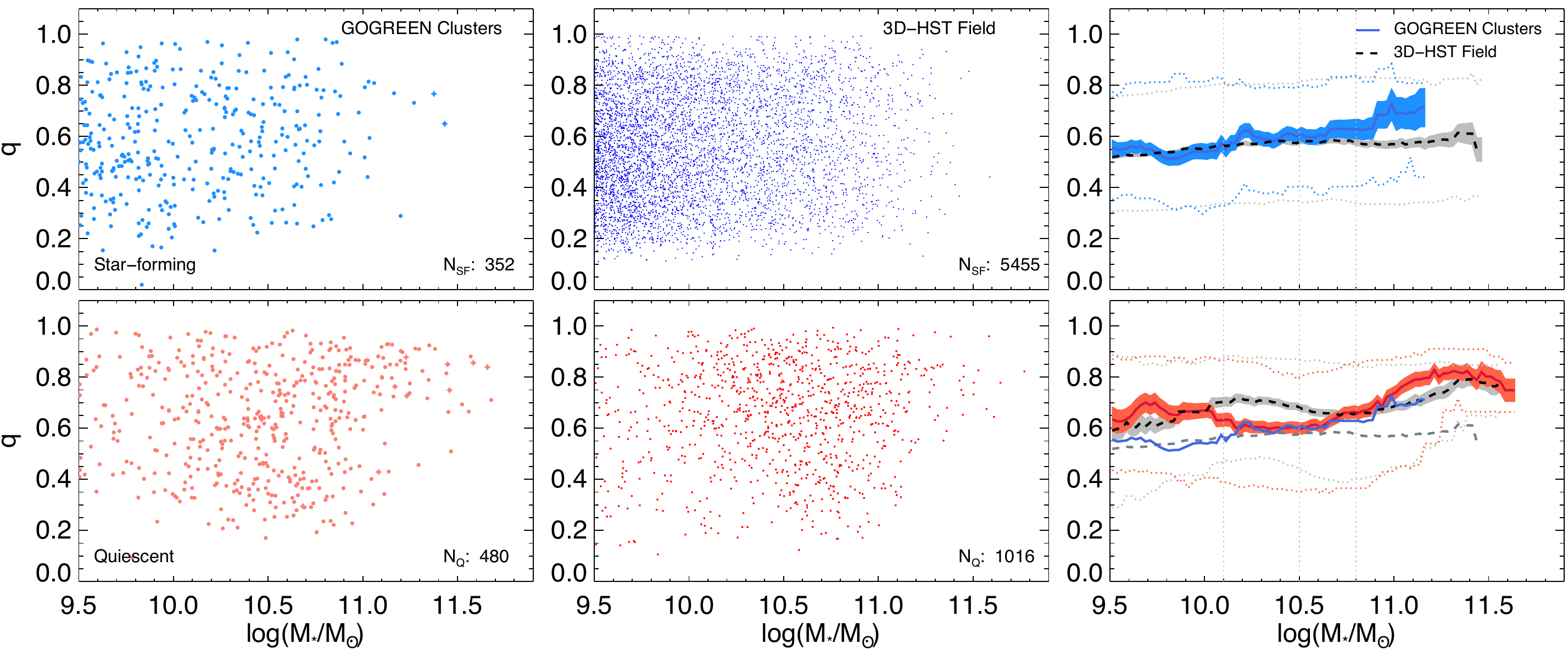}      
  \caption{Projected axis ratio $q$ distribution of star-forming (top) and quiescent (bottom) galaxies as a function of mass in clusters and the field.  The cluster BCGs are marked with a star symbol. The right column shows the $q$ -- mass relation in clusters and the field as a running median in stellar mass bins of 0.2 dex.  Only bins with $\geq 10$ galaxies in the cluster and field samples are shown.  The solid line and black dashed line correspond to the median axis ratio for clusters and field, respectively. The shaded regions correspond to the standard error of the median $q$ ($1.253 \sigma/\sqrt{N}$). The dotted lines correspond to the running 16th and 84th percentiles of the $q$ distributions. The vertical dotted lines correspond to the masses we used to divide the samples into four mass bins in Section~\ref{subsec:Reconstructing the intrinsic shapes from the projected axis ratio distributions}. The median $q$ of both star-forming and quiescent galaxies in clusters and the field both show a mass dependence, with high mass galaxies being rounder. The median relations of the star-forming galaxies in clusters and the field are plotted in the bottom panel as blue solid (cluster) and grey dashed (field) for comparison. The axis ratio distributions of quiescent galaxies in clusters and the field are distinct, most remarkably at $\log(M_{*} /{\rm M}_{\odot}) \sim 10.0 - 10.6$ ($\sim4.0 \sigma$) and $\log(M_{*} /{\rm M}_{\odot}) \sim 11.0 - 11.3$ ($\sim2.4 \sigma$). See Section~\ref{subsec:Projected axis ratio in clusters and the field} for details.}
    \label{fig_mqr}
\end{figure*}

\section{Results}
\label{sec:Results}

\subsection{Projected axis ratios in clusters and the field}          
\label{subsec:Projected axis ratio in clusters and the field}
Figure~\ref{fig_mqr} shows the axis ratio of the star-forming and quiescent population in clusters and the field as a function of mass. Clusters show a higher relative abundance of quiescent galaxies compared to star-forming galaxies than the field, with an overall quenched fraction of $f_{\rm{Q,clus}} = (N_{\rm{Q}} / N_{\rm{SF}} + N_{\rm{Q}}) = 0.58$ down to our mass limit compared to the field quenched fraction of $f_{\rm{Q,field}} = 0.16$. This confirms the enhanced quenched fraction in GOGREEN clusters, relative to the field, found by \citet{vanderBurgetal2020}. The rightmost panel of Figure~\ref{fig_mqr} shows the $q$--mass relation in clusters and the field as a running median and percentiles in stellar mass bins of 0.2 dex. We can see that the median $q$ of both star-forming galaxies and quiescent galaxies in clusters and the field show a mass dependence. For the cluster sample, the median $q$ increases from $0.55 \pm 0.02$ ($0.69\pm 0.04$) at $\log(M_{*} /{\rm M}_{\odot}) \sim 9.7$ to $0.73 \pm 0.10$ ($0.83 \pm 0.02$) at $\log(M_{*} /{\rm M}_{\odot}) \sim 11.2$ for star-forming (quiescent) galaxies. 

The median axis ratios of star-forming and quiescent galaxies in both clusters and the field are significantly different, which suggests that there are fundamental differences in the intrinsic shapes between star-forming and quiescent galaxies. For the field, these differences in median $q$ are seen at all masses. Star-forming galaxies in the field show a lower median $q$ at a fixed stellar mass than quiescent galaxies, consistent with the finding that most star-forming field galaxies are disks at this redshift range \citep[e.g.,][]{vanderWeletal2014b}. On the other hand, the median $q$ of star-forming and quiescent galaxies in clusters show a difference only at low ($\log(M_{*} /{\rm M}_{\odot}) \leq 10.1$) and high ($\log(M_{*} /{\rm M}_{\odot}) \geq 11.0$) masses.

From Figure~\ref{fig_mqr}, we can also see that massive \textit{quiescent} galaxies with $\log(M_{*} /{\rm M}_{\odot}) \geq 11$ in both clusters and the field are not only rounder than their low mass counterparts, they also have a narrower $q$ distribution, reflected by their percentiles (dotted lines).  Note that this is not an effect merely due to low number statistics, as there are 59 and 116 massive cluster and field galaxies, respectively.  A similar change is also seen in the axis ratio distributions of local and intermediate-redshift quiescent galaxies \citep[e.g.,][]{vanderWeletal2009, Holdenetal2012, Changetal2013a}.  We will focus more on their distributions in the next sections.

There are some intriguing differences between the axis ratio distributions in clusters and the field.  The medians and percentiles of the $q$ distributions of star-forming galaxies are largely consistent with each other, although there may be a weak indication that massive star-forming galaxies ($\log(M_{*} /{\rm M}_{\odot}) \gtrsim 11.0$) in clusters show a higher median $q$ ($\sim1.2 \sigma$ difference).  On the other hand, we note that the distributions of quiescent galaxies in clusters and the field show differences roughly in the following two mass ranges: $\log(M_{*} /{\rm M}_{\odot}) \sim 10.0 - 10.6$ and $\log(M_{*} /{\rm M}_{\odot}) \sim 11.0 - 11.3$, but in the opposite sense.  For $\log(M_{*} /{\rm M}_{\odot}) \sim 10.0 - 10.6$, the median $q$ are offset to lower values in clusters compared to the field, with a $\sim4.0 \sigma$ difference.  This effect can be seen in Figure~\ref{fig_mqr} as both the cluster medians and the 16th percentiles extend to lower values than the field.  On the other hand, at $\log(M_{*} /{\rm M}_{\odot}) \sim 11.0 - 11.3$, there is evidence that the median $q$ and the percentiles are offset to higher values ($\sim2.4 \sigma$ difference) in clusters compared to the field. We find similar differences if we limit the cluster sample to only spectroscopically-confirmed members, but with lower significance due to the smaller number of galaxies in the sample. In the following sections, we explore these differences more in detail using the axis ratio distributions in different mass ranges and investigate their implications using intrinsic shape reconstruction techniques.       

\begin{figure*}
  \centering
  \includegraphics[scale=0.775]{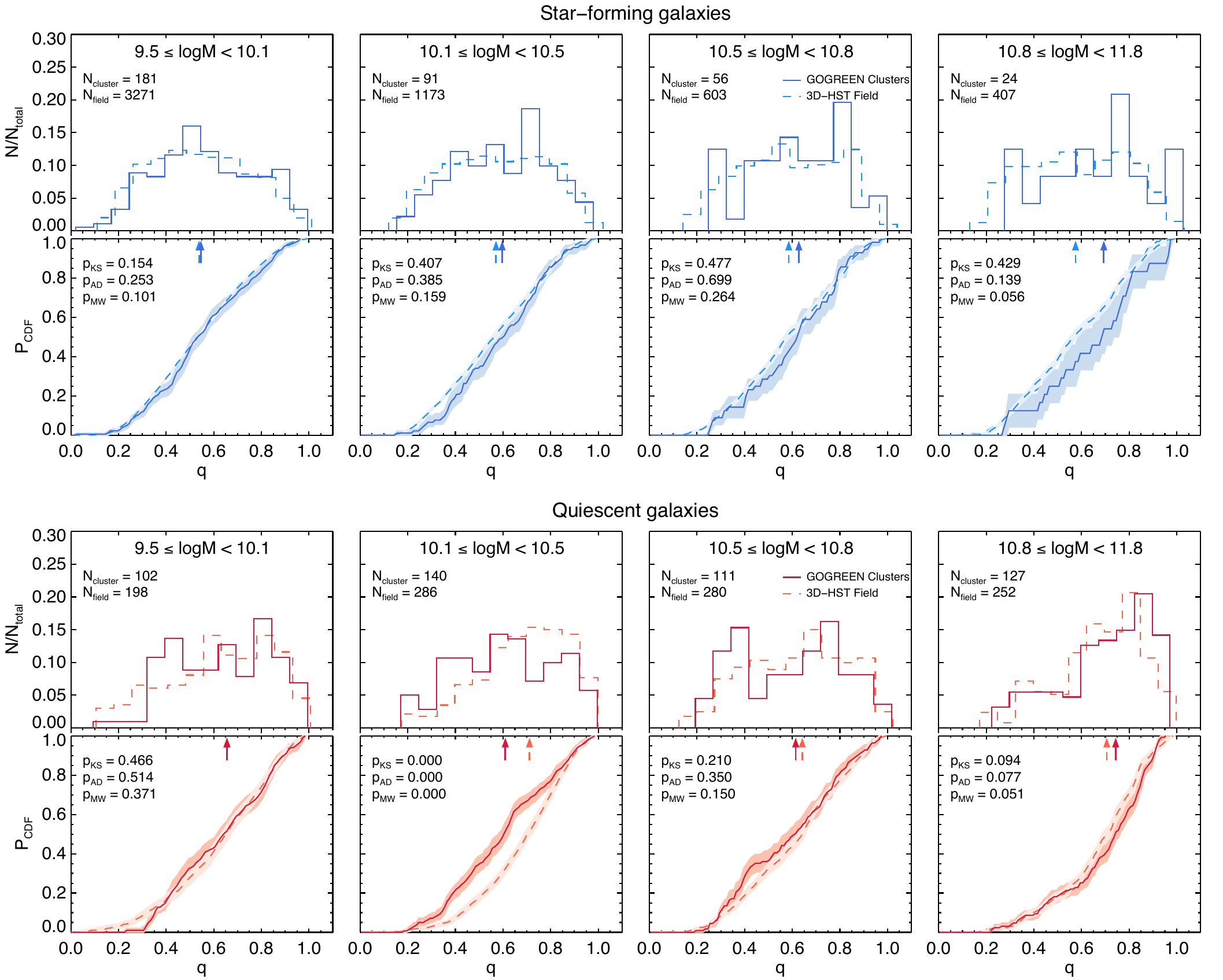}      
  \caption{Comparison of the projected axis ratio $q$ distribution of star-forming (top) and quiescent (bottom) galaxies in clusters with the field in different mass bins.  In each case the top panels show the $q$ histograms, while the bottom panels show the CDFs and the resultant $p$-values of the KS, AD and MW tests. The shaded areas shows the $1\sigma$ uncertainty of the CDF derived from the bootstrapped sample of the axis ratios. The solid and dashed arrows show the median of the distribution of the cluster and the field sample, respectively. There is no obvious difference between the $q$ distribution of star-forming galaxies in clusters and the field.  For quiescent galaxies, the $q$ distribution in clusters and the field are distinct, most prominently in the $10.1 \leq \log(M_{*}) < 10.5$ mass bin.}
    \label{fig_qhisto_pass_sf_changbin}
\end{figure*}

\subsection{Reconstructing the intrinsic shapes from the projected axis ratio distributions}     
\label{subsec:Reconstructing the intrinsic shapes from the projected axis ratio distributions}
In Figure~\ref{fig_qhisto_pass_sf_changbin} we compare the axis ratio distributions between clusters and the field in four mass bins: i) $9.5 \leq \log(M_{*}) < 10.1$, ii) $10.1 \leq \log(M_{*}) < 10.5$, iii) $10.5 \leq \log(M_{*}) < 10.8$, and iv) $10.8 \leq \log(M_{*}) < 11.8$. The bottom panels show the cumulative distribution functions (CDFs).
To take into account the measurement uncertainties of $q$, we bootstrap the observed $q$ distribution and at the same time perturb individual $q$ with its $1\sigma$ uncertainty. The shaded areas in the bottom panels show the $1\sigma$ uncertainty of the CDF derived from this bootstrapped sample. The choice of the binning is selected to match the binning used in \citet{Changetal2013} for the purposes of the $q$ distribution modeling (see Section~\ref{subsubsec:Methodology} for details). \citet{Changetal2013} adopted three mass bins with the lowest mass bin being $10.1 \leq \log(M_{*} /{\rm M}_{\odot}) < 10.5$. Since GOGREEN data allow us to go down to lower masses, here we include an additional mass bin of $9.5 \leq \log(M_{*} /{\rm M}_{\odot}) < 10.1$\footnote{Similarly, the highest mass bin in \citet{Changetal2013} only goes up to $\log(M_{*}/{\rm M}_{\odot}) = 11.5$. Here we extend it up to $\log(M_{*}/{\rm M}_{\odot}) = 11.8$. We checked that limiting it to $11.5$ gives the same conclusion.}. We have repeated the analysis with a different set of binning that have more uniform bin widths and found a consistent conclusion. We apply the Kolmogorov-Smirnov (KS) test, the Anderson-Darling (AD) test, and the Mann-Whitney U-test (MW)\footnote{The Mann-Whitney U-statistics tests the hypothesis that the two sample populations are distributed with the same median.} on the $q$ distributions, with the null hypothesis that they come from a common distribution.

Figure~\ref{fig_qhisto_pass_sf_changbin} confirms the similarities and differences between clusters and the field discussed in Section~\ref{subsec:Projected axis ratio in clusters and the field}. No obvious differences can be seen between star-forming galaxies in cluster and field, which suggests that their intrinsic shapes are likely to be similar. 
For quiescent galaxies, we see some evidence that the axis ratio distribution between cluster and field are distinct in some mass bins. Cluster galaxies in the $10.1 \leq \log(M_{*}) < 10.5$ bin show a flatter distribution with an apparent excess at low $q$ compared to the field. All three tests show a small $p$ value ($p_{\rm{KS,AD,MW}} \simeq 0.00$)\footnote{We have also assessed the significant of this difference using the half-sample method (using only half of the cluster sample) and by jackknifing the cluster sample. Both tests give small $p$ values.}.  There is some weak indication that cluster galaxies in the highest mass bin ($10.8 \leq \log(M_{*}) < 11.8$) show on average higher $q$ than those in the field ($p_{\rm{MW}} \simeq 0.05$, $p_{\rm{KS,AD}} \lesssim 0.1$). As we have shown in Section~\ref{subsec:Projected axis ratio in clusters and the field}, this is due to the high mass population ($\log(M_{*}) > 11$). There is no statistically significant difference in the other mass bins.

\subsubsection{Methodology for fitting the observed $q$ distributions}      
\label{subsubsec:Methodology}
To understand the implication of these differences, we model the projected axis ratio distributions of the \textit{quiescent population} in clusters and the field to reconstruct their intrinsic shapes.  We focus only on quiescent galaxies, as no difference can be seen for the star-forming population. We adopt the methodology used by previous works \citep[e.g.,][]{Holdenetal2012, Changetal2013, vanderWeletal2014b}, assuming the intrinsic 3D structure of a galaxy can be described by a triaxial ellipsoid.  We refer the reader to Section $5$ of \citet{Changetal2013} for a description of the relevant equations. The procedure can be briefly described as follows.

A triaxial ellipsoid can be described with three axes ($a,b,c$), with $a \geq b \geq c$. One can define two intrinsic axis ratios, $\beta = b/a$ and $\gamma = c/a$, the intrinsic ellipticity $E = (1 - \gamma)$ and the triaxiality $T = (1- \beta^2)/(1-\gamma^2)$. The triaxial ellipsoid has two axisymmetric cases; the ellipsoid is known as an oblate spheroid if $\beta = 1$ (i.e. $a=b >c$). If $\beta = \gamma$ (i.e. $a > b=c$), the ellipsoid is then known as a prolate spheroid. The goal of the modeling is to find the model galaxy population(s) (i.e., sets of triaxial ellipsoids) that best-reproduces the observed axis ratio distribution.  The model population is assumed to have Gaussian distributions of ellipticity and triaxiality.  It can, therefore, be described by four parameters ($E, \sigma_E, T, \sigma_T$), where $\sigma_E$ and $\sigma_T$ are the standard deviations of the ellipticity and triaxiality, respectively.

Assuming random viewing angles, we can compute the expected projected axis ratio distribution for such a population.  A correction is then applied to include the effects of uncertainties in the $q$ measurements \citep[see][for a description]{RixZaritsky1995}.  In practice, the projected axis ratio distribution for a model galaxy population is computed numerically by generating $100000$ galaxies with random viewing angles and input parameters according to the Gaussian distributions.  The number of galaxies is chosen so that the resolution of the axis ratio distribution of the model population is sufficient to compare with the observations. This is essentially the probability distribution function of the projected axis ratio $P(q)$ of the model population given a set of input parameters.

Star-forming galaxies are traditionally modeled with a single model population of triaxial ellipsoids \citep[e.g.,][]{vanderWeletal2014b}\footnote{\citet{Zhangetal2019} demonstrated that $R_e$ needs to be taken into account in the modeling due to the strong correlation between $q$ and $R_e$. This correlation is only seen in star-forming galaxies, not in quiescent galaxies.}. For quiescent galaxies, it is established that a single model population is not able to reproduce their axis ratio distributions. For example, \citet{Holdenetal2012} modeled the low-redshift quiescent population in SDSS and found that a single-component triaxial model cannot adequately describe the $q$ distribution, except for the massive population with $\log(M_{*} /{\rm M}_{\odot}) > 11.0$. They showed that an additional second component, composed of oblate spheroids, is needed to match the observed distributions. \citet{Changetal2013} confirmed that this is also true for quiescent galaxies in the field at higher redshifts ($1<z<2.5$). Although it originated from purely empirical needs to reproduce the axis ratio distribution, this two-component (triaxial + oblate) model is consistent with the dichotomy in local early-type galaxies discovered via stellar kinematics, \citep[i.e., the slow and fast rotators, see][for a review]{Cappellari2016}.

Following \citet{Holdenetal2012} and \citet{Changetal2013}, on top of the single model population we also adopt the two-component model. Since $\beta = 1$ in the oblate model, the triaxiality is always zero, hence the parameters that describe the model are the intrinsic axis ratio $\gamma$ and its standard deviation $\sigma_\gamma$.  To be consistent with previous works, we use $b$ and $\sigma_b$ to denote $\gamma$ and $\sigma_\gamma$\footnote{This definition is first used by \citet{Sandageetal1970}. An oblate galaxy only has two independent axes $a,b$, with the intrinsic axis ratio being $b/a$. \citet{Sandageetal1970} assumes $a=1$ without loss of generality, hence the use of $b$.}.  Hence the two-component model can be fully described by seven independent parameters ($E, \sigma_E, T, \sigma_T$, $b$, $\sigma_b$, $f_{\rm{ob}}$), where the oblate fraction $f_{\rm{ob}}$ is the fraction of oblate galaxies relative to the total model population.

Nevertheless, high redshift galaxy samples, including the cluster and field samples used in this work, are often not large enough to constrain all seven parameters simultaneously. \citet{Changetal2013} tackled this by first fitting the model to local quiescent galaxies, and assumed that the same components could be used to describe the axis ratio distributions at high redshift. They fixed the parameters for the triaxial component and only allowed the oblate parameters ($b$, $\sigma_b$, $f_{\rm{ob}}$) to vary to study the redshift evolution of these parameters. They demonstrated that using this approach can reach conclusion that is consistent with other independent analyses. Here we take a similar approach and build on the findings of \citet{Changetal2013}. Partly for this reason, we have adopted a similar mass binning as \citet{Changetal2013}. We consider three scenarios with different assumptions, summarised below:

\begin{itemize}
    \item \textbf{Case I - Fitting $E, \sigma_E, T, \sigma_T$} -- We assume the axis ratio distribution can be described by a single-component model, i.e., $f_{\rm{ob}} = 0$.  The single-component model is useful in studying the distributions at the high masses. See Section~\ref{subsubsec:The high mass bin}.           
   \item \textbf{Case II - Fitting $f_{\rm{ob}}$, $b$, $\sigma_b$} -- We assume the values of the remaining four parameters ($E, \sigma_E, T, \sigma_T$) to be the same as the best-fit values in \citet{Changetal2013}. The same assumed values are used for cluster and the field, although we find that the conclusion does not depend heavily on these assumed values (See Appendix~\ref{app:Fitting results in the four mass bins} for a discussion).\footnote{For the lowest mass bin, we use the same assumed values as the $10.1 \leq \log(M_{*} /{\rm M}_{\odot}) < 10.5$ bin in \citet{Changetal2013}.}
   \item \textbf{Case III - Fitting $f_{\rm{ob}}$ only} -- We assume the values of the remaining six parameters to be the same as the best-fit values in \citet{Changetal2013}. 
\end{itemize}

The best-fit model population is determined using a maximum likelihood estimation method. To reduce computation time, we first generate a model grid within the parameter space being considered in each case. The spacing of the grid and the assumed values for the remaining parameters can be found in Table~\ref{tab_assumepara_summary}. We then compute the likelihood for each model population. For each $q$ in the observed distribution, we calculate the probability of observing a galaxy with this particular value of $q$ according to the probability distribution function $p(q)$ of the model population. The total log-likelihood $\ln(L)$ of the model is then computed by summing the log-probability of all $q$ in the observed axis ratio distributions. The best-fit model population is then taken as the one with the highest likelihood.

\begin{figure*}
  \centering
  \includegraphics[scale=0.62]{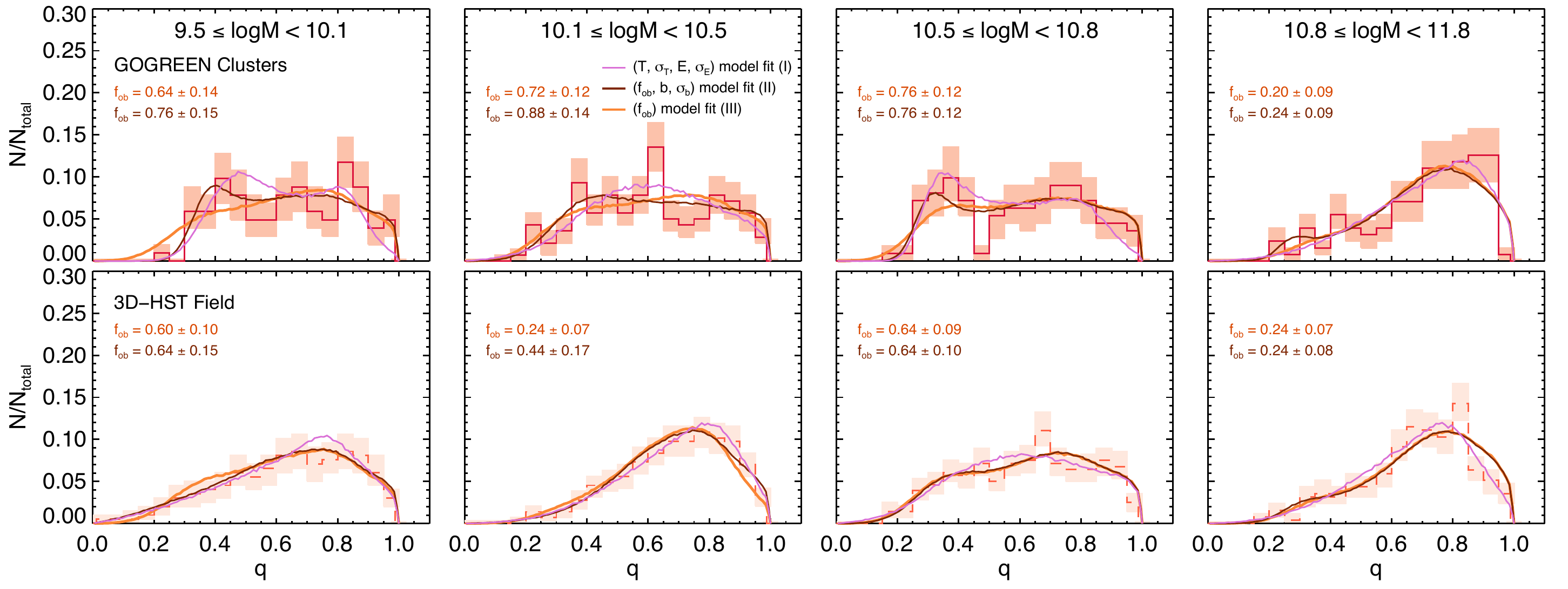}      
  \caption{The best-fit intrinsic shape models for the quiescent galaxy population in clusters (top) and the field (bottom) in different mass bins. The solid (dashed) line in each panel corresponds to the observed axis ratio distribution for the clusters (field), while the shaded area represents the $1\sigma$ variation of the distribution derived from the bootstrapped sample. A different binning width than in Figure~\ref{fig_qhisto_pass_sf_changbin} is used here to highlight the variation. The purple line in each panel corresponds to the best-fit model of case I, where a single-component model is used. The brown line corresponds to the best-fit model of case II, where $f_{\rm{ob}}$, $b$, $\sigma_b$ are free parameters. The orange line corresponds to the best-fit model of case III, where only $f_{\rm{ob}}$ is being fitted. The best-fit case II and III models for the $10.1 \leq \log(M_{*}) < 10.5$ mass bin both show a higher $f_{\rm{ob}}$ in clusters than the field, which suggests that clusters have a higher fraction of disk-like galaxies than the field at this redshift.}
    \label{fig_qhisto_pass_clus_field_changbin}
\end{figure*}

The uncertainty of the fitted parameters is derived by fitting the bootstrapped sample. The $1\sigma$ variation of the best-fit parameters of the bootstrapped sample is taken as the uncertainty. Examples of the corner plots of the fitting of the bootstrapped sample are shown in Appendix~\ref{app:Fitting results in the four mass bins}. The best-fit parameters can be found in Table~\ref{tab_fittedpara_summary}. We provide $p_{\rm{KS}}$, $p_{\rm{MW}}$, the reduced $\chi^2$ and the $p_{\rm{\chi^2}}$ values in Table~\ref{tab_fittedpara_summary} as rough goodness-of-fit indicators. The Akaike information criterion (AIC) value of the best-fit models is also provided in Table~\ref{tab_fittedpara_summary}.

\subsubsection{Evidence for a higher fraction of oblate quiescent galaxies in clusters}
\label{subsubsec:Higher fraction of oblate galaxies in clusters}
Figure~\ref{fig_qhisto_pass_clus_field_changbin} shows the result of the modeling for the four mass bins. The shaded regions show the $1\sigma$ variation of the axis ratio distribution derived from the bootstrapped sample, which we used to derive the uncertainty of the best-fit parameters.

We find that the observed axis ratio distribution of the field sample in all mass bins can be reasonably described by a single-component triaxial model (Case I). This is consistent with previous findings. \citet{Changetal2013} also found that the single-component model cannot be ruled out with only the use of high redshift data. On the other hand, there is some evidence that the single-component model is not able to accurately reproduce the shape of the distribution of the cluster sample in some of the mass bins (see Table~\ref{tab_fittedpara_summary} for the $p_{\rm{\chi^2}}$ values). This is intriguing, as the axis ratio distribution of local quiescent galaxies in this mass range is known to be not well described by a single component due to the existence of an oblate component \citep[e.g.,][]{Holdenetal2012}.

Assuming the triaxial component as found in \citet{Changetal2013} and fitting the oblate component parameters $f_{\rm{ob}}$, $b$, $\sigma_b$ (Case II), we find that the main differences between the cluster and field distribution lie in the fraction of oblate galaxies in the total population. We find tentative evidence that the cluster distribution in the three lower mass bins has a higher oblate fraction than the field, with the $10.1 \leq \log(M_{*} /{\rm M}_{\odot}) < 10.5$ mass bin showing the largest difference ($f_{\rm{ob,cluster}} = 0.88 \pm 0.14$  vs. $f_{\rm{ob,field}} = 0.44 \pm 0.17$). The best-fit value of the intrinsic axis ratio $b$ of the cluster sample is low in all four mass bins, with $b=0.25 - 0.35$, which is consistent with the value found in \citet{Changetal2013} for their $1<z<2.5$ sample ($b\sim0.29$). The best-fit $b$ for the field models are also consistent with this value, except in the two lower mass bins where both $b$ and $\sigma_b$ are poorly constrained (see Appendix~\ref{app:Fitting results in the four mass bins} for the corner plots). 

Fixing all parameters to the field values and only allowing $f_{\rm{ob}}$ to vary also gives a similar conclusion (Case III). The cluster sample has a much higher oblate fraction of $f_{\rm{ob,cluster}} = 0.72 \pm 0.06$ compared to the field $f_{\rm{ob,field}} = 0.24 \pm 0.07$ in the mass bin $10.1 \leq \log(M_{*} /{\rm M}_{\odot}) < 10.5$. However, we note that the best-fit model of the field in this bin is not a good representation of the data according to the  $p_{\rm{KS}}$ and $p_{\rm{MW}}$ values, presumably due to the limitation of the assumed models.

\begin{figure*}
  \centering
  \includegraphics[scale=0.62]{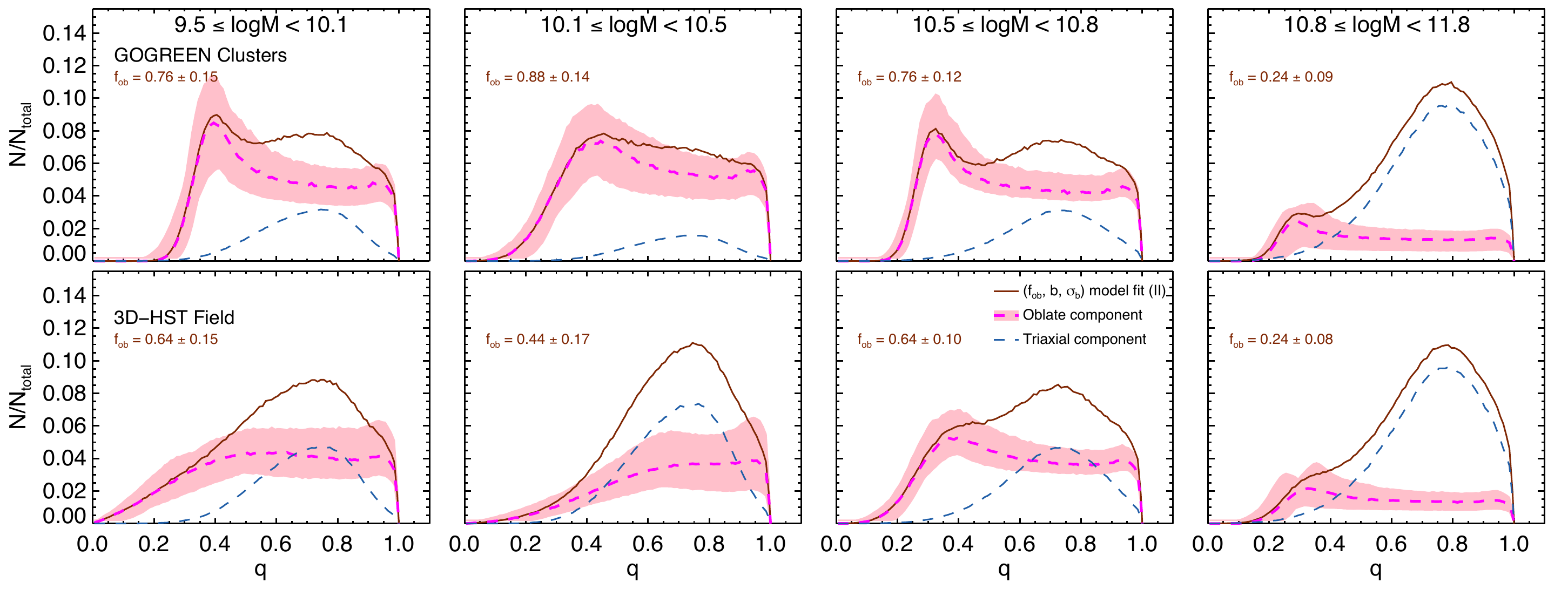}
  \caption{The axis ratio distributions for each of the oblate and triaxial components that contribute to the total distribution in the best-fit intrinsic shape models (Case II) for the quiescent galaxy population in clusters (top) and the field (bottom) in different mass bins. The brown line in each panel corresponds to the best-fit model of case II, also shown in Figure~\ref{fig_qhisto_pass_clus_field_changbin}. The magenta (blue) dashed line corresponds to the model axis ratio distributions of the oblate (triaxial) component. The shaded magneta area represents the $1\sigma$ variation of the fitted oblate component parameters derived from the bootstrapped sample. For simplicity only the variation of the oblate component is shown. Different y-scale than in Figure~\ref{fig_qhisto_pass_clus_field_changbin} is used to show the components. With a low best-fit $b$ value, the oblate component dominates the low-$q$ region of the cluster distribution.}
    \label{fig_qhisto_sep_component_changbin_4bin}
\end{figure*}

Figure~\ref{fig_qhisto_sep_component_changbin_4bin} shows the axis ratio distributions for each of the oblate and triaxial components separately, that contribute to the total distribution in the best-fit $f_{\rm{ob}}$, $b$, $\sigma_b$ models (Case II). In each panel, the brown line corresponds to the same best-fit model in Figure~\ref{fig_qhisto_pass_clus_field_changbin}, and the shaded magenta area represents the $1\sigma$ variation of the fitted oblate component parameters derived from the bootstrapped sample. We can see that the effect of having a larger $f_{\rm{ob}}$ of a low $b$ oblate population to the overall distribution. It results in a larger low-$q$ contribution relative to the total population and gives rise to a broader $q$ distribution that better describes the flatter shape of the cluster distributions, especially in the mass bin $10.1 \leq \log(M_{*} /{\rm M}_{\odot}) < 10.5$. The difference between the cluster and the field sample we described in Section~\ref{subsec:Projected axis ratio in clusters and the field} is therefore consistent with the existence of a larger population of flattened, oblate quiescent galaxies in clusters.

\subsubsection{The relation between axis ratio and S\'ersic index}     
\label{subsubsec:The relation between axis ratio and Sersic index}
We show that the main difference between the axis ratio distribution of the cluster and the field quiescent population is consistent with the existence of a larger population of flattened, oblate quiescent galaxies in clusters.
With such a low $b$ value, these oblate quiescent galaxies resemble disk-dominated galaxies as in the majority of the star-forming population. In the literature, the S\'ersic index is commonly used to separate disk- and bulge-dominated galaxies with the $n=2.5$ division.  Although we do not rely heavily on S\'ersic index in this work, here we examine the relation between axis ratio and S\'ersic index of the cluster sample as a consistency check.

\begin{figure}
  \centering
  \includegraphics[scale=0.525]{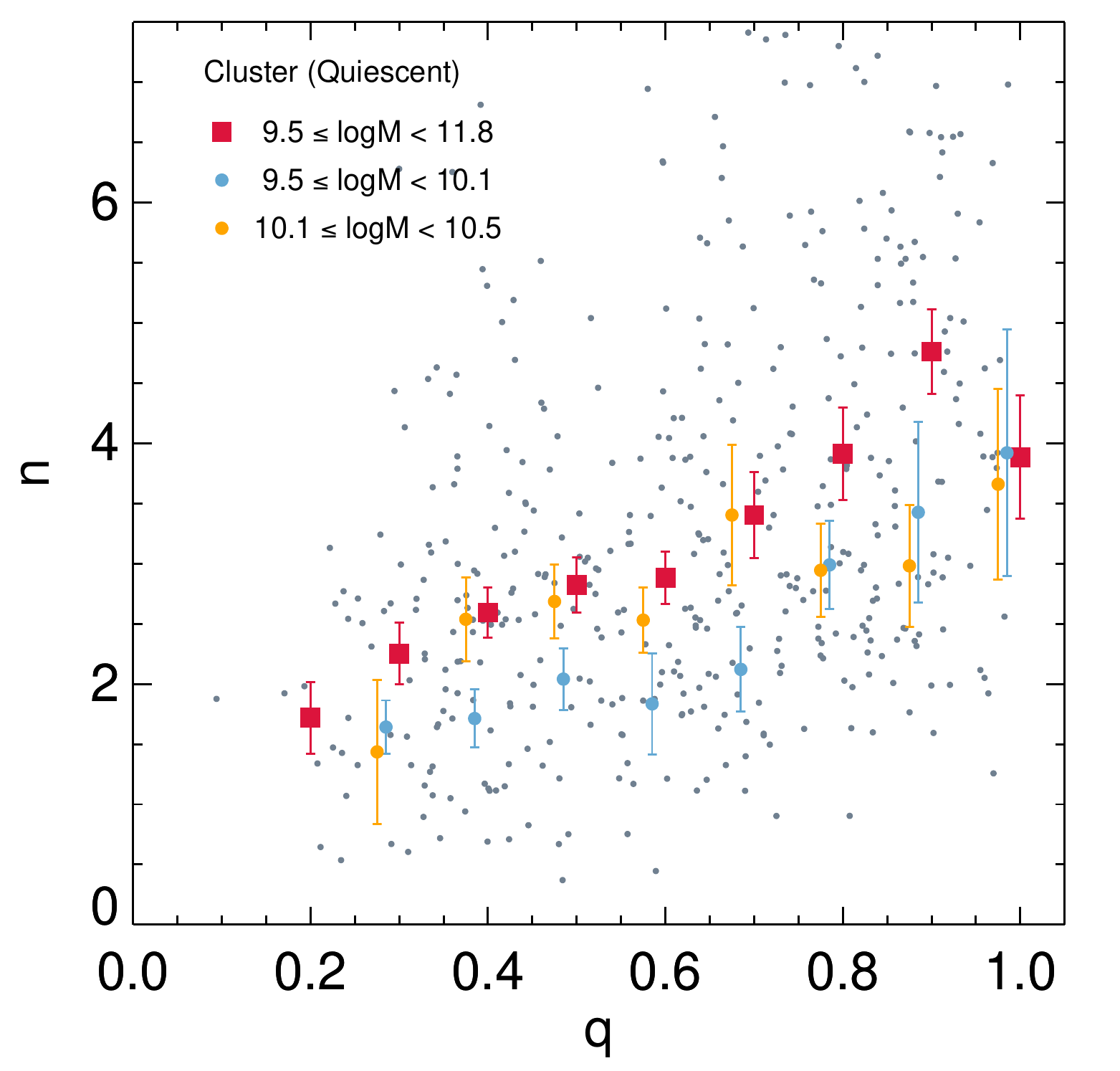}      
  \caption{The S\'ersic index of the quiescent galaxies in the cluster sample as a function of axis ratio. Grey points correspond to individual galaxies. Large red circles and error bars corresponds to the median $n$ of the entire sample and its standard error. Light blue and orange circles correspond to the median $n$ in the lowest and the second lowest mass bin, respectively. There is a positive trend between $q$ and $n$, as expected. The trend is mass dependent, with low-mass galaxies having on average lower values of $n$ for a given $q$.}
    \label{fig_nqr}
\end{figure}

Figure~\ref{fig_nqr} shows the axis ratio of the quiescent galaxies in the cluster sample as a function of their S\'ersic index.  We see a strong trend between the two, with low $q$ galaxies typically having on average smaller values of $n$. The trend is also mass-dependent, with low-mass galaxies having lower values of $n$ for a given $q$ bin. The excess population of $q\lesssim0.4$ galaxies in clusters have, on average, $n \lesssim 2.5$, and are consistent with the traditional selection of disky galaxies using S\'ersic indices. The strong trend supports our modeling results and interpretation that the difference in the axis ratio distribution originates from a larger population of flattened, disk-like galaxies in clusters compared to the field. 

The fact that we see an excess population of disk-like quiescent galaxies in clusters suggests that morphological transformation and quenching in clusters does not operate in the same way as in the field. In Section~\ref{sec:Discussion}, we discuss this further and explore possible implications together with other quantities.

\subsubsection{Properties of the massive quiescent galaxies in $10.8 \leq \log(M_{*} /{\rm M}_{\odot}) < 11.8$}                     
\label{subsubsec:The high mass bin}
In Section~\ref{subsec:Projected axis ratio in clusters and the field} we showed that massive quiescent galaxies $\log(M_{*} /{\rm M}_{\odot}) \geq 11$ in both clusters and the field are not only rounder than their low mass counterparts, they also have a narrower $q$ distribution, i.e. there is a lack of low-$q$ galaxies. There is also evidence that the median $q$ is offset to higher values in clusters compared to the field. Here we explore the underlying reason by examining the fitting results of the highest mass bin $10.8 \leq \log(M_{*} /{\rm M}_{\odot}) < 11.8$.\footnote{We note that our mass bin starts from $\log(M_{*} /{\rm M}_{\odot}) = 10.8$. We have also repeated the fits using a mass bin of $11.0 \leq \log(M_{*} /{\rm M}_{\odot}) < 11.8$ (59 galaxies) and found consistent results.}

There is a stark difference between the best-fit models of the highest mass bin and those of the three lower mass bins, as one can see in Figure~\ref{fig_qhisto_sep_component_changbin_4bin}. To reconstruct the shape of the axis ratio distribution of these massive galaxies, the best-fit case II (and also case III) models have much lower $f_{\rm{ob}}$ values compared to the three lower mass bins ($\sim 2.3 (1.7) \sigma$ for the cluster (field) sample). In fact, for both the cluster and the field sample, the $f_{ob}$s in the highest mass bin are statistically consistent with 0. This suggests that the contribution of a possible second oblate component is small. Indeed, the single-component triaxial model (case I) can describe the axis ratio distribution of the massive galaxies in both the cluster and field sample well.  

In section~\ref{subsec:Reconstructing the intrinsic shapes from the projected axis ratio distributions} we note that there is some weak indication that clusters have higher median $q$ compared to the field. From the best-fit parameters, the best-fit case I model of the cluster has a trixiality of $T = 0.36$ and an ellipticity of $E = 0.48$, compared to the field value of $T = 0.48$ and $E = 0.53$. Nevertheless, they are within $1\sigma$ uncertainty. The best-fit $\sigma_T$ and $\sigma_E$ are consistent with each other in clusters and the field. Therefore, with our sample there is no evidence of a difference in the intrinsic shape distribution of massive galaxies in clusters and the field.


\subsection{Caveats}                                          
\label{subsec:Effects of dust and stellar population gradients}
Dust obscuration, in particular dust lanes or dust gradients within the galaxy, can impact the axis ratio measurements and thus potentially affect the axis ratio distributions. It is also possible that the dust obscuration hides the disk structure of the galaxies, making them appear rounder \citep{vanderWeletal2014}. This has a larger effect on star-forming galaxies than quiescent galaxies due to their higher dust content. Previous studies have shown that the dust content of quiescent galaxies, probed via the rest-frame $V-J$ color, is low and does not correlate strongly with their axis ratio \citep{Changetal2013}. Here we check if this is also the case in clusters at this redshift range.

\begin{figure}
  \centering
  \includegraphics[scale=0.805]{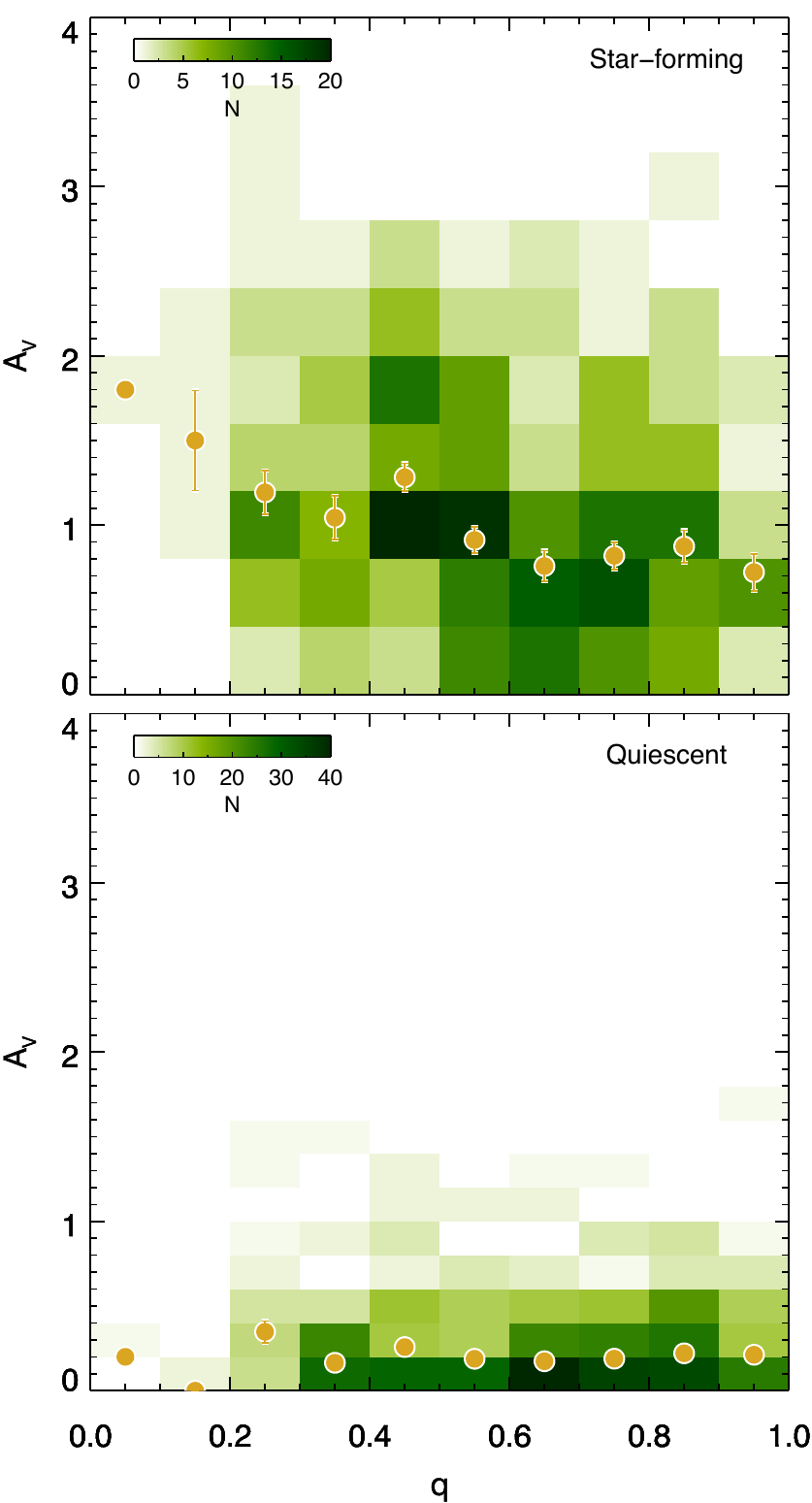}
  \caption{The dust extinction $A_V$ of the star-forming (top) and quiescent (bottom) galaxies in the cluster sample as a function of $q$. Orange circles and error bars correspond to the mean $A_V$ and its standard error. The green shading reflects the number of galaxies in a particular bin. The quiescent galaxies in our sample have low dust content. The dust extinction of the star-forming galaxies, on the other hand, shows a negative correlation with $q$.}
    \label{fig_Avq}
\end{figure}

Figure~\ref{fig_Avq} shows the dust extinction $A_{V}$ of star-forming and quiescent galaxies in the cluster sample, derived via SED fitting with \textsc{FAST} in steps of 0.2, as a function of $q$. We confirm that the quiescent galaxies in our cluster sample have, in general, a low dust content.  There is no obvious correlation between the axis ratio and the dust extinction. On the other hand, the star-forming population have a larger variation in $A_{V}$ at a certain $q$. Galaxies that have lower $q$ show higher dust extinction values on average, which is expected for an inclined / edge-on disk population.  A similar correlation can also be seen if the $V-J$ color is used instead. We have also checked that the field sample shows the similar correlations as the cluster sample. This suggests that the difference between the cluster and field axis ratios we see in the quiescent population is unlikely to be driven only by dust, but is due to the difference in their intrinsic shapes.

In addition, one potential caveat is related to the fact that our axis ratio measurements are measured from galaxy surface brightness profiles.  Luminosity-weighted structural measurements are not always a reliable measure for the mass distribution of a galaxy due to radial variation in stellar population properties such as age and metallicity. \citet{Suessetal2019a} derived mass-weighted structural parameters for galaxies in the field at $1.0<z<2.5$ and found that most galaxies, no matter star-forming or quiescent, show negative color / mass-to-light ratio ($M/L$) gradients with a strong redshift evolution, resulting in smaller mass-weighted sizes than the luminosity weighted ones. Although their work did not focus on axis ratios, they found that these gradients can account for most of the evolution in the mass-size relation. Similar effects have been observed in clusters at high redshift. \citet{Chanetal2018} showed that there are strong negative color gradients in quiescent galaxies in three clusters at $z\sim1.5$ that are consistent with being a combination of age and metallicity gradients. These gradients are even stronger in evolved clusters compared to the field at a similar redshift. The existence of these negative $M/L$ gradients suggests that the older stellar population in the bulge (which has a larger $M/L$) could be outshone by younger and bright stars in disks, which can bias the luminosity-weighted axis ratios to lower values.

Our \textit{HST} dataset does not have the necessary imaging bands to derive mass-weighted structural parameters \citep[see, e.g.][for a discussion of the methodologies and their requirements]{Chanetal2016, Suessetal2019a}, therefore we are not able to examine the effects of stellar population gradients on the axis ratio distributions. Instead we can use the results in \citet{Chanetal2018} as a reference. In addition to sizes, they compared both the luminosity-weighted and mass-weighted axis ratio distributions between clusters and the field at $z\sim1.5$\footnote{\citet{Chanetal2018} also found evidence that the $q$ distribution in the evolved clusters show an excess of low axis ratio ($q < 0.4$) compared to the field, albeit with low number statistics.}. They found that the mass-weighted axis ratios are on average smaller, with a broader mass-weighted distribution extending to lower $q$ than the luminosity-weighted $q$ distribution \citep[see Section 6.1.3 in][for more details]{Chanetal2018}. This implies that the low $q$ excess and flat $q$ distributions we see in these clusters are likely to be from genuine oblate structures instead of hidden bulges. Nevertheless, fully ruling out the possibility that the difference we see is caused by stellar population gradients would require a detailed analysis on the mass-weighted properties and color gradients in both clusters and the field.

Another potential caveat is that the axis ratio distribution modeling relies heavily on the assumption that the galaxy population is observed from random viewing angles. This assumption breaks down when the galaxies are intrinsically aligned with respect to a certain direction. There have been reports that cluster galaxies may be aligned radially towards the center of the cluster in low-redshift clusters \citep[e.g.,][]{Huangetal2018, Georgiouetal2019}. We therefore examined the radial alignment signal in our sample and measured the alignment of the galaxies with respect to the center of the cluster. The procedure and result are discussed in Appendix~\ref{app:Intrinsic alignments in cluster galaxies}.

We find that the average radial and tangential alignments for the cluster sample within $1R_{200}$ are consistent with zero. Examining the alignment signal as a function of cluster-centric radius, there is a weak evidence that the average radial alignment is positive in the region close to the cluster center ($\sim1.4 \sigma$ for $R< 0.2 R_{200}$). We conclude that the potential intrinsic alignments in the sample are unlikely to affect our results.

\begin{table*}
  \caption{Initial parameters and setup used for the model fitting}                      
  \centering
  \label{tab_assumepara_summary}
  \begin{tabular}{lccccccccc cccc}
  \hline
  \hline
  Type    & Mass ranges &            &                \multicolumn{7}{c}{Initial parameters}                            &   \multicolumn{4}{c}{Fitting grid spacing}   \\
              &                      &            & $f_{\rm{ob}}$  & $b$ &  $\sigma_b$  & $T$  & $\sigma_T$  & $E$\textsuperscript{a} &   $\sigma_E$   &    &  &   &  \\

  \hline
   \multicolumn{10}{c}{Case I -- Fitting $E, \sigma_E, T, \sigma_T$} &   $\Delta_T$ & $\Delta_{\sigma_T}$ & $\Delta_E$ & $\Delta_{\sigma_E}$  \\  [0.5ex]   
  \hline 
  Cluster/Field  &  $9.5 \leq \log(M_{*} /{\rm M}_{\odot}) < 10.1$      &      &  0  &  -  &  -  &  -  &   -  &  -  & -  & 0.04 & 0.02 & 0.01 & 0.02 \\ [0.5ex]  
  Cluster/Field  &  $10.1 \leq \log(M_{*} /{\rm M}_{\odot}) < 10.5$    &      &  0  &  -  &  -   &  - &   -  &  -  & -  & 0.04 & 0.02 & 0.01 & 0.02 \\ [0.5ex]  
  Cluster/Field  &  $10.5 \leq \log(M_{*} /{\rm M}_{\odot}) < 10.8$    &      &  0  &  -   &  -  &  - &   -  &  -  & -  & 0.04 & 0.02 & 0.01 & 0.02 \\ [0.5ex] 
  Cluster/Field  &  $10.8 \leq \log(M_{*} /{\rm M}_{\odot}) < 11.8$    &      &  0  &  -   &  -  &  - &   -  &  -  & -  & 0.04 & 0.02 & 0.01 & 0.02 \\ [0.5ex] 
  \hline  
   \multicolumn{10}{c}{Case II -- Fitting $f_{\rm{ob}}$, $b$, $\sigma_b$}  & $\Delta_{f_{\rm{ob}}}$ & $\Delta_{b}$ & $\Delta_{\sigma_b}$ & \\    [0.5ex] 
  \hline 
  Cluster/Field  &  $9.5 \leq \log(M_{*} /{\rm M}_{\odot}) < 10.1$      &      &  -  &  -  &  -  &   0.48  &   0.08  &  0.49  & 0.12 & 0.04 & 0.01 & 0.01 & -  \\ [0.5ex]
  Cluster/Field  &  $10.1 \leq \log(M_{*} /{\rm M}_{\odot}) < 10.5$      &      &  -  &  -  &  -  &   0.48  &   0.08  &  0.49  & 0.12 & 0.04 & 0.01 & 0.01 & -  \\ [0.5ex]
  Cluster/Field  &  $10.5 \leq \log(M_{*} /{\rm M}_{\odot}) < 10.8$    &      &  -  &  -  &  -   &  0.68 &   0.08  &  0.45  & 0.16 & 0.04 & 0.01 & 0.01 & - \\ [0.5ex]  
  Cluster/Field  &  $10.8 \leq \log(M_{*} /{\rm M}_{\odot}) < 11.8$    &      &  -  &  -   &  -  &  0.64 &   0.08  &  0.41  & 0.19 & 0.04 & 0.01 & 0.01 & - \\ [0.5ex]
  \hline
   \multicolumn{10}{c}{Case III -- Fitting $f_{\rm{ob}}$ only}  & $\Delta_{f_{\rm{ob}}}$ &   &   & \\    [0.45ex] 
  \hline 
  Cluster/Field  &  $9.5 \leq \log(M_{*} /{\rm M}_{\odot}) < 10.1$      &      &  -  &  0.28  &  0.09  &  0.48  &   0.08  &  0.49  & 0.12 & 0.04& -& -& - \\ [0.5ex] 
  Cluster/Field  &  $10.1 \leq \log(M_{*} /{\rm M}_{\odot}) < 10.5$      &      &  -  &  0.28  &  0.09  &  0.48  &   0.08  &  0.49  & 0.12 & 0.04& -& -& - \\ [0.5ex]  
  Cluster/Field  &  $10.5 \leq \log(M_{*} /{\rm M}_{\odot}) < 10.8$    &      &  -  &  0.28  &  0.08   &  0.68 &   0.08  &  0.45  & 0.16 & 0.04& -& -& - \\ [0.5ex]  
  Cluster/Field  &  $10.8 \leq \log(M_{*} /{\rm M}_{\odot}) < 11.8$    &      &  -  &  0.29  &  0.07  &  0.64 &   0.08  &  0.41  & 0.19  & 0.04& -& -& -\\ [0.5ex] 
  \hline 
\end{tabular}
\end{table*}

\begin{table*}
  \caption{Best-fit parameters of the models}                      
  \centering
  \label{tab_fittedpara_summary}
  \addtolength{\tabcolsep}{-1.1pt}
  \begin{tabular}{lccccccccc ccccc}
  \hline
  \hline
  Type    & Mass ranges &            &                \multicolumn{7}{c}{Best-fit parameters}                            &     \multicolumn{5}{c}{GoF}  \\
              &                      &            & $f_{\rm{ob}}$  & $b$ &  $\sigma_b$  & $T$  & $\sigma_T$  & $E$\textsuperscript{a} &  $\sigma_E$    & $p_{\rm{KS}} $   &  $p_{\rm{MW}} $  & $p_{\chi^2}$ & $\chi^2 / \nu$  & AIC \\
              &                      &            &                   &      &     &       &       &     &            &    &    &    &    &   \\
  \hline
   \multicolumn{15}{c}{Case I -- Fitting $E, \sigma_E, T, \sigma_T$ }   \\  [0.5ex]   
  \hline 
  Cluster  &  $ 9.5 \leq \log(M_{*} /{\rm M}_{\odot}) < 10.1$  &  &  -  &  -  & - &  $0.36^{+0.52}_{-0.00}$  &   $0.00^{+0.14}_{-0.00}$  &   $0.63^{+0.04}_{-0.17}$  &  $0.06^{+0.24}_{-0.06}$  &   $0.03$   &   $0.06$  &      $0.02$ & $2.13$     &  897.6  \\ [0.5ex]
  Cluster  &  $ 10.1 \leq \log(M_{*} /{\rm M}_{\odot}) < 10.5$  &  &  -  &  -  & - &  $0.88^{+0.08}_{-0.52}$  &   $0.00^{+0.22}_{-0.00}$  &   $0.53^{+0.13}_{-0.04}$  &  $0.22^{+0.10}_{-0.12}$  &   $0.63$   &   $0.49$  &     $0.02$ & $1.99$    &  1233.27    \\ [0.5ex]
  Cluster  &  $ 10.5 \leq \log(M_{*} /{\rm M}_{\odot}) < 10.8$  &  &  -  &  -  & - &  $0.36^{+0.08}_{-0.00}$  &   $0.10^{+0.06}_{-0.10}$  &   $0.72^{+0.02}_{-0.02}$  &  $0.02^{+0.24}_{-0.02}$  &   $0.12$   &   $0.04$  &     $0.76$ &  $0.66$   &  978.7    \\ [0.5ex]
  Cluster  &  $ 10.8 \leq \log(M_{*} /{\rm M}_{\odot}) < 11.8$  &  &  -  &  -  & - &  $0.36^{+0.04}_{-0.00}$  &   $0.02^{+0.02}_{-0.02}$  &   $0.48^{+0.05}_{-0.03}$  &  $0.22^{+0.04}_{-0.04}$  &   $0.80$   &   $0.34$  &     $0.18$ &  $1.38$   &  1076.6   \\ [0.5ex]

  Field  &  $ 9.5 \leq \log(M_{*} /{\rm M}_{\odot}) < 10.1$  &  &  -  &  -  & - &  $0.48^{+0.44}_{-0.12}$  &   $0.02^{+0.38}_{-0.02}$  &   $0.64^{+0.05}_{-0.17}$  &  $0.30^{+0.06}_{-0.06}$  &   $0.75$   &   $0.34$  &      $0.89$ & $0.57$     &  1757.0   \\ [0.5ex]
  Field  &  $ 10.1 \leq \log(M_{*} /{\rm M}_{\odot}) < 10.5$  &  &  -  &  -  & - &  $0.44^{+0.24}_{-0.08}$  &   $0.04^{+0.34}_{-0.04}$  &   $0.50^{+0.02}_{-0.04}$  &  $0.20^{+0.02}_{-0.02}$  &   $0.89$   &   $0.47$  &     $0.98$ & $0.35$     &  2433.0   \\ [0.5ex]
  Field  &  $ 10.5 \leq \log(M_{*} /{\rm M}_{\odot}) < 10.8$  &  &  -  &  -  & - &  $0.92^{+0.00}_{-0.32}$  &   $0.00^{+0.12}_{-0.00}$  &   $0.51^{+0.07}_{-0.04}$  &  $0.24^{+0.04}_{-0.02}$  &   $0.32$   &   $0.16$  &     $0.39$ & $1.06$    &  2464.3    \\ [0.5ex]
  Field  &  $ 10.8 \leq \log(M_{*} /{\rm M}_{\odot}) < 11.8$  &  &  -  &  -  & - &  $0.48^{+0.04}_{-0.08}$  &   $0.02^{+0.02}_{-0.02}$  &   $0.53^{+0.03}_{-0.03}$  &  $0.20^{+0.04}_{-0.02}$  &   $0.31$   &   $0.10$  &     $0.06$  & $1.67$    &  2142.4  \\ [0.5ex]

  \hline
   \multicolumn{15}{c}{Case II -- Fitting $f_{\rm{ob}}$, $b$, $\sigma_b$ }     \\    [0.5ex] 
  \hline 
  Cluster  &  $ 9.5 \leq \log(M_{*} /{\rm M}_{\odot}) < 10.1$  &  &  $0.76^{+0.16}_{-0.16}$  &   $0.35^{+0.02}_{-0.02}$  &   $0.03^{+0.03}_{-0.03}$  &  -  &  -  &  -  & - &   $0.31$   &   $0.20$  &      $0.18$ &  $1.37$   &  894.4  \\ [0.5ex]
  Cluster  &  $ 10.1 \leq \log(M_{*} /{\rm M}_{\odot}) < 10.5$  &  &  $0.88^{+0.12}_{-0.20}$  &   $0.32^{+0.04}_{-0.03}$  &   $0.09^{+0.03}_{-0.03}$  &  -  &  -  &  -  & - &   $0.88$   &   $0.41$  &     $0.06$ & $1.66$    & 1230.5  \\ [0.5ex]
  Cluster  &  $ 10.5 \leq \log(M_{*} /{\rm M}_{\odot}) < 10.8$  &  &  $0.76^{+0.16}_{-0.08}$  &   $0.28^{+0.02}_{-0.01}$  &   $0.02^{+0.02}_{-0.02}$  &  -  &  -  &  -  & - &   $0.78$   &   $0.40$   &    $0.93$ & $0.46$    & 971.1  \\ [0.5ex]
  Cluster  &  $ 10.8 \leq \log(M_{*} /{\rm M}_{\odot}) < 11.8$  &  &  $0.24^{+0.08}_{-0.12}$  &   $0.25^{+0.06}_{-0.03}$  &   $0.01^{+0.03}_{-0.01}$  &  -  &  -  &  -  & - &   $0.22$   &   $0.08$   &    $0.35$ & $1.11$     & 1074.9 \\ [0.5ex]
  
  Field  &  $ 9.5 \leq \log(M_{*} /{\rm M}_{\odot}) < 10.1$  &  &  $0.64^{+0.16}_{-0.12}$  &   $0.24^{+0.08}_{-0.08}$  &   $0.21^{+0.08}_{-0.10}$  &  -  &  -  &  -  & - &   $0.85$   &   $0.37$   &       $0.90$  &  $0.57$    &  1755.0 \\ [0.5ex]
  Field  &  $ 10.1 \leq \log(M_{*} /{\rm M}_{\odot}) < 10.5$  &  &  $0.44^{+0.16}_{-0.12}$  &   $0.45^{+0.05}_{-0.11}$  &   $0.22^{+0.07}_{-0.08}$  &  -  &  -  &  -  & - &   $0.39$   &   $0.16$   &      $0.82$   &  $0.64$    & 2434.0  \\ [0.5ex]
  Field  &  $ 10.5 \leq \log(M_{*} /{\rm M}_{\odot}) < 10.8$  &  &  $0.64^{+0.12}_{-0.08}$  &   $0.29^{+0.03}_{-0.03}$  &   $0.08^{+0.02}_{-0.03}$  &  -  &  -  &  -  & - &   $0.94$   &   $0.46$   &      $0.64$   &  $0.83$   & 2458.6  \\ [0.5ex]
  Field  &  $ 10.8 \leq \log(M_{*} /{\rm M}_{\odot}) < 11.8$  &  &  $0.24^{+0.08}_{-0.08}$  &   $0.27^{+0.05}_{-0.07}$  &   $0.05^{+0.03}_{-0.05}$  &  -  &  -  &  -  & - &   $0.12$   &   $0.25$   &      $0.06$  &  $1.70$   &  2143.7 \\ [0.5ex]

  \hline
   \multicolumn{15}{c}{Case III -- Fitting $f_{\rm{ob}}$ only}     \\  [0.5ex] 
  \hline 
  Cluster  &  $ 9.5 \leq \log(M_{*} /{\rm M}_{\odot}) < 10.1$  &  &  $0.64^{+0.12}_{-0.16}$  &  -  &  -  &  -  & -    & -  & - &   $0.28$   &   $0.15$    &      $0.38$  &  $1.07$  &  893.6  \\ [0.5ex]  
  Cluster  &  $ 10.1 \leq \log(M_{*} /{\rm M}_{\odot}) < 10.5$  &  &  $0.72^{+0.12}_{-0.08}$  &  -  &  -  &  -  & -    & -  & - &   $0.47$   &   $0.50$   &        $0.07 $  &  $ 1.57$   & 1228.4    \\ [0.5ex]  
  Cluster  &  $ 10.5 \leq \log(M_{*} /{\rm M}_{\odot}) < 10.8$  &  &  $0.76^{+0.16}_{-0.08}$  &  -  &  -  &  -  & -    & -  & - &   $0.50$   &   $0.34$   &        $0.89$  &  $0.55$    &  972.0    \\ [0.5ex]  
  Cluster  &  $ 10.8 \leq \log(M_{*} /{\rm M}_{\odot}) < 11.8$  &  &  $0.20^{+0.12}_{-0.08}$  &  -  &  -  &  -  & -    & -  & - &   $0.39$   &   $0.17$    &        $0.43$  &  $1.02$    & 1072.0   \\ [0.5ex]  
  
  Field  &  $ 9.5 \leq \log(M_{*} /{\rm M}_{\odot}) < 10.1$  &  &  $0.60^{+0.04}_{-0.12}$  &  -  &  -  &  -  & -    & -  & - &   $0.63$   &   $0.25$    &        $0.70$  &  $0.79$    &  1756.3   \\ [0.5ex]  
  Field  &  $ 10.1 \leq \log(M_{*} /{\rm M}_{\odot}) < 10.5$  &  &  $0.24^{+0.08}_{-0.08}$  &  -  &  -  &  -  & -    & -  & - &   $0.02$   &   $0.00$    &       $0.36$  &  $1.09$   &  2435.4  \\ [0.5ex]  
  Field  &  $ 10.5 \leq \log(M_{*} /{\rm M}_{\odot}) < 10.8$  &  &  $0.64^{+0.08}_{-0.08}$  &  -  &  -  &  -  & -    & -  & - &   $0.96$   &   $0.46$   &        $0.73$  &  $0.76$   &  2454.7   \\ [0.5ex]  
  Field  &  $ 10.8 \leq \log(M_{*} /{\rm M}_{\odot}) < 11.8$  &  &  $0.24^{+0.08}_{-0.08}$  &  -  &  -  &  -  & -    & -  & - &   $0.10$   &   $0.20$   &        $0.09$  &  $1.54$   &  2140.3   \\ [0.5ex]  
  \hline
\end{tabular}
\end{table*}

\section{Discussion}
\label{sec:Discussion}
The goal of this work is to examine the effect of environment on galaxy structural properties.  We find that the axis ratio distributions of quiescent galaxies in clusters and the field are distinct.  By modeling the axis ratio distribution in different mass bins, we find evidence that quiescent galaxies in clusters have a higher fraction of flattened, oblate galaxies than the field in the intermediate mass range. The most massive cluster galaxies, with $10.8 \leq \log(M_{*} /{\rm M}_{\odot}) < 11.8$, have a low fraction of oblate galaxies, and those in clusters exhibit a lower ellipticity than the field.  Here we discuss the implications of these results.  We begin with the result of the massive galaxies in Section~\ref{subsec:Evolution of the massive galaxies in clusters}.  We then discuss the result of the intermediate mass range in Section~\ref{subsec:The effect of environmental quenching on galaxy structure} in the context of a simple toy accretion model. In Section~\ref{subsec:Morphological transformation in context of the early mass quenching scenario}, we explore the implication of our results in the context of the ``early mass-quenching'' scenario, discussed in \citet{vanderBurgetal2020}. In Section~\ref{The age variation in the oblate quiescent population} we combine our results with the measured stellar age of a subset of the population to explore the underlying physical mechanism that drives environmental quenching.

\subsection{Evolution of the massive quiescent galaxies in clusters}                      
\label{subsec:Evolution of the massive galaxies in clusters}
The fact that massive galaxies, in both clusters and the field, have significantly different axis ratio distributions than their low-mass counterparts has been observed in previous studies at lower redshifts. \citet{vanderWeletal2009} showed that there is a lack of low-$q$ galaxies ($q<0.6$) in the local massive quiescent population at $\log(M_{*} /{\rm M}_{\odot}) \geq 11.0$. \citet{Holdenetal2012} reported a similar transition exists at $z\sim0.7$. Similar effects has been found in clusters locally and at intermediate redshifts \citep[e.g.,][]{Vulcanietal2011}. We show that this is also true at $1.0 < z < 1.4$.

These results are often interpreted as evidence that massive quiescent galaxies experienced repeated major and minor mergers, which make them appear rounder gradually. In simulations, it is shown that the importance of mergers increases as a function of mass \citep[e.g.][]{WangKauffmannetal2008, DeLuciaetal2010, Quetal2017}. This picture is also largely supported by the observed kinematics of these low-$z$ massive quiescent galaxies \citep[i.e., the slow rotators, e.g.][]{Emsellemetal2011}, their observed number density evolution, and the merger rates \citep[e.g.,][]{Manetal2016}. The offset to higher $q$ values in the cluster population may therefore be a consequence of massive galaxies in clusters having experienced more mergers than the field at the epoch of observation.  

This can either be due to a) merger rates in clusters are (or were) elevated compared to the field, or b) they are in a more advanced evolutionary stage compared to their field counterparts. Observations have shown that major mergers can play an important role in mass assembly of the brightest cluster galaxies at $z\sim1$ \citep[e.g.,][]{Lidmanetal2013}, although this is less likely for satellite galaxies. Nevertheless, we note that the massive cluster population we see is likely composed of galaxies that were central galaxies for most of their lifetime \citep[e.g.][]{DeLuciaetal2012}, hence the merger could have happened before or when the galaxy was being accreted.  On the other hand, a recent stellar kinematics study of massive galaxies ($\log(M_{*} /{\rm M}_{\odot}) \geq 11.0$) at $0.6 <z< 1.0$ found that only those in the densest environments are primarily slow rotators \citep{Coleetal2020}. They suggest that slow rotators are being built in dense environments first through repeated minor mergers and hence they are more kinematically evolved compared to the field. We note that this relation, known as the kinematic morphology-density relation \citep{Cappellarietal2011}, has been studied extensively in dense environments in the local Universe. Several studies showed that the fraction of massive slow rotators increases with galaxy number density and the locations of the slow rotators are strongly correlated with peak densities in groups and clusters \citep[e.g.][]{DEugenioetal2013, Houghtonetal2013, Grahametal2019}, although there are studies suggesting that this relation is driven by the stellar mass distribution with the environment, not environment itself \citep[e.g][]{Broughetal2017, Vealeetal2017}. Although we are not able to distinguish the effects of major and minor mergers, our result supports the picture that mergers are a crucial component in the evolution of the massive galaxies in clusters at $1.0 < z < 1.4$.

\subsection{The effect of environmental quenching on galaxy structure}                           
\label{subsec:The effect of environmental quenching on galaxy structure}
In this section we aim to combine the results of the axis ratio distribution modeling with the quenched fractions in our samples to quantify the extent of morphological transformation. It is essential to explore the relationship between environmental quenching and morphological transformation at this redshift, as the dominant environmental mechanism(s) needs to be able to explain the morphological mix or the morphological signatures of the population. Here we consider a simple accretion model and compute the expected axis ratio distributions under various assumptions. We then compare these distributions with the observed $q$ distribution of the quiescent galaxies in clusters.

We start by computing the quenched fraction $f_{\rm{Q}}$ in the cluster and field samples for the four mass bins, following the method of \citet{vanderBurgetal2020}. Note that the field $f_{\rm{Q}}$ value here is different from their work, as the adopted field sample comes from CANDELS/3D-\textit{HST} as opposed to UltraVISTA \citep{Muzzinetal2013c} in \citet{vanderBurgetal2020}.  Over the whole mass range we considered in this work ($9.5 \leq \log(M_{*} /{\rm M}_{\odot}) < 11.8$), the quenched fraction in clusters is more than three times higher than in the field ($f_{\rm{Q,clus}} = 0.58$ vs. $f_{\rm{Q,field}} = 0.16$).

Assuming that mass quenching occurs in the same way in clusters as in the field, we can compute the fraction of the excess quenched galaxies, i.e., quenched via environmental processes, in the cluster quiescent sample in each mass bin\footnote{Note that $f_{\rm{EQ}}$ describes the fraction of environmentally quenched galaxies in the quiescent galaxy population. It is different from the Quenched Fraction Excess (QFE) in \citet{vanderBurgetal2020}, which describes the fraction of galaxies that would have been star-forming in the field but are quenched by the environment.}: $f_{\rm{EQ}} = (f_{\rm{Q,clus}} - f_{\rm{Q,field}}) / f_{\rm{Q,clus}}$. The quantity $f_{\rm{EQ}}$ varies in different mass bins, with the lowest mass bins having the largest value.

We first investigate the effect of having such an excess population on the field quiescent axis ratio distributions in each mass bin. To do this, we compute the expected axis ratio distributions of a galaxy population with a fraction $f_{\rm{EQ}}$ of ``accreted galaxies''. Star-forming galaxies, randomly drawn from the field star-forming $q$ distributions ($P_{\rm{SF}}(q)$) of the corresponding mass bin, are added into the field quiescent $q$ distributions until the fraction of this accreted population reaches $f_{\rm{EQ}}$. Using the star-forming $q$ distributions from the field population as the parent distribution mimics the effect of having \textit{no} morphological transformation, in the sense that the accreted galaxies retain the same morphology (axis ratio) as they would have had in the field. 

\begin{figure*}
  \centering
  \includegraphics[scale=0.79]{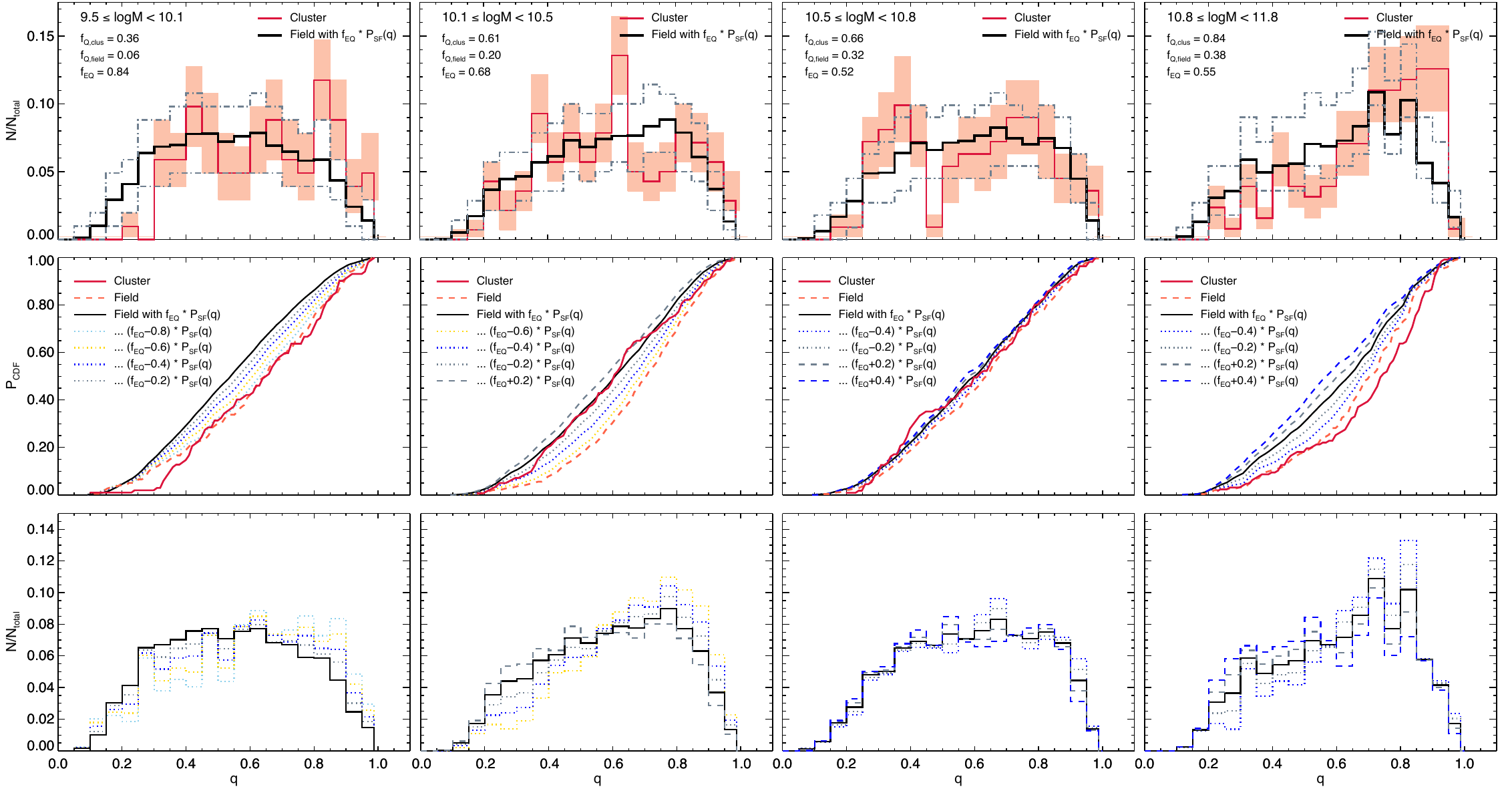}      
  \caption{The expected axis ratio distribution of the quenched galaxies in clusters from a simple accretion model.  Top: Comparison of the observed axis ratio distributions in clusters (red line) to the expected axis ratio distribution of the model with a $f_{\rm{EQ}}$ fraction of star-forming population ($P_{\rm{SF}}(q)$) (black line) in different mass bins. Grey dot-dashed lines correspond to the $1\sigma$ variation of the expected distribution.  Middle: The cumulative distribution function of the expected axis ratio distribution of models with different accreted star-forming fractions. Bottom: The axis ratio distribution of the models with different accreted star-forming fractions. See Section~\ref{subsec:The effect of environmental quenching on galaxy structure} for details.}
    \label{fig_f_eq_model}
\end{figure*}

The top row of Figure~\ref{fig_f_eq_model} shows the result of this accretion model in different mass bins. The black line corresponds to the distribution with $f_{\rm{EQ}}$ of star-forming population mixed in, while the grey lines correspond to the $1\sigma$ variation of the expected distribution derived from bootstrapping, in which the bootstrapped samples contain the same number of galaxies as the cluster quiescent sample.  Comparing with the observed $q$ distribution in the clusters, we find that the expected distribution of the accretion model matches the overall shape of the cluster distribution in the two intermediate mass bins $10.1 \leq \log(M_{*} /{\rm M}_{\odot}) < 10.5$ and $10.5 \leq \log(M_{*} /{\rm M}_{\odot}) < 10.8$, with $p_{\rm{KS}}$ values of $ \simeq 0.56 $ and $0.39$, respectively. The model cannot match the shape of the highest and lowest mass bins, with both bins having $p_{\rm{KS}}$ value of $ \simeq 0$.

The middle and bottom rows of Figure~\ref{fig_f_eq_model} illustrates the effect of varying the accreted star-forming fraction to the axis ratio distribution. In general, increasing the fraction of the accreted population increases the abundance of the low $q$ galaxies, resulting in a broader $q$ distribution. Varying this fraction can also be regarded as changing the amount of morphological transformation of the accreted galaxies. Since we have an independent constraint on the fraction of the environmental quenched population, if an accreted star-forming fraction that is smaller than $f_{\rm{EQ}}$ matches the data well, it suggests that part of the accreted population has transformed, such that their distribution matches closer to the field quiescent population than the star-forming ones.

Interestingly, we find that an accreted fraction that is consistent with $f_{\rm{EQ}}$ best-fits the cluster data in the two intermediate mass bins, as seen from the cumulative distributions. For $10.1 \leq \log(M_{*} /{\rm M}_{\odot}) < 10.5$, the KS test indicates that models with $f_{\rm{EQ}} \pm 0.2$ are also acceptable representations.  Hence for the full $10.1 \leq \log(M_{*} /{\rm M}_{\odot}) < 10.8$ range, our model suggests that the observed axis ratio distributions are statistically consistent with a scenario where no morphological transformation occurs after the galaxy was accreted and quenched environmentally.

For the lowest and highest mass bins, it is intriguing to see none of the model distributions are a good representation of the data. For the highest mass bin, since the cluster distribution are shifted to higher $q$ relative to the field, it is unsurprising that our model will not work. As we discussed in Section~\ref{subsec:Evolution of the massive galaxies in clusters}, mergers are likely a crucial component in the evolution of these massive galaxies which changes their axis ratio distributions. On the other hand, although the lowest mass bin has a high $f_{\rm{EQ}}$, our model fails to reproduce the observed distribution. The model may be too simplistic to reproduce the characteristics of the observed distribution, particularly in the low-$q$ region ($q < 0.3$). It is also possible that our sample size for the lowest-mass bin is simply too small.

An assumption we made is that the accreted population has the same axis ratio distribution as the star-forming population in the field at the same epoch. This might not be true if the star-forming population is accreted earlier. We repeat the analysis using the axis ratio distribution of two higher redshift samples of star-forming galaxies in the field at $1.5<z<2.0$ and $2.0<z<2.5$ ($1.4$ and $2.2$ Gyr earlier). The two samples have $1189$ and $995$ galaxies, respectively. We find that the results remain unchanged, primarily due to the fact that their axis ratio distributions are consistent with the distribution at GOGREEN redshifts ($p_{\rm{KS}} \simeq 0.3$ and $0.2$, respectively).

\begin{figure*}
  \centering
  \includegraphics[scale=0.420]{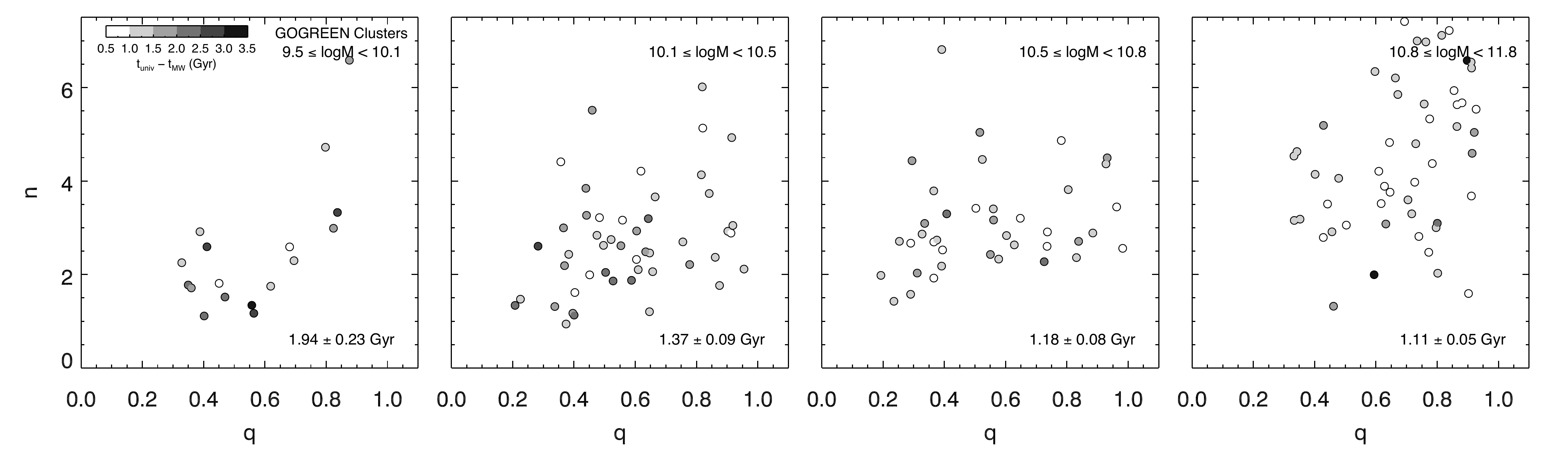}      
  \caption{The mass-weighted ages of the quiescent galaxies in clusters (in unit of cosmic time $t_{\rm{univ,z}} - t_{\rm{mw}}$) as a function of $n$ and $q$ for the four mass bins. Only galaxies that have both reliable stellar age and structural parameter measurements are shown. There is no clear trend in age with $n$ or $q$.  Disk-like galaxies (e.g., $n < 2.5$ or $q < 0.4$) show a large variation in age. \\}
    \label{fig_mwa_nq2dpts}
\end{figure*}

\subsection{Morphological transformation in context of the early mass quenching scenario}         
\label{subsec:Morphological transformation in context of the early mass quenching scenario}
\citet{vanderBurgetal2020} demonstrated that the shape of the stellar mass function (SMF) in star-forming and quiescent galaxies is indistinguishable between cluster and field at $1.0 <z < 1.4$.  This leads to the attractive explanation that galaxies in clusters quench through the same mass-quenching process as those in the field, but at an earlier time and at a higher rate. Under this ``early mass-quenching'' scenario, they found that a difference in formation time of $\gtrsim1$ Gyr can result in a quenched fraction difference that is consistent with the data. This difference in formation time would manifest itself in the age difference between the cluster and field quiescent population \citep[See the discussion in][for more details on the relationship between the two quantities]{Webbetal2020}. Although cluster galaxies are on average older than the field \citep{Webbetal2020}, the observed difference ($0.31_{^{-0.33}}^{_{+0.51}}$ Gyr) is inconsistent with the required difference in formation time, as \citet{vanderBurgetal2020} also pointed out.

It is also unclear how to reconcile our finding of a higher fraction of oblate galaxies in the cluster population with this ``early mass quenching'' scenario. Observational studies suggest that morphological transformation is a prerequisite for quenching of star-formation in central galaxies in the field, presumably as a result of the compaction phase that the galaxies underwent before quenching through internal feedback processes \citep[e.g.,][]{Tacchellaetal2015, Zolotovetal2015, Barroetal2017}. The compaction, which originated from mergers or disk instability, leads to the formation of a bulge while the disk slowly fades due to the declining star formation. The strong association between morphological properties and quiescence \citep[e.g.,][]{Langetal2014, Whitakeretal2017} is, therefore, a signature of this ``mass quenching'' process.  If the same process is responsible for the cluster population, we would naively expect the cluster population to have fewer disk-like oblate galaxies than in the field, given that the cluster population had a head-start. This seems to go against our findings except for the highest mass bin. This suggests that at least part of the environmentally quenched population originated from a different process than mass-quenching.

We note that other quenching processes that have similar effects to the morphologies are also unlikely to be directly responsible for the ``environmentally quenched'' oblate population.  An example is major mergers\footnote{\citet{vanderBurgetal2020} included a merger-quenching recipe as implemented by \citet{Pengetal2010} in the early mass-quenching model. They found that the inclusion of merger quenching has no significant effects on the SMF or the quenched fractions.}, which are expected to lead to formation of spheroids and quench galaxies through gas funnelling and triggering central starbursts \citep{Hopkinsetal2009a, Hopkinsetal2010}. Hence if the excess quenching is due to an elevated merging rate in clusters, we would also expect to see fewer disk-like galaxies.

\subsection{The age variation in the oblate quiescent population}         
\label{The age variation in the oblate quiescent population}
Here we explore the age variation in the cluster quiescent population by combining our results with the measured stellar ages. We utilize the mass-weighted age measurements from \citet{Webbetal2020}, derived using SED fitting of the GOGREEN spectroscopy and photometry.  We consider the mass-weighted age in units of cosmic time ($t_{\rm{univ,z}} - t_{\rm{mw}}$, i.e. the formation time, younger galaxies having a larger/later formation time), taking into account the redshift difference between the clusters in the sample. We refer the readers to \citet{Webbetal2020} for the methodologies. Limited by the FOV of the \textit{HST} images, there are $175$ quiescent cluster members in the \citet{Webbetal2020} sample that have both structural parameters and stellar age measurements.      

Figure~\ref{fig_mwa_nq2dpts} shows the mass-weighted ages of the sample as a function of $n$ and $q$ in different mass bins. The ages are correlated with mass, with the median formation time and its standard error decrease from the lowest mass bin ($1.94 \pm 0.23$ Gyr) to the highest mass bin ($1.11 \pm 0.05$ Gyr). We find that the ages do not show a significant trend in $q$ alone.  While there are formation times that are as late as $2.7$ Gyr at low $q$, the median formation time and its standard error at low $q$ ($q < 0.4, 1.31\pm0.10$ Gyr) for the whole sample are consistent with those at high $q$ ($q \geq 0.4, 1.21 \pm 0.05$ Gyr). On the other hand, we find tentative evidence that the median formation time is higher for low $n$ galaxies ($n < 2.5, 1.45 \pm 0.10$) compared to high $n$ ones ($n \geq 2.5, 1.15 \pm 0.04$) ($\sim2 \sigma$ difference). We also find similar results by excluding the galaxies in the highest mass bin.

The lack of a strong trend in mass-weighted age with $n$ or $q$ suggests that disk-like galaxies comprise objects with a mix of ages that are not significantly different from the bulk of the population.  Even if we select a strictly `disky' sample with both $n<2.5$ and $q<0.4$, the median formation time is $1.42 \pm 0.14$ Gyr with a standard deviation of $0.44$ Gyr, which shows that it comprises both young and old galaxies.  Hence not all disk-like galaxies in the cluster sample were recently quenched. Instead, some of them are formed and quenched early and remained a disk until the epoch of observation.  

The age variation we see implies that the quenching process that produces the disk excess has been occurring since high redshift. One possibility is that the quenching may happen when or even before the galaxy was accreted into the cluster. \citet{Fossatietal2017} studied the fraction of galaxies that were quenched by environmental processes in the five CANDELS/3D-\textit{HST} fields and showed that satellite galaxies are efficiently environmentally quenched in haloes of all masses at this redshift. Groups in GOGREEN also show higher quenched fraction relative to the field, mostly at the high-mass end (Reeves et al. 2021, in prep.). Since a cluster grows not only by accreting field central galaxies but also smaller haloes \citep{McGeeetal2009}, part of the quiescent population in the clusters will be galaxies that have gone through this `pre-processing' stage. \citet{Fossatietal2017} reported that the inferred quenching time of the satellites is consistent with them being quenched by a gas exhaustion ``starvation''-like mechanism, similar to the ``over-consumption'' model proposed by \citet{McGeeetal2014}. 

In the ``over-consumption'' model, galaxies are quenched as they essentially run out of fuel to sustain star formation. Outflows from star formation are expected to shorten the quenching time, an effect that in turn depends on the star-formation rate. Although there are no studies yet examining the effect of such a model on galaxy morphologies, as galaxies are primarily quenched via gas exhaustion, they will likely retain their disk-like structure and not undergo a drastic morphological change. Simulation studies following the evolution of disk galaxies in group environments also show that their morphological evolution is dependent on the initial inclination of the disc, and that central bulges are not produced or enhanced from interactions with the group environment alone \citep{Villalobosetal2012}.

It remains to be seen whether pre-processing can fully explain the excess quenching we see in clusters at this redshift range.  The effects of pre-processing have been established in clusters at local and intermediate redshifts \citep[e.g.,][]{Bianconietal2018, OlaveRojasetal2018, vanderBurgetal2018}. In addition, the observed halo mass dependence of the galaxy ages favours a model with pre-processing at $1 < z < 1.5$ (Reeves et al. 2021, in prep.). Nevertheless, the effect is expected to be weaker at high redshift and is thought to be negligible in group-size halos above $z\sim1.5$ \citep{Fossatietal2017}.

\subsection{Caveats - more complex morphological history?}         
\label{Caveats - more complex morphological history}
Throughout this work we have treated morphological transformation as a one-way process, in the sense that galaxies only transform from disks to spheroids.  While this might be true on a global level, the morphologies of individual galaxies can often have more complex morphological histories that switch back and forth between the two general types.  Simulations have shown that morphological transformation is a complex interplay between different processes, including mergers, disk instabilities, and gas accretion \citep[e.g.,][]{Brennanetal2015}, and specifically in the case of clusters, gravitational perturbations such as tidal shocking at pericenter passages \citep{Joshietal2020}.

Given a significant accretion of gas or stars, it is possible for a quiescent spheroid to regrow a disk and appear as a quiescent disk subsequently. \citet{DeLuciaetal2011} studied the rates of disk regrowth in bulge-dominated galaxies in simulations and found that disk regrowth is important for intermediate and low mass galaxies at high redshifts.  Although such a population has not been confirmed at this redshift, there is evidence that some local galaxies have experienced disk-regrowth at a certain stage. For example, number density studies by \citet{Grahametal2015} and \citet{delaRosaetal2016} suggested that a population of the spheroids at high redshift may have regrown a disk and have been hiding in plain sight as the bulges of local spirals and S0 galaxies. 

We cannot exclude the possibility that the excess quiescent disk population we see in clusters is due to disk-regrowth, presumably from tidal interactions or minor mergers that predominantly deposit materials in the outer part of the galaxy \citep[e.g.,][]{RodriguezGomezetal2016}. The implication of such a hypothesis is, however, worth exploring. To do that requires morphology and age data at different redshifts to track the changes in morphology in both environments over time.  One way is to look into the internal gradients of the quiescent population, as one would expect the disk-regrown galaxies to show an age difference between their inner structure and the regrown disk.  Another way is to study the morphologies of the quiescent population as a function of their local environment, e.g. local surface density or cluster-centric radius. For example, previous studies have found evidence of a relation between average intrinsic ellipticity of the quiescent cluster population and cluster-centric radius \citep[e.g.][]{DEugenioetal2015}. This analysis would help us to locate where the quiescent disks preferentially reside in the cluster and their relationship with the environment. Unfortunately, given the FOV of the HST dataset we used in this work, we do not have a large enough sample\footnote{In the GOGREEN cluster sample, only $27\%~(47\%$) of the quiescent (star-forming) galaxies are located in regions with $R \gtrsim 0.5 R_{200}$.} to perform a similar analysis as a function of cluster-centric radius.


\section{Summary and Conclusions}                                    
\label{sec:Conclusion}

In this work we have presented the axis ratio distributions of $832$ galaxies in $11$ clusters at $1.0<z<1.4$ from the Gemini Observations of Galaxies in Rich Early ENvironments (GOGREEN) Survey.  We compare their distributions with a sample of $6471$ galaxies in the field, taken from the CANDELS/3D-\textit{HST} survey to investigate the effect of the environment on galaxy structural properties. Our results can be summarized as follows:

\begin{itemize}
\item Star-forming and quiescent galaxies in clusters show different axis ratio distributions down to the mass limit of $\log(M / {\rm M}_{\odot}) = 9.5$, similar to the field.  The median $q$ of both star-forming galaxies and quiescent galaxies in clusters and the field increases with mass.

\item Massive quiescent galaxies with $\log(M / {\rm M}_{\odot}) \geq 11$ in both clusters and the field are on average rounder and have a narrower $q$ distribution than their low mass counterparts.

\item Comparing the axis ratio distribution of star-forming galaxies between cluster and field, we find that they are consistent with each other in all mass bins.

\item The axis ratio distributions of quiescent galaxies in clusters and the field are significantly distinct. For the $10.1 \leq \log(M / {\rm M}_{\odot}) < 10.5$ mass bin, cluster galaxies show a flatter axis ratio distribution with an apparent excess at low $q$ compared to the field.

\item We model the axis ratio distributions of the quiescent population in clusters and the field under different assumptions of their intrinsic shapes, following the methodology of \citet{Changetal2013}. We find some evidence that a single component (triaxial only) model is not able to reproduce the observed axis ratio distribution of the cluster galaxies in some mass bins. The axis ratio distribution is instead well-described by a two-component (triaxial + oblate) model. We find tentative evidence that the cluster distribution in the three lower mass bins has a higher oblate fraction than the field, with the $10.1 \leq \log(M / {\rm M}_{\odot}) < 10.5$ mass bin showing the largest difference ($f_{\rm{ob,cluster}} = 0.88 \pm 0.14$  vs. $f_{\rm{ob,field}} = 0.44 \pm 0.17$).

\item Our modeling shows that both the cluster and field distributions in the highest mass bin are well-described by a single component model. The contribution of a second oblate component is small, resulting in their distinct distribution shapes. 

\item We confirm that two potential sources of bias, the intrinsic shape alignment and dust extinction in cluster galaxies, are unlikely to affect our results. The intrinsic alignment signal in the cluster sample is consistent with zero, and the quiescent galaxies have a low dust content that does not exhibit an axis ratio dependence.

\item Combining the results of the axis ratio distributions and the quenched fractions of the cluster and the field samples, we show that, using a simple accretion model, the observed axis ratio distributions are statistically consistent with a scenario where no morphological transformation occurs for the environmentally quenched population for the $10.1 \leq \log(M / {\rm M}_{\odot}) < 10.5$ and $10.5 \leq \log(M / {\rm M}_{\odot}) < 10.8$ mass bins. However, the model fails to reproduce the observed axis ratio distributions in the lowest and highest mass bins.
\end{itemize}

Our results indicate that the environmental quenching mechanism(s) likely produce(s) a population that has a different morphological mix than those resulting from the dominant quenching mechanism in the field. We find that there is also no strong trend in the mass-weighted ages of the quiescent population with $q$ or $n$, which suggests that disk-like quiescent galaxies comprise objects with a mix of ages that are not significantly different from the bulk of the population. Our result is consistent with a scenario where the intermediate-mass galaxies are quenched by a starvation-like mechanism(s), such as the ``over-consumption'' model, that are not expected to drastically change the morphologies. The result of this work suggests that morphology continues to provide important constraints on the underlying physical mechanism that drives environmental quenching at this redshift range.

\acknowledgments

\section*{Acknowledgements}

The authors would also like to thank Arjen van der Wel for useful discussions on the structural parameter measurements of the CANDELS/3D-HST fields.

This work is supported by HST program number GO-15294, and by grant No. 80NSSC17K0019 issued through the NASA Astrophysics Data Analysis Program (ADAP) as well as NSF grants AST-1815475, AST-1517863 and AST-1518257. 
Support for program number GO-15294 was provided by NASA through a grant from the Space Telescope Science Institute, which is operated by the Association of Universities for Research in Astronomy, Incorporated, under NASA contract NAS5-26555. B.~V. acknowledges funding from the INAF main-stream funding programme (PI B.~Vulcani) and from the Italian PRIN-Miur 2017 (PI A.~Cimatti). J.~N. acknowledges support from Universidad Andres Bello internal grant DI-12-19/R.

This paper includes data gathered with the Gemini Observatory, which is operated by the Association of Universities for Research in Astronomy, Inc., under a cooperative agreement with the NSF on behalf of the Gemini partnership: the National Science Foundation (NSF. United States), the National Research Council (NRC, Canada), CONICYT (Chile), Ministerio de Ciencia, Tecnolog\'ia e Innovaci\'on Productiva (Argentina), and Minist\'erio da Ci\^encia, Tecnologia e Inova\c c\~ao (Brazil). This work is based (in part) on observations and archival data obtained with the Spitzer Space Telescope, which is operated by the Jet Propulsion Laboratory, California Institute of Technology under a contract with NASA; the ESO Telescopes at the La Silla Paranal Observatory under programme ID 097.A-0734; the 6.5 metre Magellan Telescopes located at Las Campanas Observatory, Chile; the Canada-France-Hawaii Telescope (CFHT) which is operated by the NRC of Canada, the Institut National des Sciences de l'Univers of the Centre National de la Recherche Scientifique of France, and the University of Hawaii; MegaPrime/MegaCam, a joint project of CFHT and CEA/DAPNIA; Subaru Telescope, which is operated by the National Astronomical Observatory of Japan (NAOJ) and the WIRCam, a joint project of CFHT, the Academia Sinica Institute of Astronomy and Astrophysics (ASIAA) in Taiwan, the Korea Astronomy and Space Science Institute (KASI) in Korea, Canada, France.


\bibliographystyle{aasjournal}
\bibliography{gg_passdisk} 

\begin{thebibliography}{}
\expandafter\ifx\csname natexlab\endcsname\relax\def\natexlab#1{#1}\fi

\bibitem[{{Baldry} {et~al.}(2006){Baldry}, {Balogh}, {Bower}, {Glazebrook},
  {Nichol}, {Bamford}, \& {Budavari}}]{Baldryetal2006}
{Baldry}, I.~K., {Balogh}, M.~L., {Bower}, R.~G., {et~al.} 2006, \mnras, 373,
  469

\bibitem[{{Balogh} {et~al.}(2004){Balogh}, {Baldry}, {Nichol}, {Miller},
  {Bower}, \& {Glazebrook}}]{Baloghetal2004}
{Balogh}, M.~L., {Baldry}, I.~K., {Nichol}, R., {et~al.} 2004, \apjl, 615, L101

\bibitem[{{Balogh} {et~al.}(1997){Balogh}, {Morris}, {Yee}, {Carlberg}, \&
  {Ellingson}}]{Baloghetal1997}
{Balogh}, M.~L., {Morris}, S.~L., {Yee}, H.~K.~C., {Carlberg}, R.~G., \&
  {Ellingson}, E. 1997, \apjl, 488, L75

\bibitem[{{Balogh} {et~al.}(2000){Balogh}, {Navarro}, \&
  {Morris}}]{Baloghetal2000}
{Balogh}, M.~L., {Navarro}, J.~F., \& {Morris}, S.~L. 2000, \apj, 540, 113

\bibitem[{{Balogh} {et~al.}(2016){Balogh}, {McGee}, {Mok}, {Muzzin}, {van der
  Burg}, {Bower}, {Finoguenov}, {Hoekstra}, {Lidman}, {Mulchaey}, {Noble},
  {Parker}, {Tanaka}, {Wilman}, {Webb}, {Wilson}, \& {Yee}}]{Baloghetal2016}
{Balogh}, M.~L., {McGee}, S.~L., {Mok}, A., {et~al.} 2016, \mnras, 456, 4364

\bibitem[{{Balogh} {et~al.}(2017){Balogh}, {Gilbank}, {Muzzin}, {Rudnick},
  {Cooper}, {Lidman}, {Biviano}, {Demarco}, {McGee}, {Nantais}, {Noble}, {Old},
  {Wilson}, {Yee}, {Bellhouse}, {Cerulo}, {Chan}, {Pintos-Castro}, {Simpson},
  {van der Burg}, {Zaritsky}, {Ziparo}, {Alonso}, {Bower}, {De Lucia},
  {Finoguenov}, {Lambas}, {Muriel}, {Parker}, {Rettura}, {Valotto}, \&
  {Wetzel}}]{Baloghetal2017}
{Balogh}, M.~L., {Gilbank}, D.~G., {Muzzin}, A., {et~al.} 2017, \mnras, 470,
  4168

\bibitem[{{Balogh} {et~al.}(2021){Balogh}, {van der Burg}, {Muzzin}, {Rudnick},
  {Wilson}, {Webb}, {Biviano}, {Boak}, {Cerulo}, {Chan}, {Cooper}, {Gilbank},
  {Gwyn}, {Lidman}, {Matharu}, {McGee}, {Old}, {Pintos-Castro}, {Reeves},
  {Shipley}, {Vulcani}, {Yee}, {Alonso}, {Bellhouse}, {Cooke}, {Davidson}, {De
  Lucia}, {Demarco}, {Drakos}, {Fillingham}, {Finoguenov}, {Forrest},
  {Golledge}, {Jablonka}, {Lambas Garcia}, {McNab}, {Muriel}, {Nantais},
  {Noble}, {Parker}, {Petter}, {Poggianti}, {Townsend}, {Valotto}, {Webb}, \&
  {Zaritsky}}]{Baloghetal2021}
{Balogh}, M.~L., {van der Burg}, R. F.~J., {Muzzin}, A., {et~al.} 2021, \mnras,
  500, 358

\bibitem[{{Barden} {et~al.}(2012){Barden}, {H{\"a}u{\ss}ler}, {Peng},
  {McIntosh}, \& {Guo}}]{Bardenetal2012}
{Barden}, M., {H{\"a}u{\ss}ler}, B., {Peng}, C.~Y., {McIntosh}, D.~H., \&
  {Guo}, Y. 2012, \mnras, 422, 449

\bibitem[{{Barro} {et~al.}(2017){Barro}, {Faber}, {Koo}, {Dekel}, {Fang},
  {Trump}, {P{\'e}rez-Gonz{\'a}lez}, {Pacifici}, {Primack}, {Somerville},
  {Yan}, {Guo}, {Liu}, {Ceverino}, {Kocevski}, \& {McGrath}}]{Barroetal2017}
{Barro}, G., {Faber}, S.~M., {Koo}, D.~C., {et~al.} 2017, \apj, 840, 47

\bibitem[{{Bassett} {et~al.}(2013){Bassett}, {Papovich}, {Lotz}, {Bell},
  {Finkelstein}, {Newman}, {Tran}, {Almaini}, {Lani}, {Cooper}, {Croton},
  {Dekel}, {Ferguson}, {Kocevski}, {Koekemoer}, {Koo}, {McGrath}, {McIntosh},
  \& {Wechsler}}]{Bassettetal2013}
{Bassett}, R., {Papovich}, C., {Lotz}, J.~M., {et~al.} 2013, \apj, 770, 58

\bibitem[{{Belli} {et~al.}(2015){Belli}, {Newman}, \& {Ellis}}]{Bellietal2015}
{Belli}, S., {Newman}, A.~B., \& {Ellis}, R.~S. 2015, \apj, 799, 206

\bibitem[{{Bertin} \& {Arnouts}(1996)}]{BertinArnouts1996}
{Bertin}, E., \& {Arnouts}, S. 1996, \aaps, 117, 393

\bibitem[{{Bianconi} {et~al.}(2018){Bianconi}, {Smith}, {Haines}, {McGee},
  {Finoguenov}, \& {Egami}}]{Bianconietal2018}
{Bianconi}, M., {Smith}, G.~P., {Haines}, C.~P., {et~al.} 2018, \mnras, 473,
  L79

\bibitem[{{Biviano} {et~al.}(2021){Biviano}, {van der Burg}, {Balogh},
  {Munari}, {Cooper}, {De Lucia}, {Demarco}, {Jablonka}, {Muzzin}, {Nantais},
  {Old}, {Rudnick}, {Vulcani}, {Wilson}, {Yee}, {Zaritsky}, {Cerulo}, {Chan},
  {Finoguenov}, {Gilbank}, {Lidman}, {Pintos-Castro}, \&
  {Shipley}}]{Bivianoetal2021}
{Biviano}, A., {van der Burg}, R.~F.~J., {Balogh}, M.~L., {et~al.} 2021, arXiv
  e-prints, arXiv:2104.01183

\bibitem[{{Bleem} {et~al.}(2015){Bleem}, {Stalder}, {de Haan}, {Aird}, {Allen},
  {Applegate}, {Ashby}, {Bautz}, {Bayliss}, {Benson}, {Bocquet}, {Brodwin},
  {Carlstrom}, {Chang}, {Chiu}, {Cho}, {Clocchiatti}, {Crawford}, {Crites},
  {Desai}, {Dietrich}, {Dobbs}, {Foley}, {Forman}, {George}, {Gladders},
  {Gonzalez}, {Halverson}, {Hennig}, {Hoekstra}, {Holder}, {Holzapfel},
  {Hrubes}, {Jones}, {Keisler}, {Knox}, {Lee}, {Leitch}, {Liu}, {Lueker},
  {Luong-Van}, {Mantz}, {Marrone}, {McDonald}, {McMahon}, {Meyer}, {Mocanu},
  {Mohr}, {Murray}, {Padin}, {Pryke}, {Reichardt}, {Rest}, {Ruel}, {Ruhl},
  {Saliwanchik}, {Saro}, {Sayre}, {Schaffer}, {Schrabback}, {Shirokoff},
  {Song}, {Spieler}, {Stanford}, {Staniszewski}, {Stark}, {Story}, {Stubbs},
  {Vanderlinde}, {Vieira}, {Vikhlinin}, {Williamson}, {Zahn}, \&
  {Zenteno}}]{Bleemetal2015}
{Bleem}, L.~E., {Stalder}, B., {de Haan}, T., {et~al.} 2015, \apjs, 216, 27

\bibitem[{{Boselli} \& {Gavazzi}(2006)}]{BoselliGavazzi2006}
{Boselli}, A., \& {Gavazzi}, G. 2006, \pasp, 118, 517

\bibitem[{{Boselli} \& {Gavazzi}(2014)}]{BoselliGavazzi2014}
---. 2014, \aapr, 22, 74

\bibitem[{{Brammer} {et~al.}(2008){Brammer}, {van Dokkum}, \&
  {Coppi}}]{Brammeretal2008}
{Brammer}, G.~B., {van Dokkum}, P.~G., \& {Coppi}, P. 2008, \apj, 686, 1503

\bibitem[{{Brammer} {et~al.}(2012){Brammer}, {van Dokkum}, {Franx},
  {Fumagalli}, {Patel}, {Rix}, {Skelton}, {Kriek}, {Nelson}, {Schmidt},
  {Bezanson}, {da Cunha}, {Erb}, {Fan}, {F{\"o}rster Schreiber}, {Illingworth},
  {Labb{\'e}}, {Leja}, {Lundgren}, {Magee}, {Marchesini}, {McCarthy},
  {Momcheva}, {Muzzin}, {Quadri}, {Steidel}, {Tal}, {Wake}, {Whitaker}, \&
  {Williams}}]{Brammeretal2012}
{Brammer}, G.~B., {van Dokkum}, P.~G., {Franx}, M., {et~al.} 2012, \apjs, 200,
  13

\bibitem[{{Brennan} {et~al.}(2015){Brennan}, {Pandya}, {Somerville}, {Barro},
  {Taylor}, {Wuyts}, {Bell}, {Dekel}, {Ferguson}, {McIntosh}, {Papovich}, \&
  {Primack}}]{Brennanetal2015}
{Brennan}, R., {Pandya}, V., {Somerville}, R.~S., {et~al.} 2015, \mnras, 451,
  2933

\bibitem[{{Brodwin} {et~al.}(2010){Brodwin}, {Ruel}, {Ade}, {Aird},
  {Andersson}, {Ashby}, {Bautz}, {Bazin}, {Benson}, {Bleem}, {Carlstrom},
  {Chang}, {Crawford}, {Crites}, {de Haan}, {Desai}, {Dobbs}, {Dudley},
  {Fazio}, {Foley}, {Forman}, {Garmire}, {George}, {Gladders}, {Gonzalez},
  {Halverson}, {High}, {Holder}, {Holzapfel}, {Hrubes}, {Jones}, {Joy},
  {Keisler}, {Knox}, {Lee}, {Leitch}, {Lueker}, {Marrone}, {McMahon}, {Mehl},
  {Meyer}, {Mohr}, {Montroy}, {Murray}, {Padin}, {Plagge}, {Pryke},
  {Reichardt}, {Rest}, {Ruhl}, {Schaffer}, {Shaw}, {Shirokoff}, {Song},
  {Spieler}, {Stalder}, {Stanford}, {Staniszewski}, {Stark}, {Stubbs},
  {Vanderlinde}, {Vieira}, {Vikhlinin}, {Williamson}, {Yang}, {Zahn}, \&
  {Zenteno}}]{Brodwinetal2010}
{Brodwin}, M., {Ruel}, J., {Ade}, P.~A.~R., {et~al.} 2010, \apj, 721, 90

\bibitem[{{Brough} {et~al.}(2017){Brough}, {van de Sande}, {Owers},
  {d'Eugenio}, {Sharp}, {Cortese}, {Scott}, {Croom}, {Bassett}, {Bekki},
  {Bland-Hawthorn}, {Bryant}, {Davies}, {Drinkwater}, {Driver}, {Foster},
  {Goldstein}, {L{\'o}pez-S{\'a}nchez}, {Medling}, {Sweet}, {Taranu}, {Tonini},
  {Yi}, {Goodwin}, {Lawrence}, \& {Richards}}]{Broughetal2017}
{Brough}, S., {van de Sande}, J., {Owers}, M.~S., {et~al.} 2017, \apj, 844, 59

\bibitem[{{Bruzual} \& {Charlot}(2003)}]{BruzualCharlot2003}
{Bruzual}, G., \& {Charlot}, S. 2003, \mnras, 344, 1000

\bibitem[{{Calzetti} {et~al.}(2000){Calzetti}, {Armus}, {Bohlin}, {Kinney},
  {Koornneef}, \& {Storchi-Bergmann}}]{Calzettietal2000}
{Calzetti}, D., {Armus}, L., {Bohlin}, R.~C., {et~al.} 2000, \apj, 533, 682

\bibitem[{{Cappellari}(2016)}]{Cappellari2016}
{Cappellari}, M. 2016, \araa, 54, 597

\bibitem[{{Cappellari} {et~al.}(2011){Cappellari}, {Emsellem}, {Krajnovi{\'c}},
  {McDermid}, {Serra}, {Alatalo}, {Blitz}, {Bois}, {Bournaud}, {Bureau},
  {Davies}, {Davis}, {de Zeeuw}, {Khochfar}, {Kuntschner}, {Lablanche},
  {Morganti}, {Naab}, {Oosterloo}, {Sarzi}, {Scott}, {Weijmans}, \&
  {Young}}]{Cappellarietal2011}
{Cappellari}, M., {Emsellem}, E., {Krajnovi{\'c}}, D., {et~al.} 2011, \mnras,
  416, 1680

\bibitem[{{Carollo} {et~al.}(2013){Carollo}, {Bschorr}, {Renzini}, {Lilly},
  {Capak}, {Cibinel}, {Ilbert}, {Onodera}, {Scoville}, {Cameron}, {Mobasher},
  {Sanders}, \& {Taniguchi}}]{Carolloetal2013}
{Carollo}, C.~M., {Bschorr}, T.~J., {Renzini}, A., {et~al.} 2013, \apj, 773,
  112

\bibitem[{{Chabrier}(2003)}]{Chabrier2003}
{Chabrier}, G. 2003, \pasp, 115, 763

\bibitem[{{Chan} {et~al.}(2016){Chan}, {Beifiori}, {Mendel}, {Saglia},
  {Bender}, {Fossati}, {Galametz}, {Wegner}, {Wilman}, {Cappellari}, {Davies},
  {Houghton}, {Prichard}, {Lewis}, {Sharples}, \& {Stott}}]{Chanetal2016}
{Chan}, J.~C.~C., {Beifiori}, A., {Mendel}, J.~T., {et~al.} 2016, \mnras, 458,
  3181

\bibitem[{{Chan} {et~al.}(2018){Chan}, {Beifiori}, {Saglia}, {Mendel}, {Stott},
  {Bender}, {Galametz}, {Wilman}, {Cappellari}, {Davies}, {Houghton},
  {Prichard}, {Lewis}, {Sharples}, \& {Wegner}}]{Chanetal2018}
{Chan}, J. C.~C., {Beifiori}, A., {Saglia}, R.~P., {et~al.} 2018, \apj, 856, 8

\bibitem[{{Chang} {et~al.}(2013{\natexlab{a}}){Chang}, {van der Wel}, {Rix},
  {Wuyts}, {Zibetti}, {Ramkumar}, \& {Holden}}]{Changetal2013a}
{Chang}, Y.-Y., {van der Wel}, A., {Rix}, H.-W., {et~al.} 2013{\natexlab{a}},
  \apj, 762, 83

\bibitem[{{Chang} {et~al.}(2013{\natexlab{b}}){Chang}, {van der Wel}, {Rix},
  {Holden}, {Bell}, {McGrath}, {Wuyts}, {H{\"a}ussler}, {Barden}, {Faber},
  {Mozena}, {Ferguson}, {Guo}, {Galametz}, {Grogin}, {Kocevski}, {Koekemoer},
  {Dekel}, {Huang}, {Hathi}, \& {Donley}}]{Changetal2013}
---. 2013{\natexlab{b}}, \apj, 773, 149

\bibitem[{{Cole} {et~al.}(2020){Cole}, {Bezanson}, {van der Wel}, {Bell},
  {D'Eugenio}, {Franx}, {Gallazzi}, {van Houdt}, {Muzzin}, {Pacifici}, {van de
  Sande}, {Sobral}, {Straatman}, \& {Wu}}]{Coleetal2020}
{Cole}, J., {Bezanson}, R., {van der Wel}, A., {et~al.} 2020, \apjl, 890, L25

\bibitem[{{Cooper} {et~al.}(2010){Cooper}, {Coil}, {Gerke}, {Newman}, {Bundy},
  {Conselice}, {Croton}, {Davis}, {Faber}, {Guhathakurta}, {Koo}, {Lin},
  {Weiner}, {Willmer}, \& {Yan}}]{Cooperetal2010b}
{Cooper}, M.~C., {Coil}, A.~L., {Gerke}, B.~F., {et~al.} 2010, \mnras, 409, 337

\bibitem[{{Cooper} {et~al.}(2012){Cooper}, {Griffith}, {Newman}, {Coil},
  {Davis}, {Dutton}, {Faber}, {Guhathakurta}, {Koo}, {Lotz}, {Weiner},
  {Willmer}, \& {Yan}}]{Cooper2012}
{Cooper}, M.~C., {Griffith}, R.~L., {Newman}, J.~A., {et~al.} 2012, \mnras,
  419, 3018

\bibitem[{{Cortese} {et~al.}(2016){Cortese}, {Fogarty}, {Bekki}, {van de
  Sande}, {Couch}, {Catinella}, {Colless}, {Obreschkow}, {Taranu}, {Tescari},
  {Barat}, {Bland-Hawthorn}, {Bloom}, {Bryant}, {Cluver}, {Croom},
  {Drinkwater}, {d'Eugenio}, {Konstantopoulos}, {Lopez-Sanchez}, {Mahajan},
  {Scott}, {Tonini}, {Wong}, {Allen}, {Brough}, {Goodwin}, {Green}, {Ho},
  {Kelvin}, {Lawrence}, {Lorente}, {Medling}, {Owers}, {Richards}, {Sharp}, \&
  {Sweet}}]{Corteseetal2016}
{Cortese}, L., {Fogarty}, L.~M.~R., {Bekki}, K., {et~al.} 2016, \mnras, 463,
  170

\bibitem[{{de la Rosa} {et~al.}(2016){de la Rosa}, {La Barbera}, {Ferreras},
  {S{\'a}nchez Almeida}, {Dalla Vecchia}, {Mart{\'\i}nez-Valpuesta}, \&
  {Stringer}}]{delaRosaetal2016}
{de la Rosa}, I.~G., {La Barbera}, F., {Ferreras}, I., {et~al.} 2016, \mnras,
  457, 1916

\bibitem[{{De Lucia} {et~al.}(2010){De Lucia}, {Boylan-Kolchin}, {Benson},
  {Fontanot}, \& {Monaco}}]{DeLuciaetal2010}
{De Lucia}, G., {Boylan-Kolchin}, M., {Benson}, A.~J., {Fontanot}, F., \&
  {Monaco}, P. 2010, \mnras, 406, 1533

\bibitem[{{De Lucia} {et~al.}(2011){De Lucia}, {Fontanot}, {Wilman}, \&
  {Monaco}}]{DeLuciaetal2011}
{De Lucia}, G., {Fontanot}, F., {Wilman}, D., \& {Monaco}, P. 2011, \mnras,
  414, 1439

\bibitem[{{De Lucia} {et~al.}(2012){De Lucia}, {Weinmann}, {Poggianti},
  {Arag{\'o}n-Salamanca}, \& {Zaritsky}}]{DeLuciaetal2012}
{De Lucia}, G., {Weinmann}, S., {Poggianti}, B.~M., {Arag{\'o}n-Salamanca}, A.,
  \& {Zaritsky}, D. 2012, \mnras, 423, 1277

\bibitem[{{Delaye} {et~al.}(2014){Delaye}, {Huertas-Company}, {Mei}, {Lidman},
  {Licitra}, {Newman}, {Raichoor}, {Shankar}, {Barrientos}, {Bernardi},
  {Cerulo}, {Couch}, {Demarco}, {Mu{\~n}oz}, {S{\'a}nchez-Janssen}, \&
  {Tanaka}}]{Delaye2014}
{Delaye}, L., {Huertas-Company}, M., {Mei}, S., {et~al.} 2014, \mnras, 441, 203

\bibitem[{{Demarco} {et~al.}(2010){Demarco}, {Wilson}, {Muzzin}, {Lacy},
  {Surace}, {Yee}, {Hoekstra}, {Blindert}, \& {Gilbank}}]{Demarcoetal2010a}
{Demarco}, R., {Wilson}, G., {Muzzin}, A., {et~al.} 2010, \apj, 711, 1185

\bibitem[{{D'Eugenio} {et~al.}(2013){D'Eugenio}, {Houghton}, {Davies}, \&
  {Dalla Bont{\`a}}}]{DEugenioetal2013}
{D'Eugenio}, F., {Houghton}, R.~C.~W., {Davies}, R.~L., \& {Dalla Bont{\`a}},
  E. 2013, \mnras, 429, 1258

\bibitem[{{D'Eugenio} {et~al.}(2015){D'Eugenio}, {Houghton}, {Davies}, \&
  {Dalla Bont{\`a}}}]{DEugenioetal2015}
---. 2015, \mnras, 451, 827

\bibitem[{{Dressler}(1980)}]{Dressler1980}
{Dressler}, A. 1980, \apj, 236, 351

\bibitem[{{Emsellem} {et~al.}(2011){Emsellem}, {Cappellari}, {Krajnovi{\'c}},
  {Alatalo}, {Blitz}, {Bois}, {Bournaud}, {Bureau}, {Davies}, {Davis}, {de
  Zeeuw}, {Khochfar}, {Kuntschner}, {Lablanche}, {McDermid}, {Morganti},
  {Naab}, {Oosterloo}, {Sarzi}, {Scott}, {Serra}, {van de Ven}, {Weijmans}, \&
  {Young}}]{Emsellemetal2011}
{Emsellem}, E., {Cappellari}, M., {Krajnovi{\'c}}, D., {et~al.} 2011, \mnras,
  414, 888

\bibitem[{{Faltenbacher} {et~al.}(2008){Faltenbacher}, {Jing}, {Li}, {Mao},
  {Mo}, {Pasquali}, \& {van den Bosch}}]{Faltenbacheretal2008}
{Faltenbacher}, A., {Jing}, Y.~P., {Li}, C., {et~al.} 2008, \apj, 675, 146

\bibitem[{{Fasano} {et~al.}(2000){Fasano}, {Poggianti}, {Couch}, {Bettoni},
  {Kj{\ae}rgaard}, \& {Moles}}]{Fasanoetal2000}
{Fasano}, G., {Poggianti}, B.~M., {Couch}, W.~J., {et~al.} 2000, \apj, 542, 673

\bibitem[{{Fillingham} {et~al.}(2015){Fillingham}, {Cooper}, {Wheeler},
  {Garrison-Kimmel}, {Boylan-Kolchin}, \& {Bullock}}]{Fillinghametal2015}
{Fillingham}, S.~P., {Cooper}, M.~C., {Wheeler}, C., {et~al.} 2015, \mnras,
  454, 2039

\bibitem[{{Foley} {et~al.}(2011){Foley}, {Andersson}, {Bazin}, {de Haan},
  {Ruel}, {Ade}, {Aird}, {Armstrong}, {Ashby}, {Bautz}, {Benson}, {Bleem},
  {Bonamente}, {Brodwin}, {Carlstrom}, {Chang}, {Clocchiatti}, {Crawford},
  {Crites}, {Desai}, {Dobbs}, {Dudley}, {Fazio}, {Forman}, {Garmire}, {George},
  {Gladders}, {Gonzalez}, {Halverson}, {High}, {Holder}, {Holzapfel}, {Hoover},
  {Hrubes}, {Jones}, {Joy}, {Keisler}, {Knox}, {Lee}, {Leitch}, {Lueker},
  {Luong-Van}, {Marrone}, {McMahon}, {Mehl}, {Meyer}, {Mohr}, {Montroy},
  {Murray}, {Padin}, {Plagge}, {Pryke}, {Reichardt}, {Rest}, {Ruhl},
  {Saliwanchik}, {Saro}, {Schaffer}, {Shaw}, {Shirokoff}, {Song}, {Spieler},
  {Stalder}, {Stanford}, {Staniszewski}, {Stark}, {Story}, {Stubbs},
  {Vanderlinde}, {Vieira}, {Vikhlinin}, {Williamson}, \&
  {Zenteno}}]{Foleyetal2011}
{Foley}, R.~J., {Andersson}, K., {Bazin}, G., {et~al.} 2011, \apj, 731, 86

\bibitem[{{Fossati} {et~al.}(2017){Fossati}, {Wilman}, {Mendel}, {Saglia},
  {Galametz}, {Beifiori}, {Bender}, {Chan}, {Fabricius}, {Bandara}, {Brammer},
  {Davies}, {F{\"o}rster Schreiber}, {Genzel}, {Hartley}, {Kulkarni}, {Lang},
  {Momcheva}, {Nelson}, {Skelton}, {Tacconi}, {Tadaki}, {{\"U}bler}, {van
  Dokkum}, {Wisnioski}, {Whitaker}, {Wuyts}, \& {Wuyts}}]{Fossatietal2017}
{Fossati}, M., {Wilman}, D.~J., {Mendel}, J.~T., {et~al.} 2017, \apj, 835, 153

\bibitem[{{Foster} {et~al.}(2017){Foster}, {van de Sande}, {D'Eugenio},
  {Cortese}, {McDermid}, {Bland-Hawthorn}, {Brough}, {Bryant}, {Croom},
  {Goodwin}, {Konstantopoulos}, {Lawrence}, {L{\'o}pez-S{\'a}nchez}, {Medling},
  {Owers}, {Richards}, {Scott}, {Taranu}, {Tonini}, \&
  {Zafar}}]{Fosteretal2017}
{Foster}, C., {van de Sande}, J., {D'Eugenio}, F., {et~al.} 2017, \mnras, 472,
  966

\bibitem[{{Franx} {et~al.}(1991){Franx}, {Illingworth}, \& {de
  Zeeuw}}]{Franxetal1991}
{Franx}, M., {Illingworth}, G., \& {de Zeeuw}, T. 1991, \apj, 383, 112

\bibitem[{{Fumagalli} {et~al.}(2014){Fumagalli}, {Fossati}, {Hau}, {Gavazzi},
  {Bower}, {Sun}, \& {Boselli}}]{Fumagallietal2014b}
{Fumagalli}, M., {Fossati}, M., {Hau}, G. K.~T., {et~al.} 2014, \mnras, 445,
  4335

\bibitem[{{Gavazzi} {et~al.}(2001){Gavazzi}, {Boselli}, {Mayer},
  {Iglesias-Paramo}, {V{\'\i}lchez}, \& {Carrasco}}]{Gavazzietal2001}
{Gavazzi}, G., {Boselli}, A., {Mayer}, L., {et~al.} 2001, \apjl, 563, L23

\bibitem[{{Georgiou} {et~al.}(2019){Georgiou}, {Chisari}, {Fortuna},
  {Hoekstra}, {Kuijken}, {Joachimi}, {Vakili}, {Bilicki}, {Dvornik}, {Erben},
  {Giblin}, {Heymans}, {Napolitano}, \& {Shan}}]{Georgiouetal2019}
{Georgiou}, C., {Chisari}, N.~E., {Fortuna}, M.~C., {et~al.} 2019, \aap, 628,
  A31

\bibitem[{{Gonzaga} {et~al.}(2012){Gonzaga}, {et al.}, {B}, \&
  {C}}]{Gonzagaetal2012}
{Gonzaga}, S., {et al.}, {B}, \& {C}. 2012, {The DrizzlePac Handbook}

\bibitem[{{Graham} {et~al.}(2015){Graham}, {Dullo}, \&
  {Savorgnan}}]{Grahametal2015}
{Graham}, A.~W., {Dullo}, B.~T., \& {Savorgnan}, G. A.~D. 2015, \apj, 804, 32

\bibitem[{{Graham} {et~al.}(2019){Graham}, {Cappellari}, {Bershady}, \&
  {Drory}}]{Grahametal2019}
{Graham}, M.~T., {Cappellari}, M., {Bershady}, M.~A., \& {Drory}, N. 2019,
  arXiv e-prints, arXiv:1910.05139

\bibitem[{{Grogin} {et~al.}(2011){Grogin}, {Kocevski}, {Faber}, {Ferguson},
  {Koekemoer}, {Riess}, {Acquaviva}, {Alexander}, {Almaini}, {Ashby}, {Barden},
  {Bell}, {Bournaud}, {Brown}, {Caputi}, {Casertano}, {Cassata}, {Castellano},
  {Challis}, {Chary}, {Cheung}, {Cirasuolo}, {Conselice}, {Roshan Cooray},
  {Croton}, {Daddi}, {Dahlen}, {Dav{\'e}}, {de Mello}, {Dekel}, {Dickinson},
  {Dolch}, {Donley}, {Dunlop}, {Dutton}, {Elbaz}, {Fazio}, {Filippenko},
  {Finkelstein}, {Fontana}, {Gardner}, {Garnavich}, {Gawiser}, {Giavalisco},
  {Grazian}, {Guo}, {Hathi}, {H{\"a}ussler}, {Hopkins}, {Huang}, {Huang},
  {Jha}, {Kartaltepe}, {Kirshner}, {Koo}, {Lai}, {Lee}, {Li}, {Lotz}, {Lucas},
  {Madau}, {McCarthy}, {McGrath}, {McIntosh}, {McLure}, {Mobasher},
  {Moustakas}, {Mozena}, {Nandra}, {Newman}, {Niemi}, {Noeske}, {Papovich},
  {Pentericci}, {Pope}, {Primack}, {Rajan}, {Ravindranath}, {Reddy}, {Renzini},
  {Rix}, {Robaina}, {Rodney}, {Rosario}, {Rosati}, {Salimbeni}, {Scarlata},
  {Siana}, {Simard}, {Smidt}, {Somerville}, {Spinrad}, {Straughn}, {Strolger},
  {Telford}, {Teplitz}, {Trump}, {van der Wel}, {Villforth}, {Wechsler},
  {Weiner}, {Wiklind}, {Wild}, {Wilson}, {Wuyts}, {Yan}, \&
  {Yun}}]{Groginetal2011}
{Grogin}, N.~A., {Kocevski}, D.~D., {Faber}, S.~M., {et~al.} 2011, \apjs, 197,
  35

\bibitem[{{Gunn} \& {Gott}(1972)}]{GunnGott1972}
{Gunn}, J.~E., \& {Gott}, III, J.~R. 1972, \apj, 176, 1

\bibitem[{{H{\"a}ussler} {et~al.}(2007){H{\"a}ussler}, {McIntosh}, {Barden},
  {Bell}, {Rix}, {Borch}, {Beckwith}, {Caldwell}, {Heymans}, {Jahnke}, {Jogee},
  {Koposov}, {Meisenheimer}, {S{\'a}nchez}, {Somerville}, {Wisotzki}, \&
  {Wolf}}]{Haussleretal2007}
{H{\"a}ussler}, B., {McIntosh}, D.~H., {Barden}, M., {et~al.} 2007, \apjs, 172,
  615

\bibitem[{{H{\"a}u{\ss}ler} {et~al.}(2013){H{\"a}u{\ss}ler}, {Bamford}, {Vika},
  {Rojas}, {Barden}, {Kelvin}, {Alpaslan}, {Robotham}, {Driver}, {Baldry},
  {Brough}, {Hopkins}, {Liske}, {Nichol}, {Popescu}, \&
  {Tuffs}}]{Haussleretal2013}
{H{\"a}u{\ss}ler}, B., {Bamford}, S.~P., {Vika}, M., {et~al.} 2013, \mnras,
  430, 330

\bibitem[{{Hinton} {et~al.}(2016){Hinton}, {Davis}, {Lidman}, {Glazebrook}, \&
  {Lewis}}]{Hintonetal2016}
{Hinton}, S.~R., {Davis}, T.~M., {Lidman}, C., {Glazebrook}, K., \& {Lewis},
  G.~F. 2016, Astronomy and Computing, 15, 61

\bibitem[{{Holden} {et~al.}(2012){Holden}, {van der Wel}, {Rix}, \&
  {Franx}}]{Holdenetal2012}
{Holden}, B.~P., {van der Wel}, A., {Rix}, H.-W., \& {Franx}, M. 2012, \apj,
  749, 96

\bibitem[{{Holden} {et~al.}(2007){Holden}, {Illingworth}, {Franx}, {Blakeslee},
  {Postman}, {Kelson}, {van der Wel}, {Demarco}, {Magee}, {Tran}, {Zirm},
  {Ford}, {Rosati}, \& {Homeier}}]{Holdenetal2007}
{Holden}, B.~P., {Illingworth}, G.~D., {Franx}, M., {et~al.} 2007, \apj, 670,
  190

\bibitem[{{Hopkins} {et~al.}(2009){Hopkins}, {Cox}, {Younger}, \&
  {Hernquist}}]{Hopkinsetal2009a}
{Hopkins}, P.~F., {Cox}, T.~J., {Younger}, J.~D., \& {Hernquist}, L. 2009,
  \apj, 691, 1168

\bibitem[{{Hopkins} {et~al.}(2010){Hopkins}, {Croton}, {Bundy}, {Khochfar},
  {van den Bosch}, {Somerville}, {Wetzel}, {Keres}, {Hernquist}, {Stewart},
  {Younger}, {Genel}, \& {Ma}}]{Hopkinsetal2010}
{Hopkins}, P.~F., {Croton}, D., {Bundy}, K., {et~al.} 2010, \apj, 724, 915

\bibitem[{{Houghton} {et~al.}(2013){Houghton}, {Davies}, {D'Eugenio}, {Scott},
  {Thatte}, {Clarke}, {Tecza}, {Salter}, {Fogarty}, \&
  {Goodsall}}]{Houghtonetal2013}
{Houghton}, R.~C.~W., {Davies}, R.~L., {D'Eugenio}, F., {et~al.} 2013, \mnras,
  436, 19

\bibitem[{{Huang} {et~al.}(2018){Huang}, {Mandelbaum}, {Freeman}, {Chen},
  {Rozo}, \& {Rykoff}}]{Huangetal2018}
{Huang}, H.-J., {Mandelbaum}, R., {Freeman}, P.~E., {et~al.} 2018, \mnras, 474,
  4772

\bibitem[{{Joachimi} {et~al.}(2015){Joachimi}, {Cacciato}, {Kitching},
  {Leonard}, {Mandelbaum}, {Sch{\"a}fer}, {Sif{\'o}n}, {Hoekstra}, {Kiessling},
  {Kirk}, \& {Rassat}}]{Joachimietal2015}
{Joachimi}, B., {Cacciato}, M., {Kitching}, T.~D., {et~al.} 2015, \ssr, 193, 1

\bibitem[{{Johnston} {et~al.}(2014){Johnston}, {Arag{\'o}n-Salamanca}, \&
  {Merrifield}}]{Johnstonetal2014}
{Johnston}, E.~J., {Arag{\'o}n-Salamanca}, A., \& {Merrifield}, M.~R. 2014,
  \mnras, 441, 333

\bibitem[{{Joshi} {et~al.}(2020){Joshi}, {Pillepich}, {Nelson}, {Marinacci},
  {Springel}, {Rodriguez-Gomez}, {Vogelsberger}, \&
  {Hernquist}}]{Joshietal2020}
{Joshi}, G.~D., {Pillepich}, A., {Nelson}, D., {et~al.} 2020, \mnras,
  doi:10.1093/mnras/staa1668

\bibitem[{{Just} {et~al.}(2010){Just}, {Zaritsky}, {Sand}, {Desai}, \&
  {Rudnick}}]{Justetal2010}
{Just}, D.~W., {Zaritsky}, D., {Sand}, D.~J., {Desai}, V., \& {Rudnick}, G.
  2010, \apj, 711, 192

\bibitem[{{Kawinwanichakij} {et~al.}(2017){Kawinwanichakij}, {Papovich},
  {Quadri}, {Glazebrook}, {Kacprzak}, {Allen}, {Bell}, {Croton}, {Dekel},
  {Ferguson}, {Forrest}, {Grogin}, {Guo}, {Kocevski}, {Koekemoer}, {Labb{\'e}},
  {Lucas}, {Nanayakkara}, {Spitler}, {Straatman}, {Tran}, {Tomczak}, \& {van
  Dokkum}}]{Kawinwanichakijetal2017}
{Kawinwanichakij}, L., {Papovich}, C., {Quadri}, R.~F., {et~al.} 2017, \apj,
  847, 134

\bibitem[{{Kelkar} {et~al.}(2017){Kelkar}, {Gray}, {Arag{\'o}n-Salamanca},
  {Rudnick}, {Milvang-Jensen}, {Jablonka}, \& {Schrabback}}]{Kelkaretal2017}
{Kelkar}, K., {Gray}, M.~E., {Arag{\'o}n-Salamanca}, A., {et~al.} 2017, \mnras,
  469, 4551

\bibitem[{{Knebe} {et~al.}(2020){Knebe}, {G{\'a}mez-Mar{\'\i}n}, {Pearce},
  {Cui}, {Hoffmann}, {De Petris}, {Power}, {Haggar}, \&
  {Mostoghiu}}]{Knebeetal2020}
{Knebe}, A., {G{\'a}mez-Mar{\'\i}n}, M., {Pearce}, F.~R., {et~al.} 2020,
  \mnras, 495, 3002

\bibitem[{{Koekemoer} {et~al.}(2011){Koekemoer}, {Faber}, {Ferguson}, {Grogin},
  {Kocevski}, {Koo}, {Lai}, {Lotz}, {Lucas}, {McGrath}, {Ogaz}, {Rajan},
  {Riess}, {Rodney}, {Strolger}, {Casertano}, {Castellano}, {Dahlen},
  {Dickinson}, {Dolch}, {Fontana}, {Giavalisco}, {Grazian}, {Guo}, {Hathi},
  {Huang}, {van der Wel}, {Yan}, {Acquaviva}, {Alexander}, {Almaini}, {Ashby},
  {Barden}, {Bell}, {Bournaud}, {Brown}, {Caputi}, {Cassata}, {Challis},
  {Chary}, {Cheung}, {Cirasuolo}, {Conselice}, {Roshan Cooray}, {Croton},
  {Daddi}, {Dav{\'e}}, {de Mello}, {de Ravel}, {Dekel}, {Donley}, {Dunlop},
  {Dutton}, {Elbaz}, {Fazio}, {Filippenko}, {Finkelstein}, {Frazer}, {Gardner},
  {Garnavich}, {Gawiser}, {Gruetzbauch}, {Hartley}, {H{\"a}ussler},
  {Herrington}, {Hopkins}, {Huang}, {Jha}, {Johnson}, {Kartaltepe},
  {Khostovan}, {Kirshner}, {Lani}, {Lee}, {Li}, {Madau}, {McCarthy},
  {McIntosh}, {McLure}, {McPartland}, {Mobasher}, {Moreira}, {Mortlock},
  {Moustakas}, {Mozena}, {Nandra}, {Newman}, {Nielsen}, {Niemi}, {Noeske},
  {Papovich}, {Pentericci}, {Pope}, {Primack}, {Ravindranath}, {Reddy},
  {Renzini}, {Rix}, {Robaina}, {Rosario}, {Rosati}, {Salimbeni}, {Scarlata},
  {Siana}, {Simard}, {Smidt}, {Snyder}, {Somerville}, {Spinrad}, {Straughn},
  {Telford}, {Teplitz}, {Trump}, {Vargas}, {Villforth}, {Wagner}, {Wandro},
  {Wechsler}, {Weiner}, {Wiklind}, {Wild}, {Wilson}, {Wuyts}, \&
  {Yun}}]{Koekemoeretal2011}
{Koekemoer}, A.~M., {Faber}, S.~M., {Ferguson}, H.~C., {et~al.} 2011, \apjs,
  197, 36

\bibitem[{{Kriek} {et~al.}(2009){Kriek}, {van Dokkum}, {Labb{\'e}}, {Franx},
  {Illingworth}, {Marchesini}, \& {Quadri}}]{Krieketal2009}
{Kriek}, M., {van Dokkum}, P.~G., {Labb{\'e}}, I., {et~al.} 2009, \apj, 700,
  221

\bibitem[{{Labb{\'e}} {et~al.}(2005){Labb{\'e}}, {Huang}, {Franx}, {Rudnick},
  {Barmby}, {Daddi}, {van Dokkum}, {Fazio}, {Schreiber}, {Moorwood}, {Rix},
  {R{\"o}ttgering}, {Trujillo}, \& {van der Werf}}]{Labbeetal2005}
{Labb{\'e}}, I., {Huang}, J., {Franx}, M., {et~al.} 2005, \apjl, 624, L81

\bibitem[{{Lang} {et~al.}(2014){Lang}, {Wuyts}, {Somerville}, {F{\"o}rster
  Schreiber}, {Genzel}, {Bell}, {Brammer}, {Dekel}, {Faber}, {Ferguson},
  {Grogin}, {Kocevski}, {Koekemoer}, {Lutz}, {McGrath}, {Momcheva}, {Nelson},
  {Primack}, {Rosario}, {Skelton}, {Tacconi}, {van Dokkum}, \&
  {Whitaker}}]{Langetal2014}
{Lang}, P., {Wuyts}, S., {Somerville}, R.~S., {et~al.} 2014, \apj, 788, 11

\bibitem[{{Lani} {et~al.}(2013){Lani}, {Almaini}, {Hartley}, {Mortlock},
  {H{\"a}u{\ss}ler}, {Chuter}, {Simpson}, {van der Wel}, {Gr{\"u}tzbauch},
  {Conselice}, {Bradshaw}, {Cooper}, {Faber}, {Grogin}, {Kocevski},
  {Koekemoer}, \& {Lai}}]{Lani2013}
{Lani}, C., {Almaini}, O., {Hartley}, W.~G., {et~al.} 2013, \mnras, 435, 207

\bibitem[{{Larson} {et~al.}(1980){Larson}, {Tinsley}, \&
  {Caldwell}}]{Larson1980}
{Larson}, R.~B., {Tinsley}, B.~M., \& {Caldwell}, C.~N. 1980, \apj, 237, 692

\bibitem[{{Leja} {et~al.}(2019){Leja}, {Johnson}, {Conroy}, {van Dokkum},
  {Speagle}, {Brammer}, {Momcheva}, {Skelton}, {Whitaker}, {Franx}, \&
  {Nelson}}]{Lejaetal2019}
{Leja}, J., {Johnson}, B.~D., {Conroy}, C., {et~al.} 2019, \apj, 877, 140

\bibitem[{{Lidman} {et~al.}(2013){Lidman}, {Iacobuta}, {Bauer}, {Barrientos},
  {Cerulo}, {Couch}, {Delaye}, {Demarco}, {Ellingson}, {Faloon}, {Gilbank},
  {Huertas-Company}, {Mei}, {Meyers}, {Muzzin}, {Noble}, {Nantais}, {Rettura},
  {Rosati}, {S{\'a}nchez-Janssen}, {Strazzullo}, {Webb}, {Wilson}, {Yan}, \&
  {Yee}}]{Lidmanetal2013}
{Lidman}, C., {Iacobuta}, G., {Bauer}, A.~E., {et~al.} 2013, \mnras, 433, 825

\bibitem[{{Man} {et~al.}(2016){Man}, {Zirm}, \& {Toft}}]{Manetal2016}
{Man}, A.~W.~S., {Zirm}, A.~W., \& {Toft}, S. 2016, \apj, 830, 89

\bibitem[{{Matharu} {et~al.}(2019){Matharu}, {Muzzin}, {Brammer}, {van der
  Burg}, {Auger}, {Hewett}, {van der Wel}, {van Dokkum}, {Balogh}, {Chan},
  {Demarco}, {Marchesini}, {Nelson}, {Noble}, {Wilson}, \&
  {Yee}}]{Matharuetal2019}
{Matharu}, J., {Muzzin}, A., {Brammer}, G.~B., {et~al.} 2019, \mnras, 484, 595

\bibitem[{{Matharu} {et~al.}(2020){Matharu}, {Muzzin}, {Brammer}, {van der
  Burg}, {Auger}, {Hewett}, {Chan}, {Demarco}, {van Dokkum}, {Marchesini},
  {Nelson}, {Noble}, \& {Wilson}}]{Matharuetal2020}
---. 2020, \mnras, 493, 6011

\bibitem[{{McGee} {et~al.}(2009){McGee}, {Balogh}, {Bower}, {Font}, \&
  {McCarthy}}]{McGeeetal2009}
{McGee}, S.~L., {Balogh}, M.~L., {Bower}, R.~G., {Font}, A.~S., \& {McCarthy},
  I.~G. 2009, \mnras, 400, 937

\bibitem[{{McGee} {et~al.}(2014){McGee}, {Bower}, \& {Balogh}}]{McGeeetal2014}
{McGee}, S.~L., {Bower}, R.~G., \& {Balogh}, M.~L. 2014, \mnras, 442, L105

\bibitem[{{McPartland} {et~al.}(2016){McPartland}, {Ebeling}, {Roediger}, \&
  {Blumenthal}}]{McPartlandetal2016}
{McPartland}, C., {Ebeling}, H., {Roediger}, E., \& {Blumenthal}, K. 2016,
  \mnras, 455, 2994

\bibitem[{{Momcheva} {et~al.}(2016){Momcheva}, {Brammer}, {van Dokkum},
  {Skelton}, {Whitaker}, {Nelson}, {Fumagalli}, {Maseda}, {Leja}, {Franx},
  {Rix}, {Bezanson}, {Da Cunha}, {Dickey}, {F{\"o}rster Schreiber},
  {Illingworth}, {Kriek}, {Labb{\'e}}, {Ulf Lange}, {Lundgren}, {Magee},
  {Marchesini}, {Oesch}, {Pacifici}, {Patel}, {Price}, {Tal}, {Wake}, {van der
  Wel}, \& {Wuyts}}]{Momchevaetal2016}
{Momcheva}, I.~G., {Brammer}, G.~B., {van Dokkum}, P.~G., {et~al.} 2016, \apjs,
  225, 27

\bibitem[{{Moore} {et~al.}(1998){Moore}, {Lake}, \& {Katz}}]{Mooreetal1998}
{Moore}, B., {Lake}, G., \& {Katz}, N. 1998, \apj, 495, 139

\bibitem[{{Muzzin} {et~al.}(2009){Muzzin}, {Wilson}, {Yee}, {Hoekstra},
  {Gilbank}, {Surace}, {Lacy}, {Blindert}, {Majumdar}, {Demarco}, {Gardner},
  {Gladders}, \& {Lonsdale}}]{Muzzinetal2009a}
{Muzzin}, A., {Wilson}, G., {Yee}, H.~K.~C., {et~al.} 2009, \apj, 698, 1934

\bibitem[{{Muzzin} {et~al.}(2012){Muzzin}, {Wilson}, {Yee}, {Gilbank},
  {Hoekstra}, {Demarco}, {Balogh}, {van Dokkum}, {Franx}, {Ellingson}, {Hicks},
  {Nantais}, {Noble}, {Lacy}, {Lidman}, {Rettura}, {Surace}, \&
  {Webb}}]{Muzzinetal2012}
---. 2012, \apj, 746, 188

\bibitem[{{Muzzin} {et~al.}(2013{\natexlab{a}}){Muzzin}, {Marchesini},
  {Stefanon}, {Franx}, {Milvang-Jensen}, {Dunlop}, {Fynbo}, {Brammer},
  {Labb{\'e}}, \& {van Dokkum}}]{Muzzinetal2013c}
{Muzzin}, A., {Marchesini}, D., {Stefanon}, M., {et~al.} 2013{\natexlab{a}},
  \apjs, 206, 8

\bibitem[{{Muzzin} {et~al.}(2013{\natexlab{b}}){Muzzin}, {Marchesini},
  {Stefanon}, {Franx}, {McCracken}, {Milvang-Jensen}, {Dunlop}, {Fynbo},
  {Brammer}, {Labb{\'e}}, \& {van Dokkum}}]{Muzzinetal2013b}
---. 2013{\natexlab{b}}, \apj, 777, 18

\bibitem[{{Naab} {et~al.}(2009){Naab}, {Johansson}, \&
  {Ostriker}}]{Naabetal2009}
{Naab}, T., {Johansson}, P.~H., \& {Ostriker}, J.~P. 2009, \apjl, 699, L178

\bibitem[{{Newman} {et~al.}(2012){Newman}, {Ellis}, {Bundy}, \&
  {Treu}}]{Newmanetal2012}
{Newman}, A.~B., {Ellis}, R.~S., {Bundy}, K., \& {Treu}, T. 2012, \apj, 746,
  162

\bibitem[{{Oke} \& {Gunn}(1983)}]{OkeGunn1983}
{Oke}, J.~B., \& {Gunn}, J.~E. 1983, \apj, 266, 713

\bibitem[{{Olave-Rojas} {et~al.}(2018){Olave-Rojas}, {Cerulo}, {Demarco},
  {Jaff{\'e}}, {Mercurio}, {Rosati}, {Balestra}, \&
  {Nonino}}]{OlaveRojasetal2018}
{Olave-Rojas}, D., {Cerulo}, P., {Demarco}, R., {et~al.} 2018, \mnras, 479,
  2328

\bibitem[{{Old} {et~al.}(2020){Old}, {Balogh}, {van der Burg}, {Biviano},
  {Yee}, {Pintos-Castro}, {Webb}, {Muzzin}, {Rudnick}, {Vulcani}, {Poggianti},
  {Cooper}, {Zaritsky}, {Cerulo}, {Wilson}, {Chan}, {Lidman}, {McGee},
  {Demarco}, {Forrest}, {De Lucia}, {Gilbank}, {Kukstas}, {McCarthy},
  {Jablonka}, {Nantais}, {Noble}, {Reeves}, \& {Shipley}}]{Oldetal2020}
{Old}, L.~J., {Balogh}, M.~L., {van der Burg}, R. F.~J., {et~al.} 2020, \mnras,
  493, 5987

\bibitem[{{Padilla} \& {Strauss}(2008)}]{PadillaStrauss2008}
{Padilla}, N.~D., \& {Strauss}, M.~A. 2008, \mnras, 388, 1321

\bibitem[{{Papovich} {et~al.}(2018){Papovich}, {Kawinwanichakij}, {Quadri},
  {Glazebrook}, {Labb{\'e}}, {Tran}, {Forrest}, {Kacprzak}, {Spitler},
  {Straatman}, \& {Tomczak}}]{Papovichetal2018}
{Papovich}, C., {Kawinwanichakij}, L., {Quadri}, R.~F., {et~al.} 2018, \apj,
  854, 30

\bibitem[{{Peng} {et~al.}(2010){Peng}, {Lilly}, {Kova{\v c}}, {Bolzonella},
  {Pozzetti}, {Renzini}, {Zamorani}, {Ilbert}, {Knobel}, {Iovino}, {Maier},
  {Cucciati}, {Tasca}, {Carollo}, {Silverman}, {Kampczyk}, {de Ravel},
  {Sanders}, {Scoville}, {Contini}, {Mainieri}, {Scodeggio}, {Kneib}, {Le
  F{\`e}vre}, {Bardelli}, {Bongiorno}, {Caputi}, {Coppa}, {de la Torre},
  {Franzetti}, {Garilli}, {Lamareille}, {Le Borgne}, {Le Brun}, {Mignoli},
  {Perez Montero}, {Pello}, {Ricciardelli}, {Tanaka}, {Tresse}, {Vergani},
  {Welikala}, {Zucca}, {Oesch}, {Abbas}, {Barnes}, {Bordoloi}, {Bottini},
  {Cappi}, {Cassata}, {Cimatti}, {Fumana}, {Hasinger}, {Koekemoer},
  {Leauthaud}, {Maccagni}, {Marinoni}, {McCracken}, {Memeo}, {Meneux}, {Nair},
  {Porciani}, {Presotto}, \& {Scaramella}}]{Pengetal2010}
{Peng}, Y.-j., {Lilly}, S.~J., {Kova{\v c}}, K., {et~al.} 2010, \apj, 721, 193

\bibitem[{{Pintos-Castro} {et~al.}(2019){Pintos-Castro}, {Yee}, {Muzzin},
  {Old}, \& {Wilson}}]{PintosCastroetal2019}
{Pintos-Castro}, I., {Yee}, H.~K.~C., {Muzzin}, A., {Old}, L., \& {Wilson}, G.
  2019, \apj, 876, 40

\bibitem[{{Poggianti} {et~al.}(2013){Poggianti}, {Calvi}, {Bindoni},
  {D'Onofrio}, {Moretti}, {Valentinuzzi}, {Fasano}, {Fritz}, {De Lucia},
  {Vulcani}, {Bettoni}, {Gullieuszik}, \& {Omizzolo}}]{Poggiantietal2013}
{Poggianti}, B.~M., {Calvi}, R., {Bindoni}, D., {et~al.} 2013, \apj, 762, 77

\bibitem[{{Poggianti} {et~al.}(2016){Poggianti}, {Fasano}, {Omizzolo},
  {Gullieuszik}, {Bettoni}, {Moretti}, {Paccagnella}, {Jaff{\'e}}, {Vulcani},
  {Fritz}, {Couch}, \& {D'Onofrio}}]{Poggiantietal2016}
{Poggianti}, B.~M., {Fasano}, G., {Omizzolo}, A., {et~al.} 2016, \aj, 151, 78

\bibitem[{{Postman} {et~al.}(2005){Postman}, {Franx}, {Cross}, {Holden},
  {Ford}, {Illingworth}, {Goto}, {Demarco}, {Rosati}, {Blakeslee}, {Tran},
  {Ben{\'\i}tez}, {Clampin}, {Hartig}, {Homeier}, {Ardila}, {Bartko},
  {Bouwens}, {Bradley}, {Broadhurst}, {Brown}, {Burrows}, {Cheng}, {Feldman},
  {Golimowski}, {Gronwall}, {Infante}, {Kimble}, {Krist}, {Lesser}, {Martel},
  {Mei}, {Menanteau}, {Meurer}, {Miley}, {Motta}, {Sirianni}, {Sparks}, {Tran},
  {Tsvetanov}, {White}, \& {Zheng}}]{Postmanetal2005}
{Postman}, M., {Franx}, M., {Cross}, N.~J.~G., {et~al.} 2005, \apj, 623, 721

\bibitem[{{Pulsoni} {et~al.}(2018){Pulsoni}, {Gerhard}, {Arnaboldi}, {Coccato},
  {Longobardi}, {Napolitano}, {Moylan}, {Narayan}, {Gupta}, {Burkert},
  {Capaccioli}, {Chies-Santos}, {Cortesi}, {Freeman}, {Kuijken}, {Merrifield},
  {Romanowsky}, \& {Tortora}}]{Pulsonietal2018}
{Pulsoni}, C., {Gerhard}, O., {Arnaboldi}, M., {et~al.} 2018, \aap, 618, A94

\bibitem[{{Qu} {et~al.}(2017){Qu}, {Helly}, {Bower}, {Theuns}, {Crain},
  {Frenk}, {Furlong}, {McAlpine}, {Schaller}, {Schaye}, \&
  {White}}]{Quetal2017}
{Qu}, Y., {Helly}, J.~C., {Bower}, R.~G., {et~al.} 2017, \mnras, 464, 1659

\bibitem[{{Rix} \& {Zaritsky}(1995)}]{RixZaritsky1995}
{Rix}, H.-W., \& {Zaritsky}, D. 1995, \apj, 447, 82

\bibitem[{{Roberts} \& {Parker}(2020)}]{RobertsParker2020}
{Roberts}, I.~D., \& {Parker}, L.~C. 2020, \mnras, 495, 554

\bibitem[{{Rodriguez-Gomez} {et~al.}(2016){Rodriguez-Gomez}, {Pillepich},
  {Sales}, {Genel}, {Vogelsberger}, {Zhu}, {Wellons}, {Nelson}, {Torrey},
  {Springel}, {Ma}, \& {Hernquist}}]{RodriguezGomezetal2016}
{Rodriguez-Gomez}, V., {Pillepich}, A., {Sales}, L.~V., {et~al.} 2016, \mnras,
  458, 2371

\bibitem[{{Ryden}(2004)}]{Ryden2004}
{Ryden}, B.~S. 2004, \apj, 601, 214

\bibitem[{{Saglia} {et~al.}(2010){Saglia}, {S{\'a}nchez-Bl{\'a}zquez},
  {Bender}, {Simard}, {Desai}, {Arag{\'o}n-Salamanca}, {Milvang-Jensen},
  {Halliday}, {Jablonka}, {Noll}, {Poggianti}, {Clowe}, {De Lucia},
  {Pell{\'o}}, {Rudnick}, {Valentinuzzi}, {White}, \&
  {Zaritsky}}]{Sagliaetal2010}
{Saglia}, R.~P., {S{\'a}nchez-Bl{\'a}zquez}, P., {Bender}, R., {et~al.} 2010,
  \aap, 524, A6

\bibitem[{{Sandage} {et~al.}(1970){Sandage}, {Freeman}, \&
  {Stokes}}]{Sandageetal1970}
{Sandage}, A., {Freeman}, K.~C., \& {Stokes}, N.~R. 1970, \apj, 160, 831

\bibitem[{{Sersic}(1968)}]{Sersic1968}
{Sersic}, J.~L. 1968, {Atlas de galaxias australes}

\bibitem[{{Shankar} {et~al.}(2013){Shankar}, {Marulli}, {Bernardi}, {Mei},
  {Meert}, \& {Vikram}}]{Shankaretal2013}
{Shankar}, F., {Marulli}, F., {Bernardi}, M., {et~al.} 2013, \mnras, 428, 109

\bibitem[{{Sheen} {et~al.}(2017){Sheen}, {Smith}, {Jaff{\'e}}, {Kim}, {Yi},
  {Duc}, {Nantais}, {Candlish}, {Demarco}, \& {Treister}}]{Sheenetal2017}
{Sheen}, Y.-K., {Smith}, R., {Jaff{\'e}}, Y., {et~al.} 2017, \apjl, 840, L7

\bibitem[{{Sif{\'o}n} {et~al.}(2015){Sif{\'o}n}, {Hoekstra}, {Cacciato},
  {Viola}, {K{\"o}hlinger}, {van der Burg}, {Sand}, \&
  {Graham}}]{Sifonetal2015}
{Sif{\'o}n}, C., {Hoekstra}, H., {Cacciato}, M., {et~al.} 2015, \aap, 575, A48

\bibitem[{{Skelton} {et~al.}(2014){Skelton}, {Whitaker}, {Momcheva}, {Brammer},
  {van Dokkum}, {Labb{\'e}}, {Franx}, {van der Wel}, {Bezanson}, {Da Cunha},
  {Fumagalli}, {F{\"o}rster Schreiber}, {Kriek}, {Leja}, {Lundgren}, {Magee},
  {Marchesini}, {Maseda}, {Nelson}, {Oesch}, {Pacifici}, {Patel}, {Price},
  {Rix}, {Tal}, {Wake}, \& {Wuyts}}]{Skeltonetal2014}
{Skelton}, R.~E., {Whitaker}, K.~E., {Momcheva}, I.~G., {et~al.} 2014, \apjs,
  214, 24

\bibitem[{{Stalder} {et~al.}(2013){Stalder}, {Ruel}, {{\v S}uhada}, {Brodwin},
  {Aird}, {Andersson}, {Armstrong}, {Ashby}, {Bautz}, {Bayliss}, {Bazin},
  {Benson}, {Bleem}, {Carlstrom}, {Chang}, {Cho}, {Clocchiatti}, {Crawford},
  {Crites}, {de Haan}, {Desai}, {Dobbs}, {Dudley}, {Foley}, {Forman}, {George},
  {Gettings}, {Gladders}, {Gonzalez}, {Halverson}, {Harrington}, {High},
  {Holder}, {Holzapfel}, {Hoover}, {Hrubes}, {Jones}, {Joy}, {Keisler}, {Knox},
  {Lee}, {Leitch}, {Liu}, {Lueker}, {Luong-Van}, {Mantz}, {Marrone},
  {McDonald}, {McMahon}, {Mehl}, {Meyer}, {Mocanu}, {Mohr}, {Montroy},
  {Murray}, {Natoli}, {Nurgaliev}, {Padin}, {Plagge}, {Pryke}, {Reichardt},
  {Rest}, {Ruhl}, {Saliwanchik}, {Saro}, {Sayre}, {Schaffer}, {Shaw},
  {Shirokoff}, {Song}, {Spieler}, {Stanford}, {Staniszewski}, {Stark}, {Story},
  {Stubbs}, {van Engelen}, {Vanderlinde}, {Vieira}, {Vikhlinin}, {Williamson},
  {Zahn}, \& {Zenteno}}]{Stalderetal2013}
{Stalder}, B., {Ruel}, J., {{\v S}uhada}, R., {et~al.} 2013, \apj, 763, 93

\bibitem[{{Suess} {et~al.}(2019){Suess}, {Kriek}, {Price}, \&
  {Barro}}]{Suessetal2019a}
{Suess}, K.~A., {Kriek}, M., {Price}, S.~H., \& {Barro}, G. 2019, \apj, 877,
  103

\bibitem[{{Tacchella} {et~al.}(2015){Tacchella}, {Carollo}, {Renzini},
  {Schreiber}, {Lang}, {Wuyts}, {Cresci}, {Dekel}, {Genzel}, {Lilly},
  {Mancini}, {Newman}, {Onodera}, {Shapley}, {Tacconi}, {Woo}, \&
  {Zamorani}}]{Tacchellaetal2015}
{Tacchella}, S., {Carollo}, C.~M., {Renzini}, A., {et~al.} 2015, Science, 348,
  314

\bibitem[{{Tenneti} {et~al.}(2020){Tenneti}, {Kitching}, {Joachimi}, \& {Di
  Matteo}}]{Tennetietal2020}
{Tenneti}, A., {Kitching}, T.~D., {Joachimi}, B., \& {Di Matteo}, T. 2020,
  arXiv e-prints, arXiv:2002.12238

\bibitem[{{Tremblay} \& {Merritt}(1996)}]{TremblayMerritt1996}
{Tremblay}, B., \& {Merritt}, D. 1996, \aj, 111, 2243

\bibitem[{{Trujillo} {et~al.}(2006){Trujillo}, {F{\"o}rster Schreiber},
  {Rudnick}, {Barden}, {Franx}, {Rix}, {Caldwell}, {McIntosh}, {Toft},
  {H{\"a}ussler}, {Zirm}, {van Dokkum}, {Labb{\'e}}, {Moorwood},
  {R{\"o}ttgering}, {van der Wel}, {van der Werf}, \& {van
  Starkenburg}}]{Trujilloetal2006b}
{Trujillo}, I., {F{\"o}rster Schreiber}, N.~M., {Rudnick}, G., {et~al.} 2006,
  \apj, 650, 18

\bibitem[{{Valentinuzzi} {et~al.}(2010){Valentinuzzi}, {Poggianti}, {Saglia},
  {Arag{\'o}n-Salamanca}, {Simard}, {S{\'a}nchez-Bl{\'a}zquez}, {D'onofrio},
  {Cava}, {Couch}, {Fritz}, {Moretti}, \& {Vulcani}}]{Valentinuzzietal2010b}
{Valentinuzzi}, T., {Poggianti}, B.~M., {Saglia}, R.~P., {et~al.} 2010, \apjl,
  721, L19

\bibitem[{{van der Burg} {et~al.}(2018){van der Burg}, {McGee}, {Aussel},
  {Dahle}, {Arnaud}, {Pratt}, \& {Muzzin}}]{vanderBurgetal2018}
{van der Burg}, R. F.~J., {McGee}, S., {Aussel}, H., {et~al.} 2018, \aap, 618,
  A140

\bibitem[{{van der Burg} {et~al.}(2013){van der Burg}, {Muzzin}, {Hoekstra},
  {Lidman}, {Rettura}, {Wilson}, {Yee}, {Hildebrandt}, {Marchesini},
  {Stefanon}, {Demarco}, \& {Kuijken}}]{vanderBurgetal2013}
{van der Burg}, R.~F.~J., {Muzzin}, A., {Hoekstra}, H., {et~al.} 2013, \aap,
  557, A15

\bibitem[{{van der Burg} {et~al.}(2020){van der Burg}, {Rudnick}, {Balogh},
  {Muzzin}, {Lidman}, {Old}, {Shipley}, {Gilbank}, {McGee}, {Biviano},
  {Cerulo}, {Chan}, {Cooper}, {De Lucia}, {Demarco}, {Forrest}, {Gwyn},
  {Jablonka}, {Kukstas}, {Marchesini}, {Nantais}, {Noble}, {Pintos-Castro},
  {Poggianti}, {Reeves}, {Stefanon}, {Vulcani}, {Webb}, {Wilson}, {Yee}, \&
  {Zaritsky}}]{vanderBurgetal2020}
{van der Burg}, R. F.~J., {Rudnick}, G., {Balogh}, M.~L., {et~al.} 2020, \aap,
  638, A112

\bibitem[{{van der Wel} {et~al.}(2009){van der Wel}, {Rix}, {Holden}, {Bell},
  \& {Robaina}}]{vanderWeletal2009}
{van der Wel}, A., {Rix}, H.-W., {Holden}, B.~P., {Bell}, E.~F., \& {Robaina},
  A.~R. 2009, \apjl, 706, L120

\bibitem[{{van der Wel} {et~al.}(2012){van der Wel}, {Bell}, {H{\"a}ussler},
  {McGrath}, {Chang}, {Guo}, {McIntosh}, {Rix}, {Barden}, {Cheung}, {Faber},
  {Ferguson}, {Galametz}, {Grogin}, {Hartley}, {Kartaltepe}, {Kocevski},
  {Koekemoer}, {Lotz}, {Mozena}, {Peth}, \& {Peng}}]{vanderWeletal2012}
{van der Wel}, A., {Bell}, E.~F., {H{\"a}ussler}, B., {et~al.} 2012, \apjs,
  203, 24

\bibitem[{{van der Wel} {et~al.}(2014{\natexlab{a}}){van der Wel}, {Franx},
  {van Dokkum}, {Skelton}, {Momcheva}, {Whitaker}, {Brammer}, {Bell}, {Rix},
  {Wuyts}, {Ferguson}, {Holden}, {Barro}, {Koekemoer}, {Chang}, {McGrath},
  {H{\"a}ussler}, {Dekel}, {Behroozi}, {Fumagalli}, {Leja}, {Lundgren},
  {Maseda}, {Nelson}, {Wake}, {Patel}, {Labb{\'e}}, {Faber}, {Grogin}, \&
  {Kocevski}}]{vanderWeletal2014}
{van der Wel}, A., {Franx}, M., {van Dokkum}, P.~G., {et~al.}
  2014{\natexlab{a}}, \apj, 788, 28

\bibitem[{{van der Wel} {et~al.}(2014{\natexlab{b}}){van der Wel}, {Chang},
  {Bell}, {Holden}, {Ferguson}, {Giavalisco}, {Rix}, {Skelton}, {Whitaker},
  {Momcheva}, {Brammer}, {Kassin}, {Martig}, {Dekel}, {Ceverino}, {Koo},
  {Mozena}, {van Dokkum}, {Franx}, {Faber}, \& {Primack}}]{vanderWeletal2014b}
{van der Wel}, A., {Chang}, Y.-Y., {Bell}, E.~F., {et~al.} 2014{\natexlab{b}},
  \apjl, 792, L6

\bibitem[{{van Dokkum} \& {Franx}(2001)}]{vanDokkumetal2001}
{van Dokkum}, P.~G., \& {Franx}, M. 2001, \apj, 553, 90

\bibitem[{{van Dokkum} {et~al.}(2010){van Dokkum}, {Whitaker}, {Brammer},
  {Franx}, {Kriek}, {Labb{\'e}}, {Marchesini}, {Quadri}, {Bezanson},
  {Illingworth}, {Muzzin}, {Rudnick}, {Tal}, \& {Wake}}]{vanDokkumetal2010}
{van Dokkum}, P.~G., {Whitaker}, K.~E., {Brammer}, G., {et~al.} 2010, \apj,
  709, 1018

\bibitem[{{Veale} {et~al.}(2017){Veale}, {Ma}, {Greene}, {Thomas}, {Blakeslee},
  {McConnell}, {Walsh}, \& {Ito}}]{Vealeetal2017}
{Veale}, M., {Ma}, C.-P., {Greene}, J.~E., {et~al.} 2017, \mnras, 471, 1428

\bibitem[{{Villalobos} {et~al.}(2012){Villalobos}, {De Lucia}, {Borgani}, \&
  {Murante}}]{Villalobosetal2012}
{Villalobos}, {\'A}., {De Lucia}, G., {Borgani}, S., \& {Murante}, G. 2012,
  \mnras, 424, 2401

\bibitem[{{Vulcani} {et~al.}(2011){Vulcani}, {Poggianti}, {Dressler}, {Fasano},
  {Valentinuzzi}, {Couch}, {Moretti}, {Simard}, {Desai}, {Bettoni},
  {D'Onofrio}, {Cava}, \& {Varela}}]{Vulcanietal2011}
{Vulcani}, B., {Poggianti}, B.~M., {Dressler}, A., {et~al.} 2011, \mnras, 413,
  921

\bibitem[{{Wang} \& {Kauffmann}(2008)}]{WangKauffmannetal2008}
{Wang}, L., \& {Kauffmann}, G. 2008, \mnras, 391, 785

\bibitem[{{Webb} {et~al.}(2020){Webb}, {Balogh}, {Leja}, {van der Burg},
  {Rudnick}, {Muzzin}, {Boak}, {Cerulo}, {Gilbank}, {Lidman}, {Old},
  {Pintos-Castro}, {McGee}, {Shipley}, {Biviano}, {Chan}, {Cooper}, {De Lucia},
  {Demarco}, {Forrest}, {Jablonka}, {Kukstas}, {McCarthy}, {McNab}, {Nantais},
  {Noble}, {Poggianti}, {Reeves}, {Vulcani}, {Wilson}, {Yee}, \&
  {Zaritsky}}]{Webbetal2020}
{Webb}, K., {Balogh}, M.~L., {Leja}, J., {et~al.} 2020, \mnras, 498, 5317

\bibitem[{{Weijmans} {et~al.}(2014){Weijmans}, {de Zeeuw}, {Emsellem},
  {Krajnovi{\'c}}, {Lablanche}, {Alatalo}, {Blitz}, {Bois}, {Bournaud},
  {Bureau}, {Cappellari}, {Crocker}, {Davies}, {Davis}, {Duc}, {Khochfar},
  {Kuntschner}, {McDermid}, {Morganti}, {Naab}, {Oosterloo}, {Sarzi}, {Scott},
  {Serra}, {Verdoes Kleijn}, \& {Young}}]{Weijmansetal2014}
{Weijmans}, A.-M., {de Zeeuw}, P.~T., {Emsellem}, E., {et~al.} 2014, \mnras,
  444, 3340

\bibitem[{{Wetzel} {et~al.}(2012){Wetzel}, {Tinker}, \&
  {Conroy}}]{Wetzeletal2012}
{Wetzel}, A.~R., {Tinker}, J.~L., \& {Conroy}, C. 2012, \mnras, 424, 232

\bibitem[{{Wetzel} {et~al.}(2013){Wetzel}, {Tinker}, {Conroy}, \& {van den
  Bosch}}]{Wetzeletal2013}
{Wetzel}, A.~R., {Tinker}, J.~L., {Conroy}, C., \& {van den Bosch}, F.~C. 2013,
  \mnras, 432, 336

\bibitem[{{Whitaker} {et~al.}(2017){Whitaker}, {Bezanson}, {van Dokkum},
  {Franx}, {van der Wel}, {Brammer}, {F{\"o}rster-Schreiber}, {Giavalisco},
  {Labb{\'e}}, {Momcheva}, {Nelson}, \& {Skelton}}]{Whitakeretal2017}
{Whitaker}, K.~E., {Bezanson}, R., {van Dokkum}, P.~G., {et~al.} 2017, \apj,
  838, 19

\bibitem[{{Williams} {et~al.}(2009){Williams}, {Quadri}, {Franx}, {van Dokkum},
  \& {Labb{\'e}}}]{Williamsetal2009}
{Williams}, R.~J., {Quadri}, R.~F., {Franx}, M., {van Dokkum}, P., \&
  {Labb{\'e}}, I. 2009, \apj, 691, 1879

\bibitem[{{Wilson} {et~al.}(2009){Wilson}, {Muzzin}, {Yee}, {Lacy}, {Surace},
  {Gilbank}, {Blindert}, {Hoekstra}, {Majumdar}, {Demarco}, {Gardner},
  {Gladders}, \& {Lonsdale}}]{Wilsonetal2009}
{Wilson}, G., {Muzzin}, A., {Yee}, H.~K.~C., {et~al.} 2009, \apj, 698, 1943

\bibitem[{{Windhorst} {et~al.}(2011){Windhorst}, {Cohen}, {Hathi}, {McCarthy},
  {Ryan}, {Yan}, {Baldry}, {Driver}, {Frogel}, {Hill}, {Kelvin}, {Koekemoer},
  {Mechtley}, {O'Connell}, {Robotham}, {Rutkowski}, {Seibert}, {Straughn},
  {Tuffs}, {Balick}, {Bond}, {Bushouse}, {Calzetti}, {Crockett}, {Disney},
  {Dopita}, {Hall}, {Holtzman}, {Kaviraj}, {Kimble}, {MacKenty}, {Mutchler},
  {Paresce}, {Saha}, {Silk}, {Trauger}, {Walker}, {Whitmore}, \&
  {Young}}]{Windhorstetal2011}
{Windhorst}, R.~A., {Cohen}, S.~H., {Hathi}, N.~P., {et~al.} 2011, \apjs, 193,
  27

\bibitem[{{Yagi} {et~al.}(2015){Yagi}, {Gu}, {Koyama}, {Nakata}, {Kodama},
  {Hattori}, \& {Yoshida}}]{Yagietal2015}
{Yagi}, M., {Gu}, L., {Koyama}, Y., {et~al.} 2015, \aj, 149, 36

\bibitem[{{Zhang} {et~al.}(2019){Zhang}, {Primack}, {Faber}, {Koo}, {Dekel},
  {Chen}, {Ceverino}, {Chang}, {Fang}, {Guo}, {Lin}, \& {Wel}}]{Zhangetal2019}
{Zhang}, H., {Primack}, J.~R., {Faber}, S.~M., {et~al.} 2019, \mnras, 484, 5170

\bibitem[{{Zolotov} {et~al.}(2015){Zolotov}, {Dekel}, {Mandelker}, {Tweed},
  {Inoue}, {DeGraf}, {Ceverino}, {Primack}, {Barro}, \&
  {Faber}}]{Zolotovetal2015}
{Zolotov}, A., {Dekel}, A., {Mandelker}, N., {et~al.} 2015, \mnras, 450, 2327

\end{thebibliography}

\appendix

\section{A. Completeness and purity of the cluster member selection}       
\label{app:Completeness and purity of the cluster member selection}
\begin{figure}
  \centering
  \includegraphics[scale=0.57]{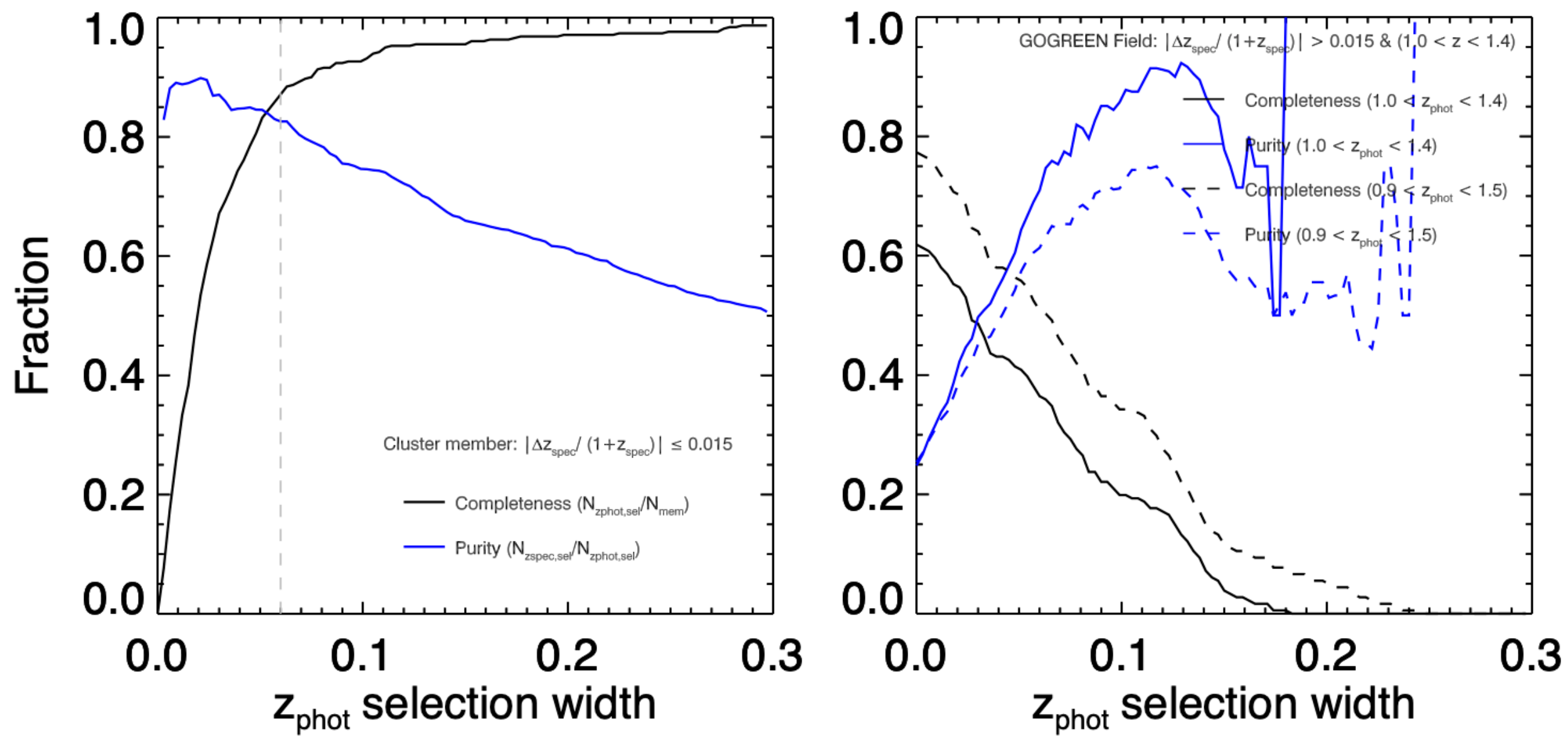}
  \caption{Completeness and purity of the sample as a function of the photometric redshift selection width in units of $\Delta z_{\rm{phot}}/(1+z_{\rm{phot}})$. Left: cluster member selection. Both completeness and purity are a strong function of the selection width. To optimize both the completeness and purity we define photometric cluster members as those with $|\Delta z_{\rm{phot}}/(1+z_{\rm{phot}})| \leq 0.06$ (gray dashed line). Right: photometric field galaxy selection using GOGREEN data. Since GOGREEN targets massive overdensities, it is difficult to select a clean photometric sample, due to contamination from cluster members. We would either end up with a low completeness sample or a sample with low purity. For this reason we choose to draw our field sample from CANDELS and 3D-HST, instead.}
  \label{fig_zselection_plot}
\end{figure}

Here we expand on the discussion in Section~\ref{subsec:Cluster membership and sample selection} about how we choose the selection criteria for cluster members.  The left panel of Figure~\ref{fig_zselection_plot} shows the completeness and purity of the resultant cluster sample as a function of the width of the photometric redshift selection (in units of $\Delta z_{\rm{phot}}/(1+z_{\rm{phot}})$ around the mean redshift of the clusters).  For simplicity, cluster members are defined as those with $|\Delta z_{\rm{spec}}/(1+z_{\rm{spec}})| \leq 0.015$ and above the mass limit of $ \log(M / {\rm M}_{\odot}) \geq 9.5$ in this test.  Note that the validity of this test is partially based on the fact that the GOGREEN spectroscopic sample is a representative subset of the photometrically selected galaxy population, for galaxies with stellar mass $ \log(M / {\rm M}_{\odot}) \geq 10.3$. The completeness and purity are a strong function of the $z_{\rm{phot}}$ selection width. Increasing this box width results in higher sample completeness at the expense of the sample purity, as the sample contains more galaxies that are photometrically selected as cluster members but are interlopers. To optimize both the completeness and purity we define photometric cluster members as those with $|\Delta z_{\rm{phot}}/(1+z_{\rm{phot}})| \leq 0.06$, which gives a completeness of $\sim85\%$ and a purity of $\sim80\%$.

As we mentioned in Section~\ref{subsec:Field comparison sample}, the number of spectroscopically confirmed GOGREEN field galaxies within the FOV of the \textit{HST} image is too small for a morphology comparison.  While, in theory, we can expand this sample with $z_{\rm{phot}}$ like the cluster members, we find that there are no good selection criteria to do so.  The right panel of Figure~\ref{fig_zselection_plot} shows the completeness and purity of the ``field'' sample as a function of the photometric redshift selection width.  The genuine field galaxies that can be used for the cluster vs. field comparison are defined as those with $|\Delta z_{\rm{spec}}/(1+z_{\rm{spec}})| > 0.015$ and are within a redshift range of $1.0 < z_{\rm{spec}} <1.4$.  In this case, the selection width refers to the width of the redshift region around the mean redshift of the cluster that was \textit{avoided} while selecting field galaxies.  Increasing the width means we select galaxies that are further from the cluster redshift, resulting in a higher purity but a less complete field sample.  Hence the trend in completeness and purity are opposite to those shown in the left panel.  Overall, the completeness (and therefore size of the resultant sample) drops sharply with the selection width.  Even imposing a wider photometric redshift cut ($0.9 < z_{\rm{phot}} <1.5$), we either end up with a low completeness sample (i.e., small sample) or a sample with low purity.  For example, using the $\Delta z_{\rm{phot}}/(1+z_{\rm{phot}}) \leq 0.06$ cut as done for the cluster member selection will result in a low completeness of $\sim50\%$ and a purity of $\sim60\%$.  We therefore use galaxies in the CANDELS/3D-\textit{HST} survey as our field comparison sample.

\section{B. Details of the simulation and the biases of the axis ratio measurements}      
\label{app:Details of the simulation and the biases of the axis ratio measurements}
In this section we describe the set-up and the result of the simulation we used to characterise the biases of our measurements.

We use a similar method as described in \citet{Chanetal2016, Chanetal2018}.  In this work, we generate a set of $20000$ simulated galaxies, uniformly distributed within a magnitude range of $19.0 \leq  $F160W$ \leq 25.0$. Each galaxy has its surface brightness profiles described by a random S\'ersic profile. The input structural parameters used to generate these profiles are taken from the parameter distribution of galaxies in the field at a similar redshift range as the GOGREEN clusters \citep{vanderWeletal2014} to mimic real galaxies. These galaxies are then convolved with the appropriate PSF and injected, 20 at a time, to random locations in the sky region of the F160W image. To avoid direct overlap with existing sources on the image, the segmentation maps from \textsc{SExtractor} are used as a reference of the sky region. We then run these 1000 images through our GALAPAGOS setup to recover the structural parameters of the simulated galaxies. The simulations were run on the images of five different clusters with a range of richness. We verified that the biases and uncertainties we obtain do not depend on the cluster used. 

The simulations are used to refine the GALAPAGOS configuration parameters. Using the bias between the input structural parameters of the simulated galaxies and the recovered ones, we then modify the GALAPAGOS configuration parameters and rerun the simulation. Through iterating this process for a few times we optimize our configuration setup by minimizing the biases in the recovered parameters.

\begin{figure*}
  \centering
  \includegraphics[scale=0.625]{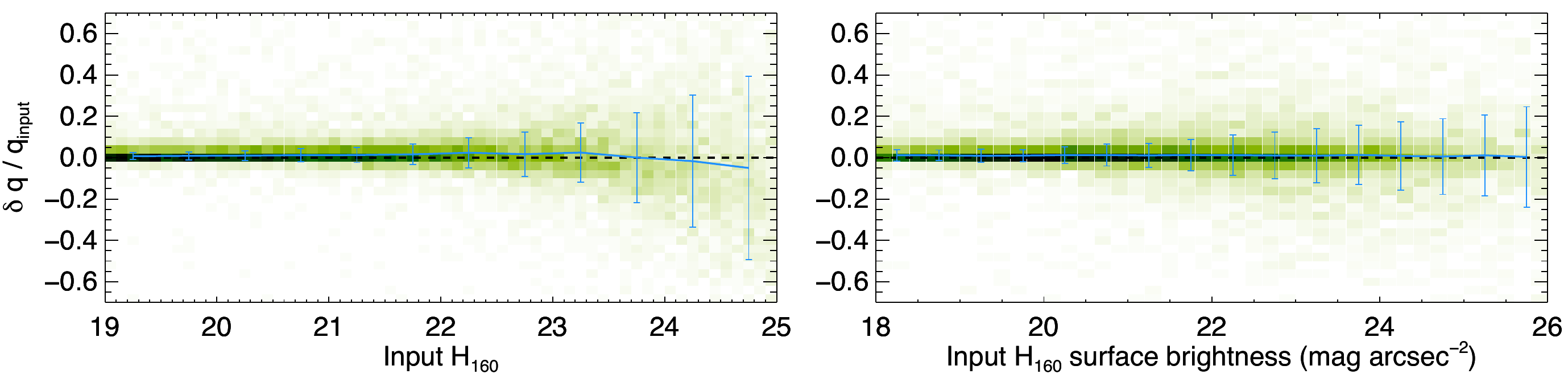}
  \caption{Fractional differences between recovered and input structural parameters as function of the input F160W mag (left) and surface brightness of the simulated galaxies (right). The blue line and the error bars correspond to the median and $1\sigma$ dispersion in different bins (0.5 mag / mag arcsec$^{-2}$ bin width). Green-shaded 2D histogram shows the number density distribution of the simulated galaxies.}
  \label{fig_bias_q_plot}
\end{figure*}

Using the simulated galaxies, we also characterize the potential biases for the derived structural parameters. The biases are derived as a function of input magnitudes, $n$, and $\log(R_e)$. We find that the biases are not only a strong function of magnitude (i.e., S/N) but also depend on $n$ and $R_e$ (albeit more weakly). Galaxies with higher $n$ at a given magnitude show higher biases and larger uncertainties \citep[see also][for discussions of these second-order effects]{Haussleretal2007, vanderWeletal2012}.  Overall the biases are only significant at faint magnitudes.  For example, the S\'ersic index $n$, the parameter that is hardest to constrain, shows an average bias and dispersion of $\sim13\%$ and $\sim24\%$ at F160W $=23$ (AB), which corresponds roughly to $\log(M / {\rm M}_{\odot}) \sim 9.5$. The effective radius $R_e$, on the other hand, show an average bias and dispersion of $\sim6\%$ and $\sim27\%$ at the same magnitude. The axis ratio $q$ shows an average bias and dispersion of $\sim2\%$ and $\sim 12\%$ at the same magnitude. Figure~\ref{fig_bias_q_plot} shows the fractional difference between the recovered and input $q$ as a function of F160W magnitude and surface brightness of the simulated galaxies. Among the three S\'ersic structural parameters, the axis ratio $q$ can typically be measured with the highest accuracy. These bias relations can be used to bias correct the structural parameter measurement for individual galaxies. Nevertheless, since the biases for $q$ is almost negligible, we have not applied the bias corrections we derived from the simulated galaxies on the cluster sample.

\section{C. Axis ratio comparison between van der Wel et al. 2014a and this work}      
\label{app:Axis ratio comparison between van der Wel et al. 2014 and this work}
The axis ratio measurements of the field sample used in this work are taken from the structural parameter catalogue of \citet{vanderWeletal2014}.  Although the \citet{vanderWeletal2014} measurements are derived with an overall consistent method using GALAPAGOS, different setups and treatment of the images may induce a bias in the cluster and field comparison. To ensure our measurements are compatible with  \citet{vanderWeletal2014}, we apply our methodology described in Section~\ref{subsec:Structural parameters} to galaxies in the redshift range of $0.9 < z < 1.5$ in the CANDELS imaging.

\begin{figure}
  \centering
  \includegraphics[scale=0.55]{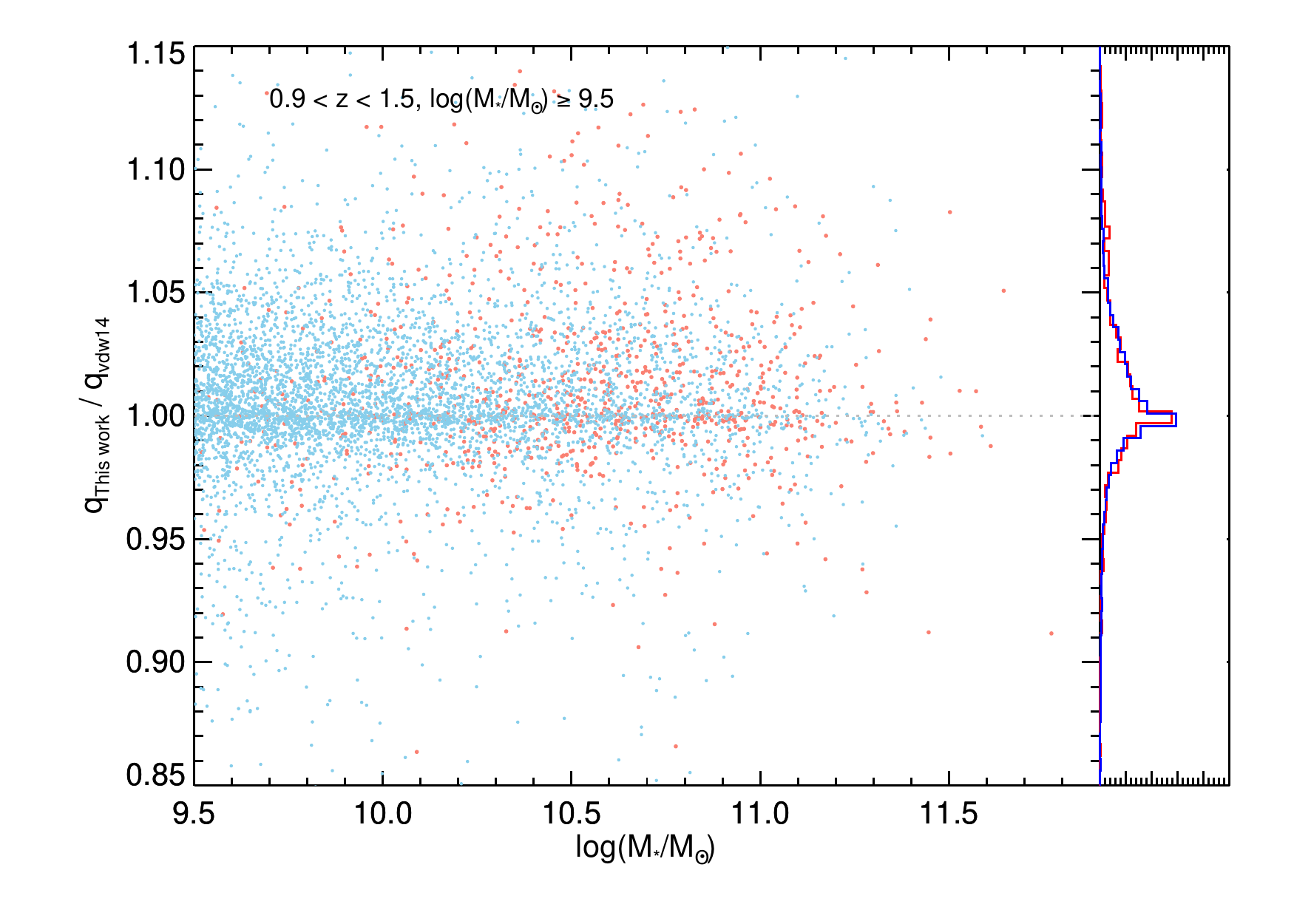}
  \caption{Comparison of axis ratio derived using the method used in this work with the \citet{vanderWeletal2014} measurements as a function of mass. Blue points correspond to star-forming galaxies and Red points correspond to the quiescent ones. The median ratio and $1\sigma$ between our measurements and \citet{vanderWeletal2014} are $1.00 \pm 0.03$ down to $\log(M_{*} /{\rm M}_{\odot}) \sim 9.5$. Both galaxy types show similar ratio distributions, as shown in the histogram on the right.}
  \label{fig_field_q_comparison}
\end{figure}

Figure~\ref{fig_field_q_comparison} shows the result of the comparison as a function of mass. Overall, our derived axis ratios are very consistent with those measured by \citet{vanderWeletal2014}.  There are a total of 6225 galaxies that have a good structural fit in both our measurements and those by \citet{vanderWeletal2014} down to $\log(M_{*} /{\rm M}_{\odot}) \sim 9.5$, the mass limit of this work. The entire sample has a median $q$ ratio \citep[ours /][]{vanderWeletal2014} and $1\sigma$ of $1.00 \pm 0.03$ (0\% bias and 3\% scatter). We also split the sample into $UVJ$ star-forming and quiescent galaxies to check if the ratio differences depend on galaxy types. The blue and red histograms in Figure~\ref{fig_field_q_comparison} show the distributions of the $q$ ratios. We find that both galaxy types show consistent ratio distributions.

The bias between the axis ratios derived using our methodologies and those by \citet{vanderWeletal2014} is small. It is even smaller than the average bias of $q$ in the cluster sample compared to simulated galaxies. Therefore, this is not a major source of uncertainty in our results. On the other hand, we note that the biases of $n$ and $R_{e}$ between our measurements and \citet{vanderWeletal2014} are non-negligible. Since we have not compared $n$ and $R_{e}$ between the clusters and the field in this work, we defer the discussion on these biases to an upcoming paper on the mass-size relations of the two samples (Chan et al. in prep).

\section{D. Results of the axis ratio distribution modeling}      
\label{app:Fitting results in the four mass bins}
\begin{figure*}
  \centering
  \includegraphics[scale=1.2]{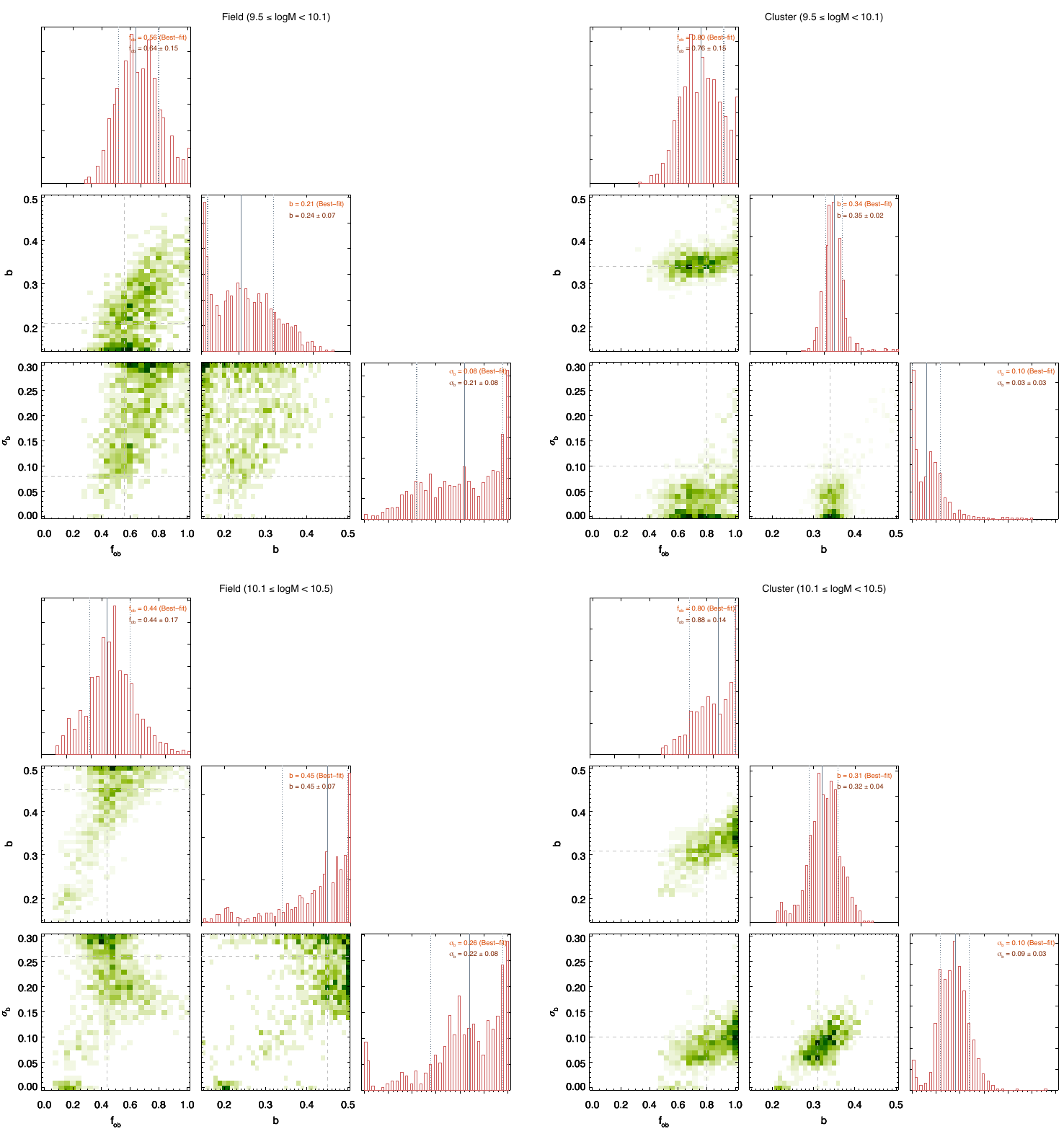}
  \caption{Corner plots of the Case II model fitting of the bootstrapped cluster and field sample in the four mass bins. The color of the green shade in each panel represents the distribution of the bootstrap results. The grey dashed lines correspond to the best-fit parameter values derived from the original sample. The vertical lines in the histogram correspond to the 16th, 50th, and the 84th percentile of the distribution.}
  \label{fig_corner_1}
\end{figure*}

\begin{figure*}
  \centering
  \includegraphics[scale=1.2]{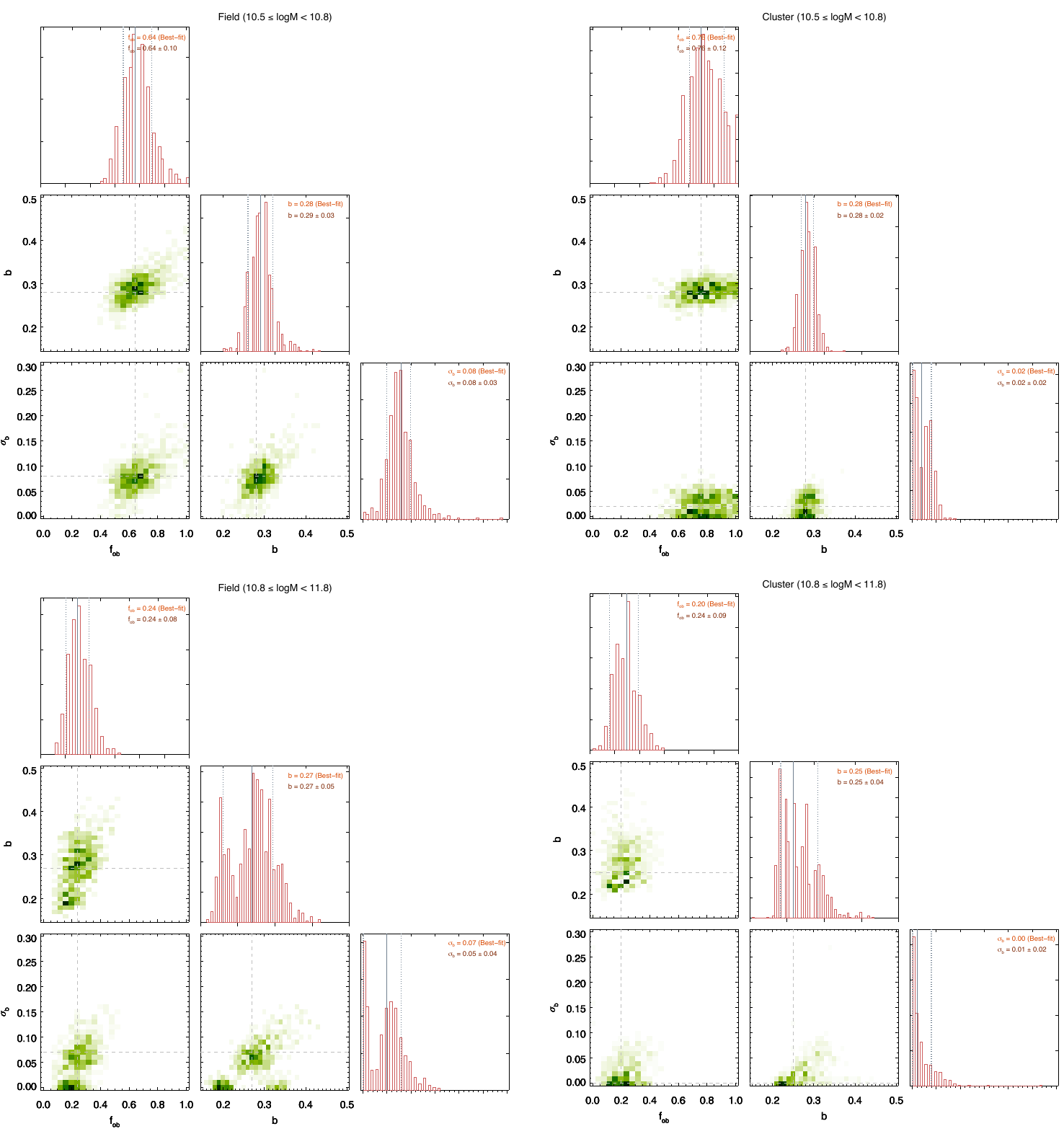}
  \caption{Corner plots of the Case II model fitting of the bootstrapped cluster and field sample in the four mass bins (cont'd).}
  \label{fig_corner_2}
\end{figure*}

Here we expand on Section~\ref{subsec:Reconstructing the intrinsic shapes from the projected axis ratio distributions} and give more details on the fitting results.  Figure~\ref{fig_corner_1} and~\ref{fig_corner_2} show the corner plots of the model fitting of the bootstrapped sample for the Case II models, where $f_{\rm{ob}}$, $b$, $\sigma_b$ are left as free parameters. The best-fit parameter values, and the median and $1\sigma$ values of the parameters derived from the bootstrapped sample are provided in the histograms. We can see that the $b$ and $\sigma_b$ parameters for the field sample in the two lower mass bins are poorly constrained. 

For the case II models we have assumed the values of the four parameters ($E, \sigma_E, T, \sigma_T$) to be the same as the best-fit values in \citet{Changetal2013}. However, we find that our results are not very sensitive to these parameters. For example, we tested that using the initial parameters for the $10.5 \leq \log(M / {\rm M}_{\odot}) < 10.8$ bin to fit the $10.1 \leq \log(M / {\rm M}_{\odot}) < 10.5$ bin gives consistent results. The best-fit parameters for the bootstrapped cluster sample are $f_{\rm{ob}} = 0.88 \pm 0.15$, $b = 0.33 \pm 0.04$, $\sigma_b = 0.09 \pm 0.03$, which is completely consistent with the result using the ``correct'' initial parameters.

\section{E. Intrinsic alignments in cluster galaxies}       
\label{app:Intrinsic alignments in cluster galaxies}
The axis ratio distribution modeling used in this work relies on the assumption that the galaxy population is observed from random viewing angles, which breaks down when the galaxies are not randomly oriented,  i.e. aligned with respect to a certain direction. The alignment of, or between, galaxies within a halo has been a subject of heated interest \citep[See][for a review]{Joachimietal2015}. There are numerous types of intra-halo galaxy alignment. In this section we focus on the alignment of the shape of the satellite galaxies, which is the only one that will impact the observed axis ratio distribution.

In numerical simulations, it is established that major axes of satellite galaxies are preferentially oriented towards the centre of mass of the halo or the central galaxy \citep[e.g.,][]{Faltenbacheretal2008, Knebeetal2020, Tennetietal2020}. The strength of this radial alignment is expected to be strongest at the cluster core and to decrease with cluster-centric radius. Observationally whether this alignment exists is still a matter of debate.  \citet{Sifonetal2015} measured the alignments of galaxies in 90 clusters at $0.05 < z < 0.55$ and detected no alignments out to $3R_{200}$. On the other hand, more recent works by \citet{Huangetal2018} and \citet{Georgiouetal2019} reported a radial alignment signal that is stronger in satellites with smaller distance to the cluster / group center, although the measured signal is a few times smaller than predicted in the simulations. The alignment studies are limited to low-redshift clusters. Here we would like to examine the radial alignment signal in our sample.  

We follow the procedure outlined in \citet{Sifonetal2015} to measure the alignment of the galaxies with respect to the center of the cluster. In this work, the center of each cluster is taken to be the location of the BCG. To quantify the alignment signal, we adopt the commonly-used ellipticity components $\epsilon_+$ and $\epsilon_{\times}$, which are defined as:

\begin{equation}
 \label{eqt:epsilon_pluscross}
 \begin{aligned}
   \epsilon_+ =  \epsilon_1 \cos 2 \theta + \epsilon_2 \sin 2 \theta , \\
   \epsilon_\times =  \epsilon_1 \sin 2 \theta - \epsilon_2 \cos 2 \theta
 \end{aligned}
\end{equation}
where $\epsilon_1$ and $\epsilon_2$ are the galaxy ellipticities in the Cartesian frame and $\theta$ is the azimuthal angle of the individual galaxy with respect to the BCG of the cluster. The ellipticity $\epsilon_1$ measures the ellipticity in the RA and Dec directions and $\epsilon_2$ in diagonal directions. The two quantities are related to the axis ratio $q$ through the following:

\begin{equation}
 \label{eqt:epsilon_12}
 \begin{aligned} 
   \epsilon_1 =  \frac{1-q}{1+q} \cos 2 \phi , \\
   \epsilon_2 =  \frac{1-q}{1+q} \sin 2 \phi
 \end{aligned}
\end{equation}
where the angle $\phi$ is the position angle of the major axis of the galaxy.  The two equations~\ref{eqt:epsilon_pluscross} show that $\epsilon_+$ and $\epsilon_{\times}$ are rotated to the frame with one axis pointing towards the radial direction from the cluster center. A positive (negative) $\epsilon_+$ therefore indicates a radial (tangential) alignment of the galaxies towards the center of the cluster. On the other hand, $\epsilon_{\times}$ measures the shape alignment at $\pm 45^{\circ}$ from the radial direction. This component is commonly used as a check for systematic effects, as $\epsilon_{\times}$ should be consistent with zero due to symmetry.

\begin{figure}
  \centering
  \includegraphics[scale=0.525]{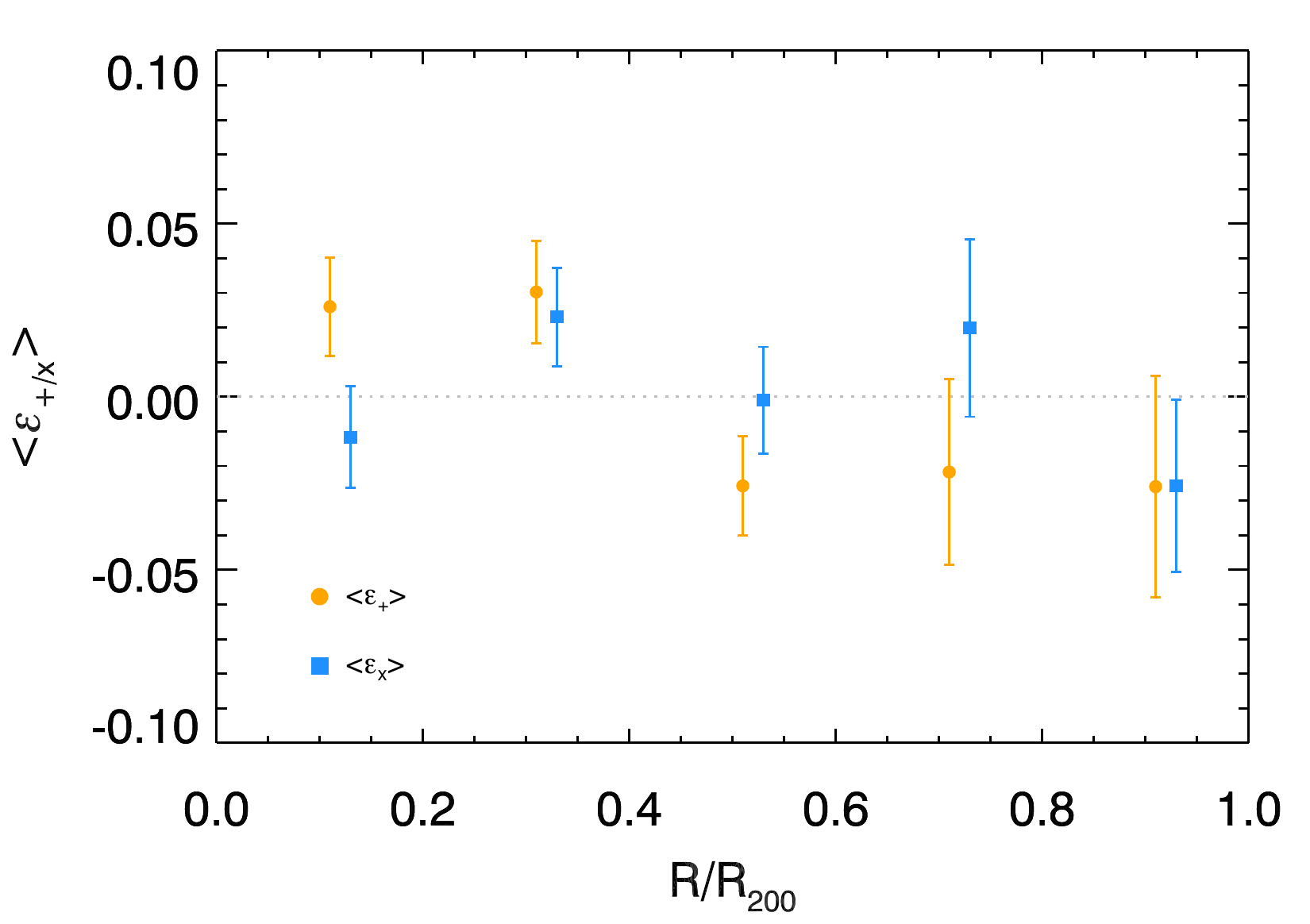}      
  \caption{Average alignments $\langle\epsilon_+\rangle$ and $\langle\epsilon_{\times}\rangle$ of the galaxies in GOGREEN clusters. The circles corresponds to $ \epsilon_+ $, while squares correspond to $ \epsilon_{\times} $. The error bars correspond to the standard error of the mean. Both alignment signals are consistent with zero.}
    \label{fig_eplusecross}
\end{figure}

Figure~\ref{fig_eplusecross} shows the average alignment of our cluster sample ($\langle\epsilon_+\rangle$ and $\langle\epsilon_{\times}\rangle$) as a function of cluster-centric radius. Given the field of view of the \textit{HST} images we can measure the alignment only out to $1R_{200}$.  We find that the average radial/tangential alignment for the entire cluster sample within $1R_{200}$ is consistent with zero, with $\langle\epsilon_+\rangle = -0.0011 \pm 0.0079$.  The cross component is also consistent with zero, with $\langle\epsilon_{\times}\rangle = 0.0018 \pm 0.0077$. Examining at the alignment signal as a function of radius, there is a weak evidence that the average radial alignment is positive in the region close to the cluster center ($\sim1.4 \sigma$ for $R< 0.2 R_{200}$).       

Since the alignment is expected to be strongest in regions close to the center, we repeat our axis ratio distribution analysis excluding galaxies in the $R< 0.2 R_{200}$ region and find that it does not affect our conclusion. The exclusion only results in an increase in uncertainty due to lower number statistics. The potential intrinsic alignments in the sample are therefore unlikely to affect our results.

\end{document}